%% file: main.tex
\documentclass[acmsmall]{acmart}

\AtBeginDocument{%
}

\input{packages}

\input{commands}

\begin{document}
\title{Diamond: End-to-End Forward-secure and Compact Authenticated Encryption for Internet of Things}
 
\author{Saif E. Nouma}
\email{saifeddinenouma@usf.edu}
\author{Gokhan Mumcu}
\email{mumcu@usf.edu}
\author{Attila A. Yavuz}
\email{attilaayavuz@usf.edu}
\affiliation{%
		\institution{University of South Florida}
		\streetaddress{3720 Spectrum Blvd, Interdisciplinary Research Building (IDR)-400}
		\city{Tampa}
		\state{Florida}
		\country{USA}
		\postcode{33612}
	}

\input{abstract.tex}
\maketitle

\input{introduction.tex}

\input{preliminaries.tex}
\input{proposed_schemes.tex}

\input{perf_analysis.tex}

\input{security_analysis.tex}

\input{conclusion.tex}

\section*{Acknowledgments}
   This work was supported by both Army Research Laboratory W911NF-24-2-0078 and National Science Foundation NSF CNS-2350213. The views and conclusions contained in this document are those of the authors and should not be interpreted as representing the official policies, either expressed or implied, of the Army Research Laboratory or the U.S. Government. The U.S. Government is authorized to reproduce and distribute reprints for Government purposes notwithstanding any copyright notation herein.

\input{appendix}
\bibliographystyle{ACM-Reference-Format} 
\bibliography{ref}

\end{document}

%% file: packages.tex
\usepackage{graphicx}
\usepackage{xcolor}
\usepackage{threeparttable}
\usepackage{multirow}
\usepackage{makecell} 
\usepackage{multicol}
\usepackage{caption}
\usepackage{subcaption}
\usepackage{enumerate}

\usepackage{adjustbox}
\usepackage{pifont}

\usepackage[noend]{algpseudocode}
\usepackage{url}
\usepackage{amsthm}

\usepackage{tabularx} 
\usepackage[normalem]{ulem}
\usepackage{xspace}
\usepackage{calc}
\usepackage{url}
\usepackage{booktabs}
\hypersetup{
	citecolor=green
}
\usepackage{wrapfig}
\usepackage{csquotes}
\usepackage{xparse}
\usepackage{listings}
\usepackage[most]{tcolorbox}

\usepackage{tikz}
\usetikzlibrary{arrows.meta, positioning, calc, decorations.pathreplacing}

\usetikzlibrary{shapes.symbols, shapes.geometric, positioning, arrows.meta}

\tikzset{
	drone/.pic={
		\fill[gray!40] (0,0) circle (3pt);
		\foreach \x/\y in {0.25/0.25, -0.25/0.25, -0.25/-0.25, 0.25/-0.25}
		\draw[gray!70, fill=gray!20] (\x,\y) circle (4pt);
	}
}

%% file: commands.tex
\newtheorem{theorem}{Theorem}

\newcommand{\Ra}{\ensuremath \stackrel{\$}{\leftarrow}{\xspace}}
\newcommand{\as}{\ensuremath {\leftarrow}{\xspace}}

\newcommand{\romanum}[1]{\uppercase\expandafter{\romannumeral #1}}

\newcommand{\prf}{\ensuremath {\texttt{PRF}}{\xspace}}
\newcommand{\prg}{\ensuremath {\texttt{PRG}}{\xspace}}
\newcommand{\fprg}{\ensuremath {\texttt{FPRG}}{\xspace}}
\newcommand{\fprgkg}{\ensuremath {\texttt{FPRG.Kg}}{\xspace}}
\newcommand{\fprgupd}{\ensuremath {\texttt{FPRG.Upd}}{\xspace}}
\newcommand{\umac}{\ensuremath {\texttt{UMAC}}{\xspace}}

\newcommand{\uhash}{\ensuremath {\texttt{UH}}{\xspace}}

\newcommand{\kg}{\ensuremath {\texttt{Kg}}{\xspace}}
\newcommand{\sign}{\ensuremath {\texttt{Sign}}{\xspace}}
\newcommand{\ver}{\ensuremath {\texttt{Ver}}{\xspace}}
\newcommand{\aver}{\ensuremath {\texttt{AVer}}{\xspace}}
\newcommand{\enc}{\ensuremath {\texttt{Enc}}{\xspace}}
\newcommand{\dec}{\ensuremath {\texttt{Dec}}{\xspace}}

\newcommand{\upd}{\ensuremath {\texttt{Upd}}{\xspace}}
\newcommand{\agg}{\ensuremath {\texttt{Agg}}{\xspace}}
\newcommand{\authenc}{\ensuremath {\texttt{AuthEnc}}{\xspace}}
\newcommand{\authenclr}{\ensuremath {\texttt{AuthEnc-LR}}{\xspace}}
\newcommand{\averdec}{\ensuremath {\texttt{AverDec}}{\xspace}}

\newcommand{\mac}{\texttt {MAC}{\xspace}}

\newcommand{\mackg}{\ensuremath {\texttt{MAC.Kg}}{\xspace}}
\newcommand{\macsign}{\ensuremath {\texttt{MAC.Sign}}{\xspace}}
\newcommand{\macver}{\ensuremath {\texttt{MAC.Ver}}{\xspace}}
\newcommand{\amac}{\ensuremath {\texttt{AMAC}}{\xspace}}

\newcommand{\famac}{\ensuremath {\texttt{FAMAC}}{\xspace}}
\newcommand{\famackg}{\ensuremath {\texttt{FAMAC.Kg}}{\xspace}}
\newcommand{\famacsign}{\ensuremath {\texttt{FAMAC.Sign}}{\xspace}}
\newcommand{\famacupd}{\ensuremath {\texttt{FAMAC.Upd}}{\xspace}}
\newcommand{\famacagg}{\ensuremath {\texttt{FAMAC.Agg}}{\xspace}}

\newcommand{\famacaver}{\ensuremath {\texttt{FAMAC.AVer}}{\xspace}}

\newcommand{\encscheme}{\ensuremath {\texttt{SE}}{\xspace}}
\newcommand{\enckg}{\ensuremath {\texttt{SE.Kg}}{\xspace}}
\newcommand{\encenc}{\ensuremath {\texttt{SE.Enc}}{\xspace}}
\newcommand{\encdec}{\ensuremath {\texttt{SE.Dec}}{\xspace}}
\newcommand{\encenclr}{\ensuremath {\texttt{SE.Enc-LR}}{\xspace}}
\newcommand{\fenc}{\ensuremath {\texttt{FSE}}{\xspace}}
\newcommand{\fenckg}{\ensuremath {\texttt{FSE.Kg}}{\xspace}}
\newcommand{\fencenc}{\ensuremath {\texttt{FSE.Enc}}{\xspace}}
\newcommand{\fencdec}{\ensuremath {\texttt{FSE.Dec}}{\xspace}}
\newcommand{\fencupd}{\ensuremath {\texttt{FSE.Upd}}{\xspace}}
\newcommand{\fencenclr}{\ensuremath {\texttt{FSE.Enc-LR}}{\xspace}}
\newcommand{\se}{\ensuremath {\texttt{SE}}{\xspace}}

\newcommand{\fse}{\ensuremath {\texttt{FSE}}{\xspace}}

\newcommand{\fseenc}{\ensuremath {\texttt{FSE.Enc}}{\xspace}}
\newcommand{\fsedec}{\ensuremath {\texttt{FSE.Dec}}{\xspace}}
\newcommand{\fseupd}{\ensuremath {\texttt{FSE.Upd}}{\xspace}}

\newcommand{\aee}{\ensuremath {\texttt{AE}}{\xspace}}

\newcommand{\faae}{\ensuremath {\texttt{FAAE}}{\xspace}}
\newcommand{\faaekg}{\ensuremath {\texttt{FAAE}\texttt{.Kg}}{\xspace}}
\newcommand{\faaeupd}{\ensuremath {\texttt{FAAE}\texttt{.Upd}}{\xspace}}
\newcommand{\faaeagg}{\ensuremath {\texttt{FAAE}\texttt{.Agg}}{\xspace}}
\newcommand{\faaeauthenc}{\ensuremath {\texttt{FAAE}\texttt{.AuthEnc}}{\xspace}}
\newcommand{\faaeverdec}{\ensuremath {\texttt{FAAE}\texttt{.AVerDec}}{\xspace}}

\newcommand{\faaea}{\ensuremath {\texttt{FAAE}_1}{\xspace}}
\newcommand{\faaeb}{\ensuremath {\texttt{FAAE}_2}{\xspace}}

\newcommand{\A}{\ensuremath {\mathcal{A}}{\xspace}}
\newcommand{\B}{\ensuremath {\mathcal{B}}{\xspace}}
\newcommand{\eucma}{\ensuremath {\texttt{EU-CMA}}{\xspace}}
\newcommand{\faeucma}{\ensuremath {\texttt{FA-EU-CMA}}{\xspace}}
\newcommand{\indcpa}{\ensuremath {\texttt{IND-CPA}}{\xspace}}
\newcommand{\indcca}{\ensuremath {\texttt{IND-CCA}}{\xspace}}
\newcommand{\intctxt}{\ensuremath {\texttt{INT-CTXT}}{\xspace}}
\newcommand{\findcpa}{\ensuremath {\texttt{F-IND-CPA}}{\xspace}}

\newcommand{\adv}{\ensuremath {\texttt{Adv}}{\xspace}}
\newcommand{\advprf}{\ensuremath {\texttt{Adv}_\texttt{\prf}}{\xspace}}
\newcommand{\advprg}{\ensuremath {\texttt{Adv}_\texttt{\prg}}{\xspace}}
\newcommand{\advfprg}{\ensuremath {\texttt{Adv}_\texttt{\fprg}}{\xspace}}
\newcommand{\advmac}{\ensuremath {\texttt{Adv}_\texttt{\mac}^\eucma}{\xspace}}
\newcommand{\advfamac}{\ensuremath {\texttt{Adv}_\texttt{\famac}^\faeucma}{\xspace}}
\newcommand{\advenc}{\ensuremath {\texttt{Adv}_\texttt{\encscheme}^\indcpa}{\xspace}}
\newcommand{\advfaaeindcpa}{\ensuremath {\texttt{Adv}_\texttt{\faae}^\findcpa}{\xspace}}
\newcommand{\advfaaefaeucma}{\ensuremath {\texttt{Adv}_\texttt{\faae}^\faeucma}{\xspace}}
\newcommand{\advfenc}{\ensuremath {\texttt{Adv}_\texttt{\fenc}^\findcpa}{\xspace}}

\newcommand{\kspace}{\mathcal {K}{\xspace}}
\newcommand{\ispace}{\mathcal {M}{\xspace}}
\newcommand{\ospace}{\mathcal {T}{\xspace}}

\newcommand{\graphenea}{\ensuremath {\texttt{Graphene}_1}{\xspace}}
\newcommand{\grapheneb}{\ensuremath {\texttt{Graphene}_2}{\xspace}}

\newcommand{\graphene}{\ensuremath {\texttt{Graphene}}{\xspace}}

\newcommand{\graphenegcm}{\ensuremath {\texttt{Graphene-GCM}}{\xspace}}

\newcommand{\offline}{\ensuremath {\texttt{-offline}}{\xspace}}
\newcommand{\online}{\ensuremath {\texttt{-online}}{\xspace}}

\newcommand{\graphenepoly}{\ensuremath {\texttt{Graphene-Poly}}{\xspace}}

\newcommand{\dmd}{\ensuremath {\texttt{Diamond}}{\xspace}}

\newcommand{\dmda}{\ensuremath {\texttt{Diamond}_1}{\xspace}}
\newcommand{\dmdb}{\ensuremath {\texttt{Diamond}_2}{\xspace}}

\newcommand{\dmdkg}{\ensuremath {\texttt{Diamond}\texttt{.Kg}}{\xspace}}
\newcommand{\dmdupd}{\ensuremath {\texttt{Diamond}\texttt{.Upd}}{\xspace}}
\newcommand{\dmdagg}{\ensuremath {\texttt{Diamond}\texttt{.Agg}}{\xspace}}
\newcommand{\dmdencmac}{\ensuremath {\texttt{Diamond}\texttt{.AuthEnc}}{\xspace}}
\newcommand{\dmdverdec}{\ensuremath {\texttt{Diamond}\texttt{.AVerDec}}{\xspace}}
\newcommand{\dmdae}{\ensuremath {\texttt{Diamond-AE}}{\xspace}}

\newtheorem{definition}{Definition}[section]

\newcommand{\vect}[1]{\ensuremath{ \boldsymbol{#1} }}

\newcommand{\algrule}[1][.2pt]{\par\vskip.5\baselineskip\hrule height #1\par\vskip.5\baselineskip}

\definecolor{grey}{rgb}{0.5,0.5,0.5}

%% file: abstract.tex
\begin{abstract}

Resource-constrained Internet of Things (IoT) devices, from medical implants to small drones, must transmit sensitive telemetry under adversarial wireless channels while operating under stringent computing and energy budgets. Authenticated Encryption (AE) is essential to ensure confidentiality, integrity, and authenticity. However, existing lightweight AE standards lack forward-security guarantees, compact tag aggregation, and offline-online (OO) optimizations required for modern high-throughput IoT pipelines.
We introduce \dmd, the first provably secure Forward-secure and Aggregate Authenticated Encryption (FAAE) framework that extends and generalizes prior FAAE constructions through a lightweight key evolution mechanism, an OO-optimized computation pipeline, and a set of performance-tier instantiations. 
\dmd~substantially reduces amortized offline preprocessing (up to $47\%$) and achieves up to an order-of-magnitude reduction in end-to-end latency for large telemetry batches.
Our comprehensive evaluation on 64-bit ARM Cortex-A72, 32-bit ARM Cortex-M4 and 8-bit AVR architectures confirms that \dmd~outperforms baseline \faae~variants in authenticated encryption throughput and end-to-end verification latency while maintaining compact tag aggregation and strong breach resilience. \dmd~outperforms NIST lightweight AE candidates for medium and large payloads, while remaining competitive for small messages when amortized across batches.
%
%
We formally prove the security of \dmd~and provide two concrete instantiations optimized for compliance and high efficiency. Our open-source release enables reproducibility and seamless integration into IoT platforms.

\ccsdesc[500]{Security and Privacy~Cryptography~Symmetric Techniques~Authenticated Encryption}

\end{abstract}

\keywords{
Lightweight Cryptography, Authenticated Encryption, Internet of Things
}

%% file: introduction.tex
\section{Introduction} \label{sec:introduction}

The rapid growth of modern cyber-physical infrastructures has driven the widespread deployment of battery-powered embedded devices operating in unmanned and low-maintenance environments, ranging from medical implants and wearable devices \cite{shah2024efficient} to smart city sensors \cite{wang2024crowdsensing}, autonomous surveillance nodes \cite{liu2024toward}, and unattended military ground sensors \cite{rettore2025military}. 
Such systems are increasingly operating autonomously and interacting with remote services, forming large-scale, distributed Internet of Things (IoT) networks for monitoring, analytics, and actuation \cite{he2022collaborative}. 

Ensuring secure and efficient communication in constrained IoT networks remains challenging due to the inherent constraints of IoT hardware, including limited computational capabilities, restricted bandwidth, and stringent energy budgets \cite{trilla2019worst, viswanathan2022challenges, yavuz2010new}. IoT applications require robust security guarantees of data confidentiality, integrity, and authenticity, particularly when transmitting sensitive mission data or control instructions. Authenticated Encryption (AE) \cite{degabriele2024sok, khan2021scalable, nouma2025lightweight} provides the aforementioned security services by combining Symmetric Encryption (SE) with Message Authentication Codes (MACs), making it suitable for wireless and low-power environments. 
%

\subsection{Problem Statement}

Although widely used AE schemes such as AES-GCM \cite{mcgrew2004security} benefit from hardware acceleration and offer high throughput on many (embedded) platforms \cite{hofemeier2012introduction}, they do not  address the unique performance and security requirements imposed by the adversarial nature of unattended IoT, nor do they fully leverage the potential of IoT infrastructures through lightweight cryptography. A desirable AE scheme must resist key compromise attacks \cite{siwakoti2023advances, wang2024survey} and achieve minimal cryptographic overhead on constrained embedded devices in terms of amount of computational processing and transmitted payload \cite{wagner2022bp}. We elaborate on these points as follows:

\textit{i) Key Compromise Resilience}:
The compromise of long-term credentials (e.g., secret keys) comprise a dominant attack vector in adversarial settings where adversaries obtain full (physical) access to embedded or resource-constrained platforms \cite{siwakoti2023advances}. 
For example, implantable glucose monitors and medical pacemakers periodically transmit physiological measurements to a patient-held gateway for clinical triage \cite{shah2024efficient}. In adversarial settings, hackers exploit vulnerabilities (e.g., in Medtronic devices) to enable remote tampering that could alter pacemaker rhythms, potentially endangering lives\footnote{\url{https://www.cisa.gov/news-events/ics-medical-advisories/icsma-23-180-01}}.
Similarly, low-power aerial drones perform reconnaissance in adversarial environments and generate mission telemetry that is stored and later relayed to a nearby edge node \cite{rettore2025military, wang2024survey}. 
%



Forward security aims to mitigate the risk of exposure of long-term secret keys \cite{boyd2021modern, wei2021communication}. In essence, it enforces key evolution such that exposure of the {\em current} secret key does not jeopardize confidentiality or authenticity of data produced in {\em prior} time periods \cite{dodis2022forward, wei2021communication}. 
In other words, secret keys are updated forward in time (e.g., via a one-way function) so that past keys remain protected even after a compromise event \cite{bellare2003forward}.
Thus, forward security is vital against several attack vectors (e.g., malware and impersonation attacks) that target IoT networks \cite{wang2024survey}. 
Forward security is especially relevant for IoT communication networks against eavesdropping attacks, where adversaries may passively intercept encrypted traffic and later attempt decryption or tag forgery of encrypted data, upon successful compromise of secret keys. 
While previous works have applied forward security in both private-key (e.g., symmetric encryption \cite{bellare2003forward, dodis2022forward} and MACs \cite{FssAggMAC_DiMa, SUHaSAFSS11}) and public-key (e.g., digital signatures \cite{cronin2003performance, nouma2024trustworthy}) settings, it remains under-explored in symmetric-key AE settings \cite{nouma2025lightweight, collins2024tight}. 

{\em ii) Minimal Bandwidth Footprint}: Several critical applications such as Over-The-Air (OTA) updates \cite{el2022secure}, secure logging \cite{nouma2026poslo, steger2017efficient} and secure offloading \cite{yao2025optimization} techniques require low uplink rates from low-end embedded devices to remote edge and cloud servers. For example, IEEE 802.15.4 networks are limited to 175 kbps uplink bandwidth, whereas LoRaWAN is limited to only 250 bps.


Aggregation is crucial to improve the efficiency of wireless communication in low-power wireless networks \cite{wagner2024and}. Aggregation compresses multiple per-message authentication tags into a single, compact aggregate authenticator via algebraic (e.g., XOR-sums \cite{wagner2024and, katz2008aggregate}) or hash-based (e.g., cryptographic hash-based \cite{FssAggMAC_DiMa, SUHaSAFSS11}) compositions.
Therefore, it reduces per-message communication overhead and verification bandwidth, enabling high-throughput verification of large message batches.
Aggregation is vital for bandwidth- and energy-constrained networks (e.g., LoRaWAN, ZigBee) where uplink rates are often limited to just a few kilobits per second.
As such, aggregation has been extensively used in digital signatures (e.g., \cite{nouma2024trustworthy}) and MACs (e.g., \cite{wagner2025mac, nouma2025lightweight, FssAggMAC_DiMa}). 
In contrast, only a few works \cite{nouma2025lightweight} propose aggregate MACs integrated into AEs.

{\em iii) Minimal Computational Overhead}:
Improving the efficiency of current AE standards is also equally critical for extending the energy lifetime of battery-powered constrained devices and for improving the End-to-End (E2E) delay of received data packets at the verifier side.
OO cryptography amortizes the computational cost across an offline precomputation phase, where expensive input-independent cryptographic operations are performed, whereas the online phase executes constant-time input-dependent operations \cite{wagner2022bp, hiller2018secure}. 
Several works investigate OO properties in public-key (e.g., digital signature \cite{samandari2025online}) and symmetric-key encryption (e.g., \cite{hiller2018secure}). 
However, we observe that leveraging OO for AEs has been overlooked, with only recent work \cite{nouma2025lightweight} precomputing costly computations during the offline phase and demonstrating substantial performance improvements compared to vanilla AE standards. In contrast, the NIST lightweight AE standard, Ascon, does not possess OO capabilities due to its integrated sponge-based structure \cite{li2023automatic}.

In the following, we review related works, with an emphasis on the aforementioned properties.

\subsection{Related Work and Limitations}

Ascon \cite{turan2024ascon} has been selected as the NIST Lightweight Cryptography (LWC) standard (NIST 800-232 standard) for both AE and hash functionalities. Although Ascon outperforms the currently deployed AE standards—AES-GCM (NIST SP 800-38D) \cite{salowey2008aes} and ChaCha20-Poly1305 (RFC 8439) \cite{nir2018chacha20}—on small microcontrollers, it lacks OO cryptography, which stems from its sponge-based AE structure and the design choices to meet lightweight and side-channel-resilience goals. In particular, its integrated AE mode precludes ciphertext-independent AE structure, and its tag verification requires full payload decryption, making it incompatible with desirable efficient verification and low end-to-end delay. 
Ascon is developed to offer a viable alternative when the Advanced Encryption Standard (AES) may not perform optimally \cite{turan2024ascon}. In this work, we provide a comprehensive benchmarking for AE standards, particularly when considering forward security, aggregation, and OO optimization.

%

AES-GCM \cite{salowey2008aes} is one of the most widely deployed authenticated-encryption (AE) standards, used in protocols such as DTLS 1.3 and IEEE 802.15.4 for constrained sensor and low-rate wireless personal-area networks (LR-WPANs). ChaCha20-Poly1305 \cite{nir2018chacha20}, is also mandated in TLS 1.3 as the preferred AE scheme for IoT-class processors (e.g., ARM Cortex-M) where hardware-accelerated AES is unavailable or inefficient. Both AES-GCM and ChaCha20-Poly1305 follow the same high-level design principle: 
{\em (i) Counter (CTR) Mode of Encryption.} providing OO Cryptography and parallelizable key-stream generation \cite{hiller2018secure}.
{\em (ii) Wegman-Carter MAC Authentication} \cite{degabriele2024sok}. instantiated with a fast universal hash function. AES-GCM employs AES-128 as its block cipher and GHASH (a binary-field polynomial hash) for authentication. In contrast, ChaCha20-Poly1305 uses the ChaCha20 stream cipher with Poly1305 as an efficient universal hash. 
Both AE standards can achieve OO property, which is especially relevant for low encryption overhead and efficient batch verification with a minimal end-to-end delay. 
Hiller et al. \cite{hiller2018secure} showcase $75.9\%$ latency reduction in encryption/decryption using AES in the CTR mode of operation. 
In practice, AES-GCM achieves high performance on platforms with specialized hardware instructions (e.g., AES-NI on Intel/AMD CPUs or embedded cryptographic accelerators). For example, the AE runtime of AES-GCM can be reduced by up to $70\%$ when harnessing OO cryptography techniques \cite{nouma2025lightweight}. ChaCha20-Poly1305, while typically slower on hardware-accelerated platforms, provides consistently fast pure-software performance across commodity hardware and low-end microcontrollers \cite{degabriele2024sok}.

Regarding minimizing bandwidth overhead, existing AE primitives lack the aggregation property of authentication tags, i.e., they do not support compressing multiple tags into a single compact authenticator via a keyless aggregation method. In contrast, aggregation in the context of Aggregate MACs (AMACs) has been studied extensively \cite{hirose2014forward, SUHaSAFSS11, FssAggMAC_DiMa, wagner2025mac, wagner2024and, katz2008aggregate, li2021cumulative}. Hash-based AMACs (e.g., \cite{SUHaSAFSS11, FssAggMAC_DiMa}) instantiate aggregation via recursive cryptographic hash calls and thereby inherit sequential immutability from the underlying cryptographic hash function but incur higher computational cost. 
Another line of work employs XOR-based aggregation (e.g., \cite{wagner2025mac, wagner2024and, li2021cumulative, hirose2014forward, eikemeier2010history, wagner2022take}) due to its constant-time and linear-algebraic simplicity. Wagner et al. \cite{wagner2024and} provide a systematic analysis of different XOR-based aggregation methods across various communication channels, offering guidance on selecting the aggregation method based on network topology and conditions. 
More recently, Wagner et al. \cite{wagner2025mac} implement aggregation methods on the Datagram Transport Layer Security (DTLS) 1.3 protocol and demonstrate up to 50\% improvement in goodput and 15\% reduction in energy usage on an embedded testbed equipped with an ARM Cortex-M4 microcontroller.
%

Existing work on forward security in symmetric-key cryptography originates with Bellare et al. \cite{bellare2003forward}, which provides formal treatments and security for forward-secure MACs and encryption but only briefly discusses AEs. In parallel, Ma et al. \cite{FssAggMAC_DiMa} proposed the first forward-secure and aggregate MAC (FssAggMAC), combining key evolution and tag aggregation with order integrity. 
Yavuz et al. \cite{SUHaSAFSS11} improved FssAggMAC by designing more efficient hash-based forward-secure aggregate authentication tailored for unmanned sensor networks. Later, Hirose et al. \cite{hirose2014forward} revisited forward-secure AMACs with generalized constructions and formal security treatment. 
Beyond MACs, forward-secure encryption has also been developed, including puncturable and revocation-friendly primitives \cite{green2015forward} as well as fast-forwardable key-evolution \cite{marson2014even, dodis2022forward}. 
Several works \cite{collins2024tight} formalize forward-secure AE and provide concrete constructions with deployment in secure messaging protocols such as Signal. However, such constructions require bidirectional communication and are not optimized for constrained IoT devices using the aforementioned aggregation and OO properties. 

To the best of our knowledge, no provably secure FAAE primitive has been proposed. Prior works address only forward-secure encryption and AMACs, none of which leverage full OO potential. 
%
There is a large gap in the state-of-the-art in achieving a breach-resilient, compact, lightweight, and precomputable authenticated encryption scheme, specifically tailored for constrained IoT environments by leveraging the aforementioned security and performance properties.

\subsection{Our Contributions}

In this work, {\em we propose \dmd, the first, to the best of our knowledge,  provably secure authenticated encryption framework that simultaneously achieves forward security, tag aggregation, and OO cryptography, ensuring breach resilience against compromise attacks, minimal energy cost, low end-to-end delay with small bandwidth usage, which we believe to be attractive features towards efficient and robust IoT networks. The desirable properties of \dmd~include the following:}
\vspace{2pt}

\noindent \textbf{$\bullet$ Provable-secure Forward-secure and Aggregate AE~Framework (FAAE)}: 
\dmd~introduces a unified authenticated encryption framework that composes forward-secure key evolution with sequential aggregate authentication via universal MACs. \dmd~uncovers synergies among forward-secure encryption and aggregate universal MACs overlooked by prior approaches while being backward compatible with current standards and implementations. Therefore, it achieves modular construction in which per-message keys are evolved using lightweight pseudo-random functions (PRFs), and authentication tags are incrementally compressible without dropping unforgeability. \dmd~achieves high efficiency and security while opening a path for alternative constructions through generalizable integrations and ease of implementation. \vspace{1pt}

\noindent \textbf{$\bullet$ High Online Computation Efficiency via OO Cryptography}:
OO cryptography amortizes input-independent cryptography computations into an offline stage, leveraging idle and power-transfer periods, thereby reducing the online latency for performance-critical use-case scenarios. Although well-studied for digital signatures \cite{samandari2025online}, OO remain largely unexamined in the \faae~context \cite{nouma2025lightweight}. \dmd~explores the synergies of various universal MACs and AEs in OO settings and achieves a significant online performance gain with only a modest increase in memory consumption.  \dmd~consistently outperforms vanilla \faae~standards during the online authenticated encryption across architectures. For example, our \dmdb~instantiation achieves up to $3.5\times$ online speedup for a batch of 1024 16-byte telemetry on 32-bit Cortex-M4 embedded processor, with only 32KB extra storage. Even under stringent computing constraints on an 8-bit AVR ATmega2560, \dmdb~exhibits high efficiency with $3.8\times$ speedup compared to \faaeb~(instantiated from the NIST lightweight standard, Ascon \cite{turan2024ascon}). In wearable medical settings, this results in significant savings in battery life during online operations, as the next batch of OO keys can be supplied when the device is recharged (e.g., during power transfer \cite{shah2024efficient}) while telemetry is uploaded. In aerial drone applications, latency reduction can improve flight safety and real-time telemetry transmission.~\vspace{1pt}

\noindent \textbf{$\bullet$ Minimal Offline Preprocessing via Lightweight Key Evolution}:
In contrast to the key evolution from cryptographic hashing in \graphene~counterpart \cite{nouma2025lightweight}, \dmd~employs a more lightweight and computation-efficient evolution mechanism from PRFs, enabling fast update of key materials. 
\dmd~achieves up to $40\%$ and $47\%$ reductions in offline preprocessing runtime compared to \graphene~counterpart on two representative embedded platforms, a 64-bit ARM Cortex-A72 processor and an 8-bit AVR ATmega2560 microcontroller, respectively.~\vspace{1pt}

\noindent \textbf{$\bullet$ Low End-to-End Verification Latency For Large Payload Batches}: 
The proposed \dmd~instantiations substantially minimize the E2E verification latency by coupling lightweight universal hashing and efficient online encryption. 
Our two instantiations, \dmda~and \dmdb, exhibit efficient E2E scaling across message sizes (16-128 bytes). For example, when the IoT device is a constrained 8-bit AVR ATmega2560 and the IoT server is a commodity hardware, \dmda~achieves up to $3\times$ lower E2E delay compared to baseline \faae~counterparts.
Our highly efficient variant \dmdb, leveraging ChaCha20-Poly1305 (RFC standard), yields even higher acceleration with $8.1\times$ lower E2E latency than \dmda~and up to $3.8\times$ lower delay than the fastest NIST lightweight \faae~counterparts across both 16-byte and 128-byte payloads. 
\vspace{1pt}


\noindent \textbf{$\bullet$ Comprehensive Benchmarks with Full-Fledged Implementation}:
\sloppy We implemented \dmd~on commodity hardware (x86\_64) and three constrained embedded devices: a 64-bit ARM Cortex-A72 processor, a 32-bit ARM Cortex-M4 microcontroller, and an 8-bit AVR ATmega2560 microcontroller. Our performance results highlight the efficiency of each instantiation, guiding its application context accordingly. To encourage public testing and reproducibility, we release our implementation at: \fcolorbox{green!40}{white}{\url{https://github.com/SaifNouma/Diamond}}. To the best of our knowledge, \dmd~is the first comprehensive open-source \faae~framework on the aforementioned platforms.

\subsection{Improvements over Preliminary Version} 
This work is an extension of our conference publication at IEEE MILCOM 2025 \cite{nouma2025lightweight}, with extended algorithmic descriptions and optimizations, formal and in-depth security analysis, and extended performance analysis. Our journal extension makes substantially expanded treatment, with more than doubling technical content, and introduces the following major improvements and contributions:
\vspace{1pt}

\textbf{1) Additional Experimental Evaluation and Analysis on Constrained Devices.}
We significantly expand the performance analysis by incorporating two more constrained platforms, a 64-bit ARM Cortex-A72 edge processor and an ultra-low-power 8-bit AVR ATmega2560, with full benchmarking on \dmd~instantiations and selected counterparts. Our extended analysis rigorously quantifies offline/online execution, their energy usage on the 8-bit platform, and end-to-end (E2E) batch verification latency. \dmd~consistently outperforms \graphene~on the selected platforms. 
\vspace{1pt}

\textbf{2) Comprehensive Algorithmic Formalization with Further Optimizations.}
We provide a comprehensive formalization of the proposed scheme (Section \ref{sec:schemes}). We elaborated the main building blocks to construct \dmd: (i) forward-secure symmetric encryption based on counter mode of operation and (ii) forward-secure and aggregate MAC based on Carter-Wegman construction, in Fig. \ref{alg:fenc} and Fig. \ref{alg:umac}, respectively. Moreover, unlike hash-based key evolution used in initial \graphene, \dmd~employ a more efficient key evolution instantiated from pseudo-random functions (\prf). 
\vspace{1pt}

\textbf{3) Formal and In-Depth Security Analysis.}
We introduce formal security definitions and prove that \dmd~achieves forward-secure confidentiality, integrity, and authenticity in the standard model (Section \ref{sec:security_analysis}), through tight reductions to the pseudo-randomness of the underlying \prf~and the universality of the hash function. This yields minimal and transparent assumptions, which are substantially stronger than those attained in the initial \graphene.

%% file: preliminaries.tex
\section{Preliminaries}
\label{sec:preliminaries}

\subsection{Notations}

The notations and acronyms are described in Table \ref{tab:notations}. 


\begin{table}[ht]
\caption{List of Notations, Acronyms, and Their Descriptions}
\label{tab:notations}
\centering
\resizebox{\textwidth}{!}{
\begin{tabular}{llll}
\hline
\textbf{Notation} & \textbf{Description} & \textbf{Acronym} & \textbf{Description} \\
\hline
$X \| Y$ & Concatenation of variables $X$ and $Y$ 
& IoT & Internet of Things \\
$|X|$ & Bit length of variable $X$ & 
PRF & Pseudo-Random Function \\
$X \xleftarrow{\$} \mathcal{S}$ & $X$ sampled uniformly at random from set $\mathcal{S}$ & 
CTR & Counter Mode (of encryption) \\
$\{0,1\}^*$ & Set of binary strings of arbitrary length & 
OO & Offline-Online Cryptography \\
$\mathbb{Z}_q$ & Finite field of order $q$ & 
[F]SE & [Forward-secure] Symmetric-key Encryption \\
$\vec{X} = \{X_1,\ldots,X_n\}$ & Collection of items; $x = |\vec{X}|$ & 
[F][A]MAC & \text{[Forward-secure]} \text{[Aggregate]} Message Authentication Codes \\
 & & 
 [F][A]AE & \text{[Forward-secure]} \text{[Aggregate]} Authenticated Encryption \\
\hline
\end{tabular}
}
\end{table}

\subsection{Building Blocks}
\label{subsec:buildingblocks}
In the following, we define the main cryptographic primitives used in our proposed scheme.
\vspace{2pt}

\vspace{2pt}
\noindent \textbf{Family of Functions with Extended Properties.} A family of functions is defined as follows:

\begin{definition}
	A family of functions is represented as ($F \as \{ f: \kspace \times \ispace \rightarrow \ospace$\}) where $\kspace, \ispace$, and $\mathcal{T}$ denote the key, input, and output space, respectively. We denote by $\kappa$, $m$, and $\tau$ the bit sizes of $\kspace, \ispace$, and $\mathcal{T}$, respectively. $f$ is defined as follows:
	\begin{enumerate}[-]
		\item $\tau \as f_K(M)$: Given a key $K \in \kspace$ and an input $M \in \ispace$, it outputs $\tau \in \ospace$.
	\end{enumerate}
\end{definition}

\begin{definition}
	A family of functions $F \as \{ f: \kspace \times \ispace \rightarrow \ospace\}$ is called a universal family of hash functions (\uhash)
	if $m \ge \tau$ and for all $M, M' \in \ispace$ with $M \neq M'$ and $K \Ra \{0,1\}^\kappa$, $\Pr[\uhash_K(M)=\uhash_k(M')] < 2^{-\tau}$. 
	Moreover, \uhash~is $\epsilon$-almost universal if $\forall M \neq M'$, $\Pr[\uhash_k(M)=\uhash_k(M')] < \epsilon$. 
\end{definition}

\begin{definition}
	A family of functions $F \as \{ f: \kspace \times \ispace \rightarrow \ospace\}$ is called a family of pseudo-random functions $\prf$ if a random function $PF \as \{\prf_K: \ispace \rightarrow \ospace, \forall K \in \kspace \}$ is {indistinguishable} from a random function uniformly selected from the set $\mathcal{F}=\{f : \ispace \rightarrow \ospace  \}$ having domain $\ispace$~and range~$\ospace$.
	%
\end{definition}

\begin{definition}
	A pseudo-random generator (\prg) is a function $\prg: \kspace \rightarrow \kspace \times \ospace$ that extends a uniformly random input $K \in \kspace$ to a longer random string $(K',Y) \in \kspace \times \ospace$. 
\end{definition}

\begin{definition}
	A forward-secure pseudo-random generator (\fprg) is a scheme initially introduced by Bellare et al. \cite{bellare2003forward} and consists of two algorithms $\fprg=(\kg, \upd)$, defined as follows: 
	\begin{enumerate}[-]
		\item $S_1 \as \fprgkg(1^{\kappa}, n)$: Given security parameter $\kappa$, it outputs an initial key $S_1$.
		\item $(S_{i+1}, K_i) \as \fprgupd(S_i)$: Given $S_i$, it returns $S_{i+1}$ and a key $K_i$ if $i \le n$ and $\perp$ otherwise. 
	\end{enumerate}
\end{definition}
Bellare et al. \cite{bellare2003forward} describes how \fprg~with $\fprgupd: \kspace  \rightarrow \kspace \times \ospace' $ can be constructed from a secure $\prf: \kspace \times \ispace \rightarrow \ospace$~via $\fprgupd(K_i) \as \prf(K_i,1) \| \ldots \| \prf(K_i,\lceil \frac{(\tau' + \kappa)}{\tau} \rceil), \forall i=1, \ldots, n$. 
\vspace{2pt}


\noindent \textbf{Message Authentication Codes (MAC).}
A MAC enables a receiver to verify the integrity of message $M$ via an authentication tag $\sigma$ after transmission, given a shared secret key $K$. Formally: 

\begin{definition}
	A Message Authentication Code~$\mac=(\kg, \sign, \ver)$ is a triple of algorithms:
	
	\begin{enumerate}[-]
		\item $k \as \mackg(1^\kappa)$: It takes the security parameter $\kappa$, and generates a secret key $K \Ra \kspace = \{0,1\}^{\kappa}$. 
		\item $\sigma \as \macsign(K, M)$: It takes $K \in \kspace$ and a message $M \in \ispace$, and outputs a tag $\sigma \in \ospace$.
		\item $b \as \macver(K, M, \sigma)$: It outputs $b=1$ if $(M,\sigma)$-pair is valid, or $b=0$ otherwise.
	\end{enumerate}
\end{definition}

\noindent \textit{Carter-Wegman Construction.} A \mac~scheme can be instantiated via Carter-Wegman \cite{etzel1999square} given an $\epsilon$-almost universal hash (\uhash) and a secure \prf. A {\em universal} \mac~(\umac) is defined as follows:
\[
\umac.\sign((K_1,K_2), M, N) = \uhash(K_1, M) + \prf(K_2, N),~\text{where } K_1, K_2 \in \kspace, M \in \ispace, \text{ and }  N \in \mathcal{N},
\]

That is, \umac~consists of a nonce-based \mac: given a message $M \in \ispace$, it computes a tag $\sigma \as \umac.\sign((K_1,K_2), M, N)$ where $\prf(K_2, N)$ can be precomputed during the {\em offline phase}. 
\vspace{2pt}


\begin{definition}
	A forward-secure and sequential-aggregate message authentication code~$\famac=(\kg, \sign, \upd, \agg, \aver)$ consists of five algorithms:
	
	\begin{enumerate}[-]
		\item $K_1 \as \famackg(1^\kappa, n)$: It takes the security parameter $\kappa$, the maximum number of generated tags $n$, and generates an initial secret key $K_1 \Ra \kspace = \{0,1\}^{\kappa}$. 
		\item $\sigma_i \as \famacsign(K_i, M_i)$: It takes $K_i \in \kspace$ and a message $M_i \in \ispace$, and outputs a tag $\sigma_i \in \ospace$.
		\item $b \as \famacaver(K_1, \vect{M_{1,i}}, \sigma_{1,i})$: Given $K_1$, it outputs $b=1$ if  $(\vect{M_{1,i}}, \sigma_{1,i})$ is valid, or $b=0$ otherwise.
		\item $K_{i+1} \as \famacupd(K_i, n)$: It takes $K_i \in \kspace$ and $n$, and outputs $K_{i+1}$ if $i < n$ and $\perp$ otherwise.
		\item $\sigma_{1, i+1} \as \famacagg(\sigma_{1,i}, \sigma_{i+1})$: Given a forward-secure and aggregate tag $\sigma_{1,i}$ and current tag $\sigma_{i+1}$, it returns a constant-size aggregate tag $\sigma_{1,i+1}$.
	\end{enumerate}
\end{definition}

\noindent \textbf{Symmetric Encryption (SE).}
A symmetric encryption (\se) ensures the confidentiality of transmitted data between two parties given a shared secret key $K$. Formally, \se~is defined as follows:

\begin{definition}
	A symmetric encryption scheme $\encscheme=(\kg, \enc, \dec)$ is a triple of algorithms:
	\begin{enumerate}[-]
		\item $K \as \enckg(1^\kappa)$: Given the security parameter $1^\kappa$, it returns a secret key $K$.
		\item $C \as \encenc(K, M)$: Given a secret key $K \in \kspace$ a message $M \in \ispace$, it returns a ciphertext $C \in \ospace$. 
		\item $M' \as \encdec(K, C)$: Given a secret key $K \in \kspace$ and a ciphertext $C \in \ospace$, it returns a plaintext $M'$. 
	\end{enumerate}
\end{definition}

\begin{definition}
	A forward-secure symmetric encryption scheme $\fenc=(\kg, \enc, \dec, \upd)$ consists of four algorithms and is presented as follows:
	\begin{enumerate}[-]
		\item $K_1 \as \fenckg(1^\kappa, n)$: Given the security parameter $\kappa \in \kspace$ and the maximum number of encryptions $n$, it returns an initial secret key $K_1$.
		\item $C_i \as \fencenc(K_i, M_i)$: Given $K_i \in \kspace$ a message $M_i \in \ispace$, it returns a ciphertext $C_i \in \ospace$. 
		\item $M_i' \as \fencdec(K, C)$: Given a secret key $K_i \in \kspace$ and a ciphertext $C_i \in \ospace$, it returns a plaintext $M_i'$. 
		\item $K_{i+1} \as \fencupd(K_i, n)$: It takes $K_i \in \kspace$ and $n$, and outputs $K_{i+1}$ if $i < n$ and $\perp$ otherwise.
	\end{enumerate}
\end{definition}

%
%
%
%



\section{Models}

\noindent \textbf{System Model.}

Our system model consists of two primary entities, as illustrated in Fig. \ref{fig:systemmodel}:

\begin{figure}[ht!]
	\centering
	\includegraphics[scale=0.5]{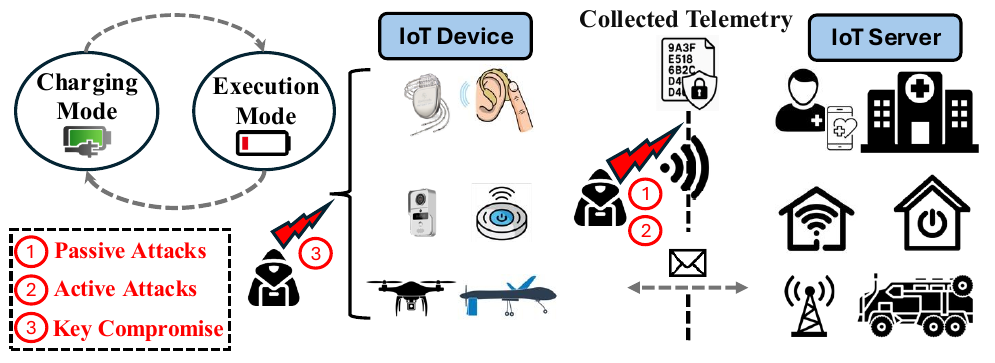}
	\caption{Our system and threat models}
	\label{fig:systemmodel}
\end{figure}

\begin{enumerate}[-]
\item \textbf{Resource-constrained IoT devices.} 
(e.g., implantable medical sensors, autonomous unmanned aerial vehicles (UAVs), and smart-city robots) operate in adversarial and bandwidth-limited environments under stringent constraints on storage, memory, computation, and energy budgets. These endpoints continuously produce high-value telemetry (e.g., biomedical signals, flight stability parameters, or navigation/obstacle avoidance traces) that must remain confidentiality- and integrity-protected \cite{yaqoob2019security, nouma2024trustworthy}. Such store-and-forward pipelines arise in numerous domains, including but not limited to
implantable devices (e.g., pacemakers) in medical IoT (IoMT) \cite{shah2024efficient} and low-power aerial drones in autonomous UAVs \cite{rettore2025military, wang2024survey}.
%
%

\item \textbf{IoT Server (Edge Verifier).} 
is a storage- and computation-resourceful entity, being a patient's phone or a clinical gateway in IoMT, or a nearby access point in smart-home and smart-city applications. Upon receiving encrypted/authenticated payloads, the server performs verification and decryption and optionally offloads the data to a cloud server for analytics and autonomous actuation. We assume that the two main entities pre-share an initial secret key.
\end{enumerate}



\noindent \textbf{FAAE Model.} A Forward-secure and sub-Aggregate Authenticated Encryption (\faae) aims to achieve the security of both \fenc~and \famac, i.e., {\em data confidentiality} of the communicated traffic and {\em data integrity and authenticity} of generated encrypted data. Formally, an \faae~scheme consists of five algorithms ($\kg, \upd, \authenc, \agg, \averdec$):

\begin{enumerate}[-]
	\item $\vect{K_1} \as \faaekg(1^\kappa, n, b)$: Given the security parameter $1^\kappa$, the maximum messages to be processed $n$, and the batch size $b$, it returns an initial secret key $\vect{K_1}$. $\vect{K_1}$ contain a single key when instantiated with a unified \aee~or multiple keys in case of generic \aee~construction via \fenc~and an \famac. 
	
	\item $\vect{K_{i+1}} \as \faaeupd(\vect{K_{i}})$: Given the current secret key $\vect{K_i}$, it returns an updated $\vect{K_{i+1}}$.
	
	\item $(C_i, \sigma_i) \as \faaeauthenc(\vect{K_i}, M_i)$: Given the current secret key $\vect{K_i}$ and a message $M_i$, it returns a ciphertext $C_i$ and its corresponding authentication tag $\sigma_i$.
	
	\item $\sigma_{i_1, i_2+1} \as \faaeagg(\sigma_{i_1,i_2}, \sigma_{i_2+1})$: Given an aggregate tag $\sigma_{i_1,i_2}$ and $\sigma_{i_2+1}$, it returns $\sigma_{i_1,i_2+1}$.
	
	\item $\vect{M_i} \as \faaeverdec(\vect{K_i}, \vect{C_{i}}, \sigma_{i,i+b-1})$: Given the current secret key $\vect{K_i}$, a batch of ciphertexts $\vect{C_{i}} := (C_i, \ldots, C_{i+b-1})$, and an aggregate tag $\sigma_{i,i+b-1}$, it returns the batch of plaintexts $\vect{M_i}:= (M_i, \ldots, M_{i+b-1})$ if and only if the pair $(\vect{C_{i}}, \sigma_{i,i+b-1})$ is valid. 
\end{enumerate}

The maximum number of messages to be processed $n$ is divided into $\lfloor \frac{n}{b} \rfloor$ epochs, each epoch contains $b$ messages.
As illustrated in Fig. \ref{fig:communication_flow}, the sender (IoT device) evolves the private key by invoking $\faaeupd(.)$ after every $\faaeauthenc(.)$ call. Then, it aggregates the current tag $\sigma_i$ with the aggregate tag in the current epoch by invoking $\faaeagg(.)$ algorithm, except for the first tag in each epoch. Overall, \faae~enable the sender to reduce the communication overhead from $\mathcal{O}(n)$ into an adjustable $\mathcal{O}{(\frac{n}{b})}$ based on the epoch size $b$. 
\vspace{2pt}

\begin{figure}[ht]
	\centering
	\begin{tikzpicture}[>=Latex, font=\small]
		\coordinate (S) at (2,0);
		\coordinate (R) at (8,0);
		
		\node[anchor=west, align=center, draw, rounded corners=2pt, inner sep=3pt] (proc) at ($(S)+(-0.5, 0)$) {
			IoT device
		};
	
		\node[anchor=west, align=center, draw, rounded corners=2pt, inner sep=3pt] (proc) at ($(R)+(-0.5, 0)$) {
			IoT Server
		};
	
		\draw[dashed] ($(S)+(0,-0.3)$) -- ($(S)+(0,-5.0)$);
		\draw[dashed] ($(R)+(0,-0.3)$) -- ($(R)+(0,-5.0)$);
		
		\node[anchor=west, align=left] (proc) at ($(S)+(0.1,-1.2)$) {%
			1. $(C_i, \sigma_i) \as \faaeauthenc(\vect{K_i}, M_i)$\\[4pt]
			2. $\vect{K_{i+1}} \as \faaeupd(\vect{K_i})$\\[4pt]
			3. $\sigma_{i,i} \as \sigma_i$
		};
		\node[anchor=west, align=left] (proc) at ($(S)+(0.5,-2.2)$) {%
			$\vdots$
		};
		\node[anchor=west, align=left] (proc) at ($(S)+(0.1,-3.5)$) {%
			1. $(C_{i}, \sigma_{i+b-1}) \as \faaeauthenc(\vect{K_{i+b-1}}, M_{i+b-1})$\\[4pt]
			2. $\vect{K_{i+b}} \as \faaeupd(\vect{K_{i+b-1}})$\\[4pt]
			3. $\sigma_{i,i+b-1} \as \faaeagg(\sigma_{i,i+b-2}, \sigma_{i+b-1})$
		};
		\draw[->] ($(S)+(0.6,-4.5)$) -- ($(R)+(-0.6,-4.5)$)
		node[midway, below, align=center] {$(\vect{C_i} \as (C_i,\ldots,C_{i+b-1}), \sigma_{i,i+b-1})$};
		\draw[decorate,decoration={brace,mirror,amplitude=6pt}]
		($(S)+(-0.6,-0.4)$) -- ($(S)+(-0.6,-4.15)$)
		node[midway,left=8pt,align=left] {$\text{$\lceil\frac{i}{b} \rceil$}^\text{th}$ epoch  \\[2pt] ($1 \le i \le n$)};
		\node[anchor=west, align=left] (proc) at ($(R)+(0.1,-5.0)$) {%
			$\vect{M_i} \as \faaeverdec(\vect{K_i}, \vect{C_i}, \sigma_{i,i+b-1})$\\[4pt]
			$(\vect{M_i} := (M_i , \ldots, M_{i+b-1}))$
		};
	\end{tikzpicture}
	\caption{\faae~Communication Flow}
	\label{fig:communication_flow}
\end{figure}

\noindent \textbf{Threat Model.} We assume a probabilistic polynomial-time (PPT) adversary $\mathcal{A}$ capable of:

\begin{enumerate}[i.]
    \item {\em Eavesdropping attacks.} monitoring and analyzing encrypted traffic over wireless channels.
    \item {\em Active attacks.} Modifying, dropping, or relaying encrypted data over communication channels. 
    \item {\em Key compromise attacks.} 
    Extracting device's credentials during a system breach (e.g., malware) or a physical compromise (e.g., drone captured).
    Upon a break-in, \A~aims to decrypt previously intercepted traffic using compromised keys and forge forward-secure and aggregate tags. 
\end{enumerate}

%% file: proposed_schemes.tex
\section{PROPOSED SCHEMES} \label{sec:schemes}

A standard \faae~scheme typically applies conventional encryption with an HMAC \cite{SUHaSAFSS11}. Its aggregation and forward security mechanisms are built on only a standard hash function \cite{FssAggMAC_DiMa}. However, this construction falls short of achieving the high efficiency and desirable properties outlined in Section \ref{sec:introduction}, such as precomputation OO capabilities for minimal online latency, flexible aggregation modes, all essential for our system model. While prior MACs have explored precomputation, these efforts were isolated, neglecting key evolution and encryption. In this work, we introduce \dmd~an improved variant of \graphene~and is designed to deliver near-optimal online computational efficiency and compactness, tailored for resource-constrained IoT environments.

Our generic \dmd~construction instantiates an \faae~scheme by combining counter-mode (CTR) encryption with a universal \mac~scheme (\umac) following Encrypt-then-MAC construction \cite{bellare2000authenticated}, thereby allowing for efficient online running time thanks to the offline precomputation of both components. \dmd~rely on \fprg~to transform  CTR-based \se~and \umac~into an \fenc~and \famac.

Given the security parameter $\kappa$ and the block size $\ell$ of the block cipher used in \fenc, we define $\prf_1: \{0,1\}^\kappa \times \{0,1\}^\ell \rightarrow \{0,1\}^\ell$ and $\prf_2: \{0,1\}^\kappa \times \{0,1\}^\kappa \rightarrow \{0,1\}^\kappa$ as two \prf~functions. For a maximum time periods of  $n$, we instantiate \fprg~based on $\prf_2$ as follows: 
\[\fprgupd(S_i) \as (\prf_2(S_i, 0), \prf_2(S_i,1)), \forall i =1,\ldots,n\]

In the following, we present the building blocks of our proposed scheme \dmd: the generic constructions of \fenc~and \famac~via CTR-based \enc~and \umac~schemes, respectively.

\subsection{Generic \fse~Construction}
Fig. \ref{alg:fenc} depicts the algorithmic description of our proposed CTR-based \fse~construction. 
\fenc~is composed of the encryption scheme (\se) and the forward-secure pseudo-random generator (\fprg), where \se~is instantiated using a \prf~in the CTR mode of operation. Our construction harnesses the OO capabilities of CTR mode and batch precomputation of forward-secure private keys, thus enabling highly efficient online runtime during encryption and decryption phases. 

\subsubsection{ Algorithmic Description} 

The key generation (\fenckg) accepts the security parameter $\kappa$ and the maximum number of encryption operations $n$. Then, it generates an initial \fprg~state $S_1$ and the private key $K_1$. \fenc~generates an initial internal state $St$ that contains the current iteration $i$ and randomly generated counter $ctr$ to be used in the CTR-based encryption. We set the \fenc~secret key $\vect{K}_1$, which contains the \fprg~internal state for the next time period and its secret key.

\begin{figure}[ht!]
	\centering
	\fbox{%
		\begin{tabularx}{0.95\columnwidth}{X|X} 
			\scriptsize
			\begin{algorithmic}[1]
				\Statex \underline{$\vect{K_1} \as \fenckg(1^\kappa, n )$}: 
				\State $S_1 \Ra \{0,1\}^\kappa$ 
				\State $(S_{2}, K_{1} ) \as \fprgupd(S_1)$
				\State $St \as (i \as 1, ctr \as \{0,1\}^{\ell})$
				\State \Return $\vect{K_1} \as (K_1, S_2)$
			\end{algorithmic}
			\algrule
			\begin{algorithmic}[1]
				\Statex \underline{$\vect{K_{i+1}} \as \fencupd(\vect{K_i})$} if $i \ge n$ then abort
				\State $(S_{i+2}, K_{i+1} ) \as \fprgupd(S_{i+1})$ \Comment{delete $\vect{K_i}$ }
				\State $St \as (i \as i+1, ctr \as ctr+\lceil m/\ell \rceil \bmod 2^\ell)$
				\State \Return $\vect{K_{i+1}} \as (K_{i+1}, S_{i+2})$
			\end{algorithmic}
			\algrule
			\begin{algorithmic}[1]
				\Statex \underline{$C_i \as \fseenc(\vect{K_i}, M_i)$}:  
				\Statex \hspace*{3pt} \underline{$\tilde{C}_i \as \textit{offline}(\vect{K_i}, ctr)$} 
				\For{$j = 1 , \ldots, \lceil m / \ell \rceil $}
				\State $\tilde{C}_j' \as \prf_1(K_i, ctr+j)$
				\EndFor
				\State $\vect{K_{i+1}} \as \fencupd(\vect{K_i})$
				\State $\tilde{C}_i \as \tilde{C}_1' \| \ldots \| \tilde{C}_{\lceil m / \ell \rceil}'$
			\end{algorithmic}
			&
			\scriptsize
			\begin{algorithmic}[1]
				\setcounter{ALG@line}{4}
				\Statex \hspace*{3pt} \underline{${C}_i \as \textit{online}(M_i, \tilde{C}_i)$} 
				\For{$j = 1 , \ldots, \lceil m / \ell \rceil $}
				\State $C_j' \as \tilde{C}_j' \oplus M_j'$
				\EndFor
				\State \Return $C_i \as C_1' \| \ldots \| C_{\lceil m / \ell \rceil}'$
			\end{algorithmic}
			\algrule
			\begin{algorithmic}[1]
				\Statex \underline{$M_i \as \fsedec(\vect{K_i}, C_i)$}: 
				\Statex \hspace*{3pt} \underline{$\tilde{M}_i \as \textit{offline}(\vect{K_i}, ctr)$} 
				\For{$j = 1 , \ldots, \lceil c / \ell \rceil $}
				\State $\tilde{M}_j' \as \prf_1(K_i, ctr+j)$
				\EndFor
				\State $\vect{K_{i+1}} \as \fencupd(\vect{K_i})$
				\State $\tilde{M}_i \as \tilde{M}_1' \| \ldots \| \tilde{M}_{\lceil c / \ell \rceil}'$
				\Statex \hspace*{3pt} \underline{${M}_i \as \textit{online}({C}_i, \tilde{M}_i)$} 
				\For{$j = 1 , \ldots, \lceil m / \ell \rceil $}
				\State $M_j' \as \tilde{M}_j' \oplus C_j'$
				\EndFor
				\State \Return $M_i \as M_1' \| \ldots \| M_{\lceil c / \ell \rceil}'$
			\end{algorithmic}
		\end{tabularx}%
	}
	\caption{CTR-based \fenc~with OO Capability}
	\label{alg:fenc}
\end{figure}

The encryption algorithm (\fencenc) accepts the private key $\vect{K_i}$ and a message $M_i$, and is decomposed into two phases:
{\em (i) offline phase.} computes a pre-ciphertext $\tilde{C}_i$ using \prf~evaluations on the incremental counter $ctr$, where the \prf~calls are determined based on the number of input chunks w.r.t \prf~block size $\ell$. Moreover, the private key is updated for the next time period. The latter ensures forward security of the \fenc~before transitioning to {\em online} stage.
{\em (ii) online phase.} computes the ciphertext $C_i$ by XORing each chunk $M_j'$ of message $M_i$ with the corresponding pre-ciphertext chunk. Note that the computational overhead of the online encryption is optimal with merely $\lceil m/\ell \rceil$ bitwise XOR operations of $\ell$-bit strings, while the offline computational overhead equals $\lceil m/\ell \rceil$ \prf~evaluations and one key update via \fseupd. The latter (\fencupd) evolves the private key $\vect{K_1}$ via \fprg~and increments the counter $ctr$ by $\lceil m/\ell \rceil$ modulo $2^\ell$.
Note that the counter operates modulo $2^{\ell}$ ($\ell \ge 128$), but counter exhaustion requires encrypting $2^{128}$ blocks, which is computationally infeasible.

The decryption algorithm (\fencdec) proceeds akin to the encryption algorithm (\fencenc) except that it accepts as input the current private key $\vect{K_i}$ and the ciphertext $C_i$ instead of the plaintext $M_i$.

%
%

\subsubsection{Instantiations}
We propose different instantiations of CTR-based \fenc~using different block ciphers. 
{\em (1) AES-128.} AES-128 is the most widely standardized block cipher, with a 128-bit block size and strong security guarantees against known cryptanalysis. Using AES-128 in CTR mode allows \fenc~to benefit from well-established hardware support, particularly the AES-NI instruction set on modern CPUs and specialized cryptographic processors on constrained IoT devices. 
{\em (2) Chacha20.} Chacha20 is a stream cipher based on ARX (add–rotate–xor) operations and is designed for efficiency in software and resistance against side-channel attacks. Chacha20 processes data in 512-bit blocks (64 bytes), and requires fewer data-dependent operations, therefore incurs lower instruction overhead compared to AES when hardware acceleration is unavailable.



\subsection{Generic \famac~Construction}
Fig. \ref{alg:umac} illustrates the algorithmic description of our proposed universal \famac~construction. It harnesses the OO properties of the universal \mac~as described in Def. \ref{subsec:buildingblocks} to efficiently compute the \mac~authentication tag during online runtime using a single \uhash~evaluation. 

\subsubsection{Algorithmic Description}
The key generation (\famackg) accepts the security parameter  $1^\kappa$, the epoch size $b$, and an aggregation flag $f_{agg}$. 
We investigate different aggregation methods in \famacagg:
{\em(1) Hash-based Accumulator}: computes a digest using a cryptographic hash function. It is an immutable aggregation (i.e., $\sigma \as H(\sigma_1 || \sigma_2)$).
{\em(2) Bitwise XOR}: is an efficient aggregation using XOR operations proportional to tag size. 
{\em(3) Modular Addition}: suitable for \umac s (e.g., Linear Congruential MAC (LC) \cite{etzel1999square}) where aggregation is performed via a modular addition ($Add_q$). This aggregation is additively homomorphic, allowing efficient batch verification. 
\famackg~generates two \fprg~states and computes initial private keys for \umac~components (i.e., \uhash~and \prf).

\begin{figure}[ht!]
	\centering
	\fbox{%
		\scriptsize
		\begin{tabularx}{0.95\columnwidth}{X|X} 
			\begin{algorithmic}[1]
				\Statex \underline{$\vect{K_1} \as \famackg(1^\kappa, n , b, f_{agg})$}: 
				\State $S_1^{\prf} \Ra \{0,1\}^\kappa$ and  $(S_{2}^{\prf}, K_{1}^{\prf} ) \as \fprgupd(S_1^{\prf})$
				\State $S_1^{\uhash} \Ra \{0,1\}^\kappa$ and  $(S_{2}^{\uhash}, K_{1}^{\uhash} ) \as \fprgupd(S_1^{\uhash})$
				\State $\vect{K_1^{\prf}} \as (K_1^{\prf}, S_2^{\prf})$ and $\vect{K_1^{\uhash}} \as (K_1^{\uhash}, S_2^{\uhash})$ 
				\State $St_1 \as (i \as 1)$ 
				\State \Return $\vect{K_1} \as (\vect{K_1^{\prf}}, \vect{K_1^{\uhash}})$
			\end{algorithmic}
			\algrule
			
			\begin{algorithmic}[1]
				\Statex \underline{$\vect{K_{i+1}} \as \famacupd(\vect{K_i})$} if $i \ge n$ then abort
				\State \textbf{if} $\vect{K_i}=(\vect{K_i}^{\prf}, \vect{K_i}^{\uhash})$ \textbf{or} $\vect{K_i}=\vect{K_i}^{\prf}$ \textbf{then}
				\State \hspace{3pt} $(S_{i+2}^\prf, K_{i+1}^\prf ) \as \fprgupd(S_{i+1}^\prf)$ \Comment{delete $\vect{K_i^{\prf}}$}
				\State \textbf{if} $\vect{K_i}=(\vect{K_i}^{\prf}, \vect{K_i}^{\uhash})$ \textbf{or} $\vect{K_i}=\vect{K_i}^{\uhash}$ \textbf{then}
				\State \hspace{3pt} $(S_{i+2}^\uhash, K_{i+1}^\uhash ) \as \fprgupd(S_{i+1}^\uhash)$ \Comment{delete $\vect{K_i^{\uhash}}$}
				\State \hspace{3pt} $St_{i+1} \as (i+1)$
				\State \Return $\vect{K_{i+1}}$
			\end{algorithmic}
			\algrule
			\begin{algorithmic}[1]
				\Statex \underline{$\sigma_{i_1, i_2+1} \as \famacagg(\sigma_{i_1,i_2}, \sigma_{i_2+1})$}: require $i_1 \bmod b = 1$ and $i_2 - i_1 < b$
				
				\State \textbf{if} $f_{agg} = H$ \textbf{then}
				$\sigma_{i_1,i_2+1} \as H (\sigma_{i_1,i_2} \| \sigma_{i_2+1} )$
				\State \textbf{if} $f_{agg} = \oplus$ \textbf{then}
				$\sigma_{i_1,i_2+1} \as \sigma_{i_1,i_2} \oplus \sigma_{i_2+1}$
				\State \textbf{if} $f_{agg} = Add_q$ \textbf{then}
				$\sigma_{i_1,i_2+1} \as \sigma_{i_1,i_2} + \sigma_{i_2+1} \bmod q$
				\State \Return $\sigma_{i_1, i_2+1}$
			\end{algorithmic}
			&
			{
				\begin{algorithmic}[1]
					\Statex \underline{$\sigma_i \as \famacsign(\vect{K_i}, M_i)$}: 
					\Statex \hspace{3pt} \underline{$\tilde{\sigma}_i \as \textit{offline}(\vect{K_i})$} 
					\State $\tilde{\sigma}_i \as \prf_2(K_i^\prf, i)$
					\State $\vect{K_{i+1}^\prf} \as \famacupd(\vect{K_i^\prf})$
					
					\Statex \hspace{3pt} \underline{${\sigma}_i \as \textit{online}(\vect{K_i}, M_i, \tilde{\sigma}_i)$} 
					\State $\bar{\sigma}_i' \as \uhash(K_i^\uhash, M_i)$
					\State $\sigma_{i} \as \famacagg(\tilde{\sigma}_i , \bar{\sigma}_i)$
					\State $\vect{K_{i+1}^\uhash} \as \famacupd(\vect{K_i^\uhash})$
					\State \Return $\sigma_i$
				\end{algorithmic}
				\algrule
				\begin{algorithmic}[1]
					\Statex \underline{$b \as \famacaver(\vect{K_i}, \vect{M_i}, \sigma_{i+b-1})$}: 
					
					\Statex \hspace{3pt} \underline{$\tilde{\sigma}_{i,i+b-1} \as \textit{offline}(\vect{K_i})$} 
					\For{$j = i , \ldots, i+b-1 $}
					\State $\tilde{\sigma}_j \as \prf_2(K_j^{\prf}, j)$
					\State $\tilde{\sigma}_{i,j} \as \famacagg(\tilde{\sigma}_{i,j-1},\tilde{\sigma}_j) $
					\State $\vect{K_{j+1}^\prf} \as \famacupd(\vect{K_j^\prf})$
					\EndFor                                        
					
					\Statex \hspace{3pt} \underline{$b \as \textit{online}(\vect{K_i}, \vect{M_i}, \tilde{\sigma}_{i,i+b-1})$} 
					\For{$j = i , \ldots, i+b-1 $}
					\State $\bar{\sigma}_{j} \as \uhash(K_j, M_{j})$
					\State $\bar{\sigma}_{i,j} \as \famacagg(\bar{\sigma}_{i,j-1},\bar{\sigma}_j) $
					\State $\vect{K_{j+1}^\uhash} \as \famacupd(\vect{K_j^\uhash})$
					\EndFor
					\State $\sigma_{i,i+b-1}' \as \famacagg(\tilde{\sigma}_{i,i+b-1}, \bar{\sigma}_{i,i+b-1})$
					\State \textbf{if} $\sigma_{i,i+b-1} = {\sigma}_{i,i+b-1}'$ \textbf{then return} $1$ else \Return $0$
			\end{algorithmic}}	
		\end{tabularx}%
	}
	\caption{Universal \famac~with OO Capability}
	\label{alg:umac}
\end{figure}

The tag generation algorithm (\famacsign) is split into two stages: 
{\em (i) offline}: accepts the private key $\vect{K_i}$ and computes the \prf~component in the underlying \umac~(i.e., $\tilde{\sigma_i}$) and updates the \prf's~private key.
{\em (ii) online}: computes the remaining computation to generate the authentication tag $\sigma_i$ and update the \uhash-related private key component. 
Unlike the CTR-based \fenc, the \famac's key update algorithm (\famacupd) can be invoked from both offline (steps 1-2 in \famacupd) and online (steps 3-5) stages of the tag generation (\famacsign), or independently after the execution of \famacsign. 

The tag verification algorithm (\famacaver) verifies a batch of received messages and a constant-size aggregate tag. During the offline stage, it computes individual \prf~outputs  and sequentially aggregates to obtain a constant-size digest $\tilde{\sigma}_{i,i+b-1}$. During the online stage, it computes individual \uhash~outputs and sequentially aggregates into $\bar{\sigma}_{i,i+b-1}$. The aggregate authentication tag is computed by aggregating the final \prf~and \uhash~outputs via \famacagg.

%
%

\subsubsection{Instantiations}

We propose two main instantiations of the universal \famac~construction, each differing in the choice of the underlying universal hash function $\uhash$, while both $\prf_1$ and $\prf_2$ are instantiated with the AES block cipher. These instantiations are designed to cover a range of efficiency and hardware/software trade-offs while preserving the theoretical security guarantees.

{\em (1) GHASH.} This variant uses the widely deployed \emph{GHASH} universal hash function \cite{mcgrew2004security} as $\uhash$. GHASH computes $\uhash(K,M)$ by interpreting each 128-bit message block as an element of the finite field $GF(2^{128})$ and evaluating a polynomial under multiplication by a secret hash subkey $K \in GF(2^{128})$. For a message $M$ split into $q = \lceil m/16 \rceil$ blocks $\{M_i'\}$, GHASH computes
$
\uhash(K,M) 
= (((M_1' \cdot K \oplus M_2')\cdot K \oplus \cdots \oplus M_q')\cdot K
$
where multiplication is performed modulo the irreducible polynomial $x^{128} + x^7 + x^2 + x + 1$. GHASH is an $\varepsilon$-almost-universal $\uhash$ with $\varepsilon = 2^{-128}$ for 16-byte blocks. It is highly efficient in software and hardware, relying solely on \texttt{XOR} operations and carry-less multiplications over $GF(2^{128})$. GHASH is the standard $\uhash$ in GCM mode of operation and is widely used in TLS, IPsec, and IEEE~802.1AE (MACsec).

{\em (2) Poly1305.} This variant employs the widely used \emph{Poly1305} universal hash function \cite{UMAC:AESPoly1305:Bernstein2005} as $\uhash$. Poly1305 computes $\uhash(K,M) = (\sum_{i=1}^{q} C_i \cdot K^{q-i} ) \pmod{2^{130}-5}$, where $K$ is 128-bit secret key derived from a \prf~and $\{C_i \as f(M)\}_{i=1}^{q=\lceil m / 16 \rceil}$ are evaluations of message blocks via a function $f$ \cite{UMAC:AESPoly1305:Bernstein2005}. Poly1305 is $\varepsilon$-almost-universal \uhash~with $\varepsilon = 2^{-103}$ for 16-byte blocks. It is highly efficient in software, leveraging 64-bit integer arithmetic and available vector instructions. This instantiation offers provable security and high performance on general-purpose CPUs and 32-bit microcontrollers (e.g., ARM Cortex M4). Poly1305 is widely used in numerous protocols (e.g., TLS 1.3) and present in standard-compliant security frameworks (e.g., OpenSSL, WireGuard).

\subsection{Generic Diamond Contruction}
\begin{figure*}[ht!]
	\centering
	\includegraphics[scale=0.5]{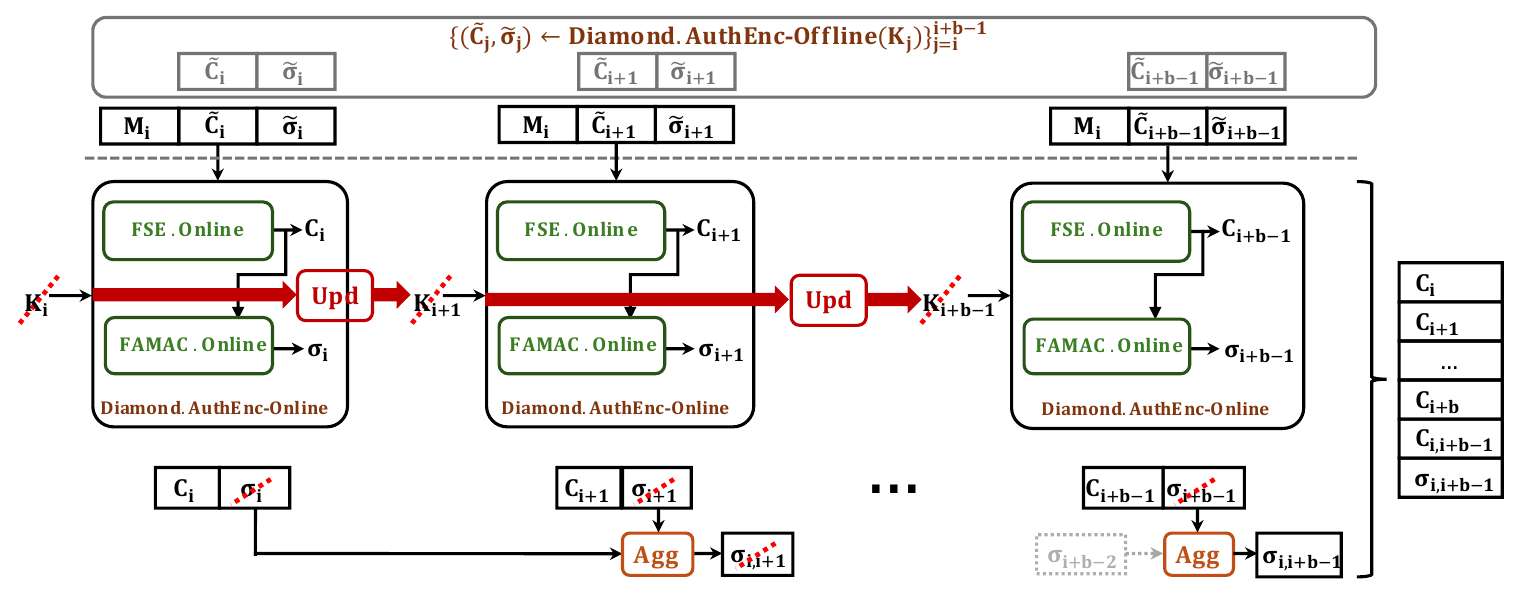}
	\vspace{-6pt}
	\caption{Overview of \dmd~Framework}
	\label{fig:graphene}
\end{figure*}

Fig. \ref{fig:graphene} and Fig. \ref{alg:graphenegeneric} illustrate the overview design and algorithmic description of our proposed \dmd~framework, respectively. It consists of an Encrypt-then-MAC construction of the CTR-based \fenc~and the universal \famac~schemes while harnessing offline batch precomputation.

\begin{figure}[ht!]
	\centering
	\fbox{%
		\begin{tabularx}{0.99\columnwidth}{X|X} 
			\scriptsize
					\begin{algorithmic}[1]
						\Statex \underline{$\vect{K_1} \as \dmdkg(1^\kappa, n , b, f_{agg})$}: 
						\State $\vect{K_1^\fenc} \as \fenckg(1^\kappa, n)$ 
						\State $\vect{K_1^\famac} \as \famackg(1^{\kappa}, n, b, f_{agg})$ 
						\State $St_i \as (i \as 1 , ctr \Ra \{0,1\}^{\kappa})$
						\State \Return $\vect{K_1} \as (\vect{K_1^\fenc}, \vect{K_1^\famac} )$
					\end{algorithmic}
					\algrule
					
					\begin{algorithmic}[1]
						\Statex \underline{$\vect{K_{i+1}} \as \dmdupd(\vect{K_i})$} if $i \ge n$ then abort
						\State $\vect{K_{i+1}}' \as \fencupd(\vect{K_i}')$
						\State $\vect{K_{i+1}}'' \as \famacupd(\vect{K_i}''')$
						\State $St_i \as (i \as i+1, ctr \as ctr+ \lceil m/\ell \rceil \bmod 2^\kappa)$
						\State \Return $\vect{K_{i+1}} \as (\vect{K_{i+1}}', \vect{K_{i+1}}'')$
					\end{algorithmic}
					\algrule
					
					\begin{algorithmic}[1]
						\Statex \underline{$\sigma_{i_1, i_2+1} \as \dmdagg(\sigma_{i_1,i_2}, \sigma_{i_2+1})$}: 
						
						\State $\sigma_{i_1, i_2+1} \as \famacagg(\sigma_{i_1,i_2}, \sigma_{i_2})$
						\State \Return $\sigma_{i_1, i_2+1}$
					\end{algorithmic}
					\algrule
					
					\begin{algorithmic}[1]
						\Statex \underline{$(C_i, \sigma_i) \as \dmdencmac(\vect{K_i}, M_i)$}: 
						\Statex \hspace{3pt} \underline{\em $(\tilde{C}_i, \tilde{\sigma}_i) \as$ offline ($\vect{K_i}$)}
						\State $\tilde{C}_i \as \fencenc\offline(\vect{K_i^\fenc})$
						\State $\tilde{\sigma}_i \as \famacsign\offline(\vect{K_i^\famac})$
						
					\end{algorithmic}
					
					&
					\scriptsize
				
				\begin{algorithmic}[1]
					\setcounter{ALG@line}{2}
					\Statex \hspace{3pt} \underline{\em $(C_i, \sigma_i) \as$ online ($\vect{K_i}, M_i, \tilde{C}_i, \tilde{\sigma}_i$)}
					\State ${C}_i \as \fencenc\online(M_i, \tilde{C_i})$
					\State ${\sigma}_i \as \famacsign\online(\vect{K_i^\famac}, C_i, \tilde{\sigma_i})$
					\State \Return  $(C_i, \sigma_i)$
				\end{algorithmic}
			\algrule
					
					\begin{algorithmic}[1]
						\Statex \underline{$\vect{M_i} \as \dmdverdec(\vect{K_i}, \vect{C_i}, \sigma_{i,i+b-1})$}: 
						
						\Statex \hspace{3pt} \underline{\em $(\vect{\tilde{M}_i}, \tilde{\sigma}_{i,i+b-1}) \as$ offline ($\vect{K_i}$)}
						\For{$j=i , \ldots, i+b-1$}
						\State $\tilde{M}_j \as \fencdec\offline(\vect{K_i^\fenc})$
						\EndFor
						\State $\vect{\tilde{M}_i} \as (\tilde{M}_i, \ldots, \tilde{M}_{i+b-1})$
						\State $\tilde{\sigma}_{i,i+b-1} \as \famacaver\offline(\vect{K_i^\famac})$
						
						\Statex \hspace{3pt} \underline{\em $\vect{M_i} \as$ online ($\vect{K_i}, \vect{C_i}, \vect{\tilde{M}_i}, \tilde{\sigma}_{i,i+b-1}$)}
						\State $b \as \famacaver\online(\vect{K_i^\famac}, \vect{C_i}, \tilde{\sigma}_{i,i+b-1})$
						\State \textbf{if} $b = 0$ \textbf{then return }  $\perp$
						\For{$j=i , \ldots, i+b-1$}
						\State ${M}_j \as \fencdec\online(C_j , \tilde{M}_j)$
						\EndFor
						\State \Return $\vect{M_i} \as (M_i, \ldots, M_{i+b-1})$
					\end{algorithmic}		
	\end{tabularx}}
	\caption{Generic \dmd~Framework}
	\label{alg:graphenegeneric}
\end{figure}

\subsubsection{Algorithmic Description}
The key generation algorithm (\dmdkg) accepts the security parameter $\kappa$, the maximum $\authenc(.)$~operations $n$, the epoch size $b$, and an aggregation flag $f_\agg$. It initializes independent key states for the forward-secure encryption scheme (\fenc) and the forward-secure MAC scheme (\famac). An internal variable $St_i = (i, ctr)$ is maintained, where $i$ is the epoch index and $ctr$ is a counter initialized at random. The initial secret key $\vect{K_1}$ is returned. 
At each iteration, the encryption and \mac~secret keys are updated using their respective update algorithms. The counter is increased by the number of consumed blocks $\lceil \frac{m}{\ell} \rceil$, and the previous secret key is securely deleted to ensure forward security.

To encrypt then authenticate a message $M_i$, \dmdencmac~incurs two stages:
\emph{(i) Offline phase.} It precomputes $\tilde{C}_i$ and $\tilde{\sigma}_i$ using \fenc~and \famac. 
\emph{(ii) Online phase.} It compute ciphertext ${C_i}$ and partial authentication tag $\bar{\sigma_i}$. Finally, it outputs the ciphertext-tag pair $(M_i,\sigma_i)$.

To perform verified decryption of a batch of ciphertexts, \dmdverdec~first performs:
\emph{(i) Offline phase.} Precompute decryption masks $\{\tilde{M}_j\}$ and the partial aggregate tag $\tilde{\sigma}_{i,i+b-1}$.
\emph{(ii) Online phase.} Verify the aggregate tag via \famacaver\online. If verification fails, output $\perp$. Otherwise, decrypt each ciphertext $C_j$ as $M_j = \fencdec\online(C_j, \tilde{M}_j)$ and return $(M_i,\ldots,M_{i+b-1})$.

%
Upon every \authenc~operation at iteration $i$, the sender securely deletes the (precomputed) keys. 

\subsubsection{Instantiations}
We present two instantiations of \dmd~by selecting AES-128 as the underlying $\prf_2$ and deriving the designs from the $\prf_1$ and the universal hash \uhash~functions embedded within the corresponding \fse~and \famac~construction, respectively:

(1) $\dmda$. This variant integrates AES-128 as the stream cipher in the \fse~algorithm and employs GHash as the universal hash function in \famac. GHash operates over the finite field $\mathbb{F}_{2^{128}}$ and provides $2^{-128}$-universality, making $\dmda$ a strong, standard-compliant instantiation with the NIST security recommendations. The key update mechanism leverages a forward-secure pseudo-random generator (\fprg) instantiated with AES-128, which provide resilience against key compromise. We selected XOR-based aggregation to provide high efficient and minimal energy usage, unlike cryptographic hashing or modular addition.

(2) $\dmdb$. This configuration integrates ChaCha20 with Poly1305 in the \fse~and \famac~algorithms, respectively, following the well-established \aee~construction \cite{langley2021rfc7539}. Poly1305 operates modulo $2^{130}-5$ and achieves $2^{-103}$-universality, offering a strong security-performance balance for embedded and real-time systems. 
Compared to $\dmda$, \dmdb~is optimized for software-centric and ARM-based platforms, which benefit from the ChaCha20 ARX design and Poly1305 high efficiency. Similar to $\dmda$, the key-update procedure relies on AES-128-based \fprg~for forward-secure key evolution, while we also opt for XOR-based aggregation akin to \dmda.

%% file: perf_analysis.tex
\section{PERFORMANCE EVALUATION}
\label{sec:perf_analysis}

This section presents a detailed performance analysis of our proposed \dmd~framework and its building blocks. We assess \dmd~with \faae~counterparts from existing AE standards.

\subsection{Evaluation Setup}

\noindent \textbf{Hardware Configuration.} \sloppy We experimentally evaluate the performance of the generic \dmd~framework instantiated with its underlying \fse~and \famac~building blocks across a heterogeneous set of embedded and commodity hardware platforms. This cross-platform evaluation allows us to quantify the efficiency and scalability of \dmd~under different computational and memory constraints, ranging from resource-constrained 8-bit microcontrollers (MCUs) to high-end general-purpose processors. The experimental platforms are summarized in Table \ref{tab:hardware_characteristics}:

\noindent {\em 1. Commodity Hardware (baseline, x86\_64). } As our reference platform, we use a desktop equipped with an Intel Core i9-9900K @ 3.6~GHz and 64GB of DDR4 memory. We consider optimizations such as AES-NI and AVX2 instruction sets, enabling hardware-accelerated evaluation of the AES-based instantiations of both \fse~and \famac~building blocks.

\noindent {\em 2. ARM Cortex A72 (edge-class, 64-bit).} 
To assess \dmd’s performance on low-end edge devices, we use a Raspberry Pi 4 Model B featuring a 64-bit quad-core ARM Cortex-A72 SoC @ 1.8 GHz and equipped with 2GB of SDRAM memory. 
This platform represents a mid-tier embedded processor (e.g., found in IoT gateways), therefore allowing the scalability assessment of \graphene~in moderately constrained environments {\em without hardware-based cryptographic accelerators}.

\noindent {\em 3. ARM Cortex-M4 (MCU-class, 32-bit).} For the MCU category, we use STM32F439ZI that integrates a 32-bit ARM Cortex-M4 @ 168MHz, with 2MB of flash and 256KB of SRAM. It features a hardware cryptographic accelerator supporting AES-{128,192,256}, SHA-256, and HMAC primitives.
We enabled hardware AES acceleration in our experiments.

\noindent {\em 4. AVR ATmega2560 (constrained MCU, 8-bit).} 
For constrained MCUs, we use an Arduino Mega~2560 board based on the 8-bit AVR ATmega2560 @ 16MHz, equipped with 256KB of flash memory and 8KB of SRAM. This platform represents low-end devices without hardware acceleration. Thus, it enables characterizing its minimal footprint and the performance of lightweight instantiations of \fse~and \famac~under high resource constraints.
\vspace{2pt}

\begin{table}[ht]
	\centering
	\caption{Hardware configurations of our selected platforms.}
	\label{tab:hardware_characteristics}
	\resizebox{0.99\textwidth}{!}{
	\begin{tabular}{lcccc}
		\toprule
		\textbf{Characteristic} 
		& \textbf{x86\_64} 
		& \textbf{Cortex-A72} 
		& \textbf{Cortex-M4} 
		& \textbf{AVR ATmega2560} \\
		\midrule
		CPU Architecture   & 64-bit CISC (Intel Core i9-9900K) & 64-bit RISC (ARMv8-A) & 32-bit RISC (ARMv7E-M) & 8-bit RISC (AVR) \\
		Core Configuration & 8 cores / 16 threads              & Quad-core              & Single-core             & Single-core \\
		Frequency          & 3.6 GHz                           & 1.8 GHz                & 168 MHz                 & 16 MHz \\
		Memory (SRAM/DRAM) & 64 GB DDR4                        & 2 GB SDRAM             & 256 KB SRAM             & 8 KB SRAM \\
		Flash Storage      & --                                & microSD / external     & 2 MB                    & 256 KB \\
		Crypto Acceleration& AES-NI, AVX2                      & None                   & AES, SHA-256, HMAC & None \\
		Power Class        & High-performance desktop           & Edge-class (IoT gateway) & Low-power MCU          & Ultra-low-power MCU \\
		\bottomrule
	\end{tabular}
	}
\end{table}

\noindent \textbf{Software Configuration.}
Throughout the different heterogeneous platforms, we carefully select standard-compliant and most efficient libraries and open-source implementations, as follows: 

\noindent {\em 1. On Commodity Hardware.} We use OpenSSL\footnote{\url{https://github.com/openssl/openssl}} to implement cryptographic primitives (e.g., cryptographic hash functions, universal hash functions, and \prf~functions). 

\noindent {\em 2. On Edge Class.} We use a cross-compiled build of OpenSSL targeting ARM Cortex-A72.

\noindent {\em 3. On MCU Class.} We use wolfSSL\footnote{\url{https://github.com/wolfSSL/wolfssl/tree/master/IDE/STM32Cube}} for Cortex-M4, given its compliance with cryptographic standards (e.g., FIPS~140-2/3, TLS~1.3) and optimized performance for embedded targets.

\noindent {\em 4. On 8-bit MCUs.} To the best of our knowledge, there exist only a limited number of cryptographic software libraries, each offering a distinct and often incomplete set of implementations for fundamental cryptographic primitives. Consequently, no available library provides a unified, performance-optimized solution in order to implement \dmd~variants.
%
%
For instance, the Arduino Cryptography Library\footnote{\url{https://github.com/rweather/arduinolibs}}
offers the essential primitives required to instantiate \graphene. However, it lacks optimizations (e.g., at the assembly level). Furthermore, its object-oriented design introduces non-negligible computational and memory overheads.
%
%
Likewise, the AVR-Crypto-Lib\footnote{\url{https://github.com/cantora/avr-crypto-lib}} was implemented
in C and Assembly, and is over a decade old. It also does not support the Poly1305 universal hash function and integrated authenticated encryptions, which are integral for our performance analysis.
%
%
The $\mu$NaCl framework\footnote{\url{https://munacl.cryptojedi.org/atmega.shtml}}
\cite{hutter2013nacl}, is also outdated by more than a decade and provides only Poly1305 as a universal hash and SHA-512 as its sole cryptographic hash function.
%
%
%
%
In parallel, Cardoso et al. \cite{cardoso2019felics} proposed a unified authenticated encryption evaluation framework that includes existing implementations from earlier libraries. 
In our work, we benchmark our deployed cryptographic primitives across the aforementioned libraries and select the most efficient implementation.
\vspace{2pt}

\noindent \textbf{Counterpart Selection.} 
To provide a fair performance evaluation, we evaluate \dmd~instantiations against initial \graphene~variants: i.e., \graphenegcm~and \graphenepoly, which we now refer to as \graphenea~and \grapheneb, respectively. 
We also included two \faae~instantiations, \faaea~and \faaeb, which incorporate the FIPS standard AES-128-GCM and the NIST lightweight standard Ascon128a~\cite{turan2024ascon} as an integrated \aee s, and SHA-256 and Ascon-hash-256 for key update, respectively.  
We selected the bitwise XOR operation as an aggregation method for all schemes. 
The cryptographic components and parameters of \dmd~and its counterparts are depicted in Table \ref{tab:scheme_components}.

\begin{table}[t]
	\centering
	\caption{Components and parameters in FAAE (Ascon), Graphene, and Diamond schemes.}
	\label{tab:scheme_components}
	\resizebox{0.95\textwidth}{!}{
		\begin{tabular}{lcccccccc}
			\toprule
			\textbf{Scheme} &
			\multicolumn{3}{c}{\textbf{$\prf_1$ (in \fse)}} &
			\multicolumn{3}{c}{\textbf{$\uhash$ (in \famac)}} &
			\textbf{Key Update} &
			\textbf{Aggregation} \\
			
			\cmidrule{1-1} \cmidrule(lr){2-4} \cmidrule(lr){5-7} \cmidrule(lr){8-8} \cmidrule(lr){9-9}
			
			& \texttt{Name} & \texttt{Type} & \texttt{Key/State} &
			\texttt{Name} & \texttt{Modulus} & \texttt{Universality} &
			\texttt{Method} &
			\texttt{Operator} \\
			
			\midrule
			
			$\boldsymbol{\textbf{FAAE}_1}$ &
			\multicolumn{3}{c}{AES-128-GCM} &
			\multicolumn{3}{c}{ (integrated)} &
			HASH(SHA-256) & XOR \\
			
			$\boldsymbol{\textbf{FAAE}_2}$ &
			\multicolumn{3}{c}{Ascon128a} &
			\multicolumn{3}{c}{ (integrated)} &
			HASH(AsconHash256) & XOR \\
			
			$\boldsymbol{\textbf{Graphene}_1}$ &
			AES-128 & SPN & 128 / 128 &
			GHASH & $2^{128}$ & $2^{-128}$ &
			HASH(SHA-256) & XOR \\
			
			$\boldsymbol{\textbf{Graphene}_2}$ &
			ChaCha20 & ARX & 256 / 512 &
			Poly1305 & $2^{130}-5$ & $2^{-103}$ &
			HASH(SHA-256) & XOR \\
			
			$\boldsymbol{\textbf{Diamond}_1}$ &
			AES-128 & SPN & 128 / 128 &
			GHASH & $2^{128}$ & $2^{-128}$ &
			FPRG(AES-128) & XOR \\
			
			$\boldsymbol{\textbf{Diamond}_2}$ &
			ChaCha20 & ARX & 256 / 512 &
			Poly1305 & $2^{130}-5$ & $2^{-103}$ &
			FPRG(AES-128) & XOR \\
			
			\bottomrule
		\end{tabular}
	}
\end{table}

\noindent \textbf{Evaluation Metrics.}
Our evaluation addresses three key research questions (RQs) to fairly evaluate \dmd~and its selected \faae~and \graphene~counterparts:

\begin{enumerate}[\textit{RQ.}1]
    \item \textit{Preprocessing Overhead and Overall Energy Drawings.} What is the cost of offline preprocessing, and what is its impact on the overall energy usage, including online and transmission costs?
    \item \textit{Online Latency.} How much can OO optimization reduce encryption throughput and E2E verification delay on constrained IoT devices and nearby IoT servers, respectively?
    \item \textit{Fairness \& Baseline.} How does \dmd~compare fairly to \graphene~and \faae~instantiations based on prior constructions and the NIST lightweight standard Ascon?
\end{enumerate}

\subsection{Performance Analysis}

By combining the existing instantiations of \fse~and \famac~with the corresponding aggregation functions, we derive multiple \dmd~variants, each offering a distinct performance advantage on at least one evaluation metric.
Subsequently, we perform a comprehensive evaluation of these \dmd~variants on diverse hardware platforms to identify the most suitable configuration.

\begin{figure}[t]
	\centering
	
	\begin{subfigure}[b]{0.31\textwidth}
		\centering
		\includegraphics[width=\textwidth]{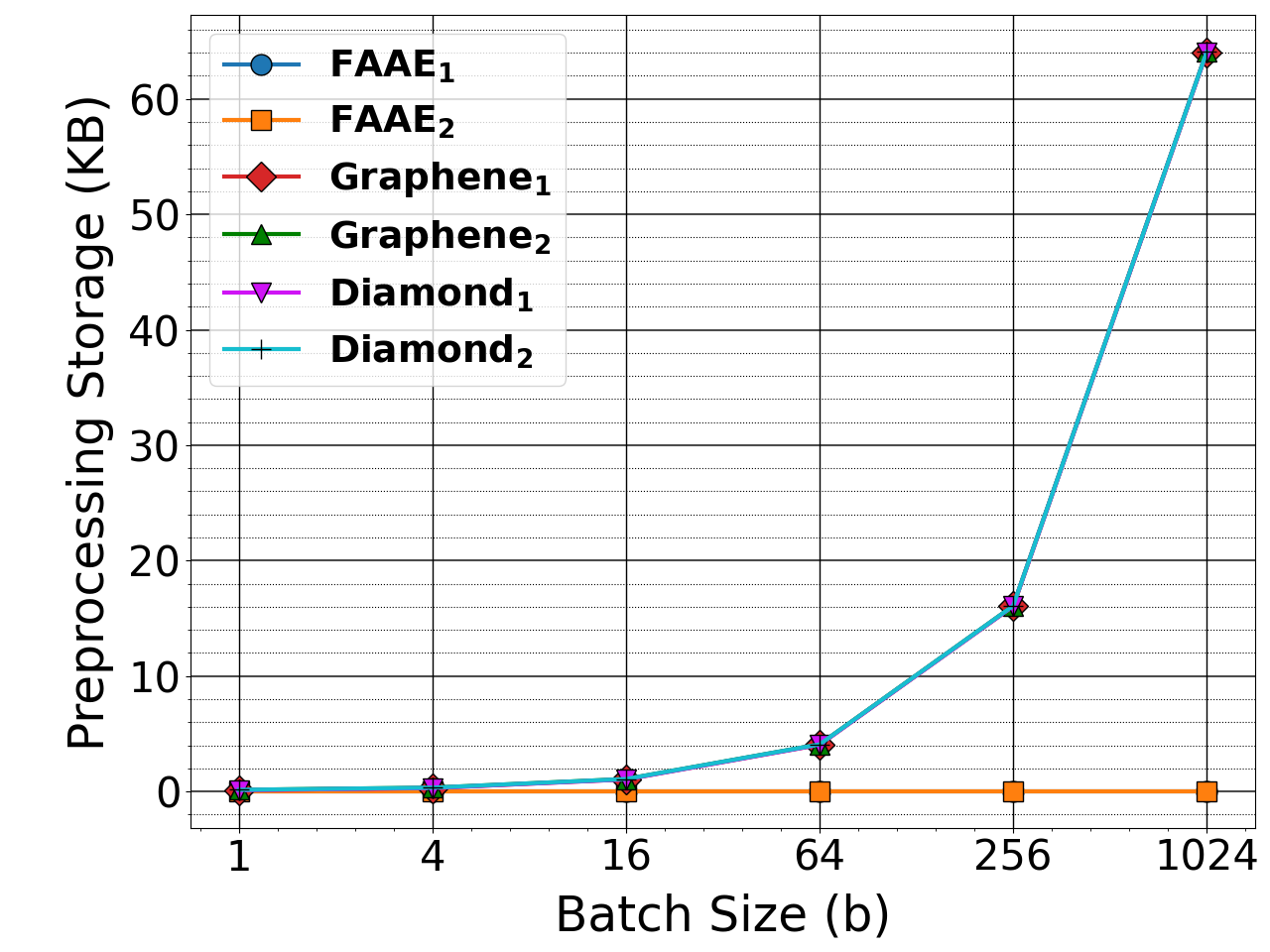}
		\caption{$\boldsymbol{m=16 }$ Bytes}
	\end{subfigure} 
	\hfill
	\begin{subfigure}[b]{0.31\textwidth}
		\centering
		\includegraphics[width=\textwidth]{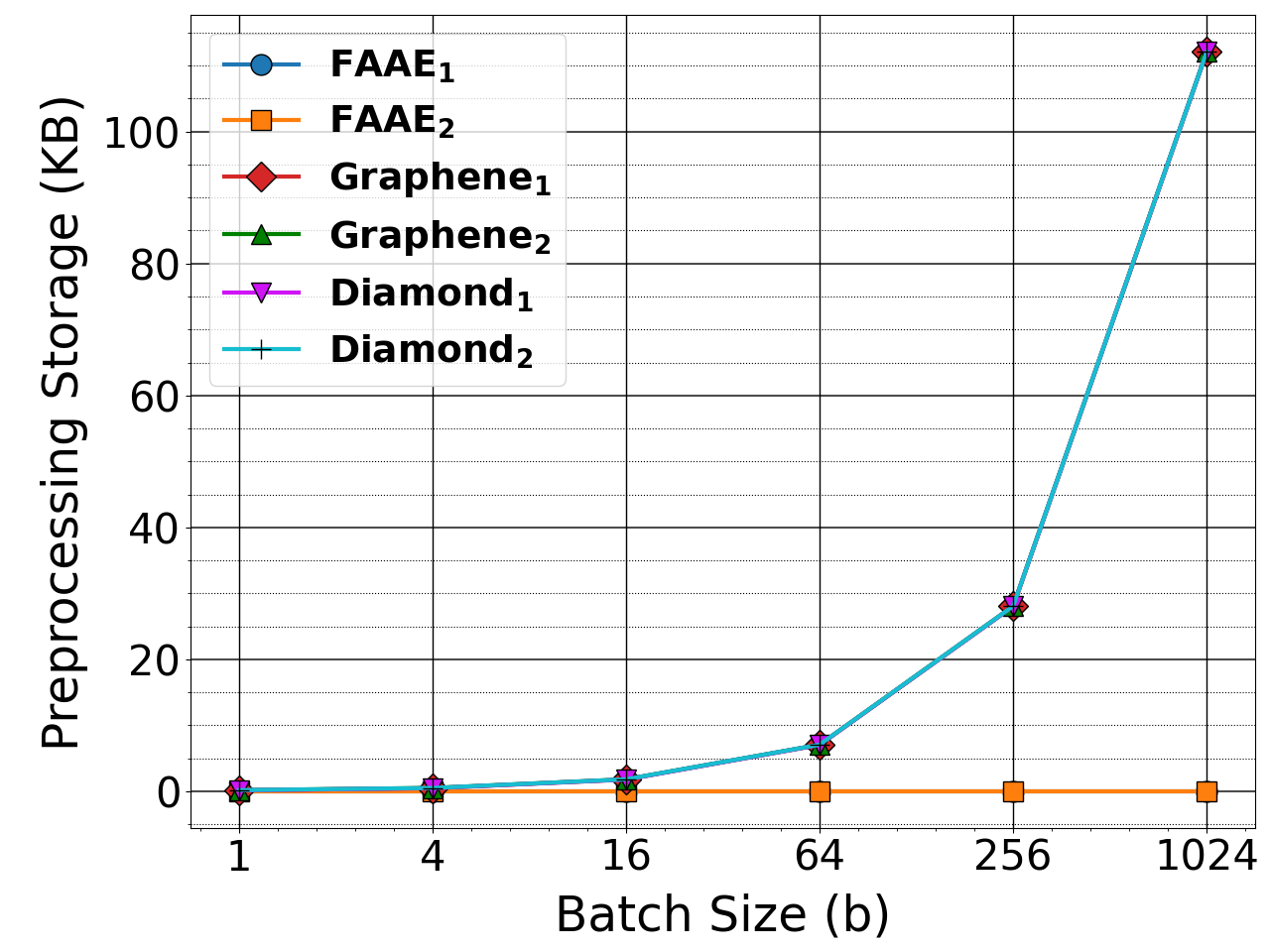}
		\caption{$\boldsymbol{m=64}$ Bytes}
	\end{subfigure} 
	\hfill
	\begin{subfigure}[b]{0.31\textwidth}
		\centering
		\includegraphics[width=\textwidth]{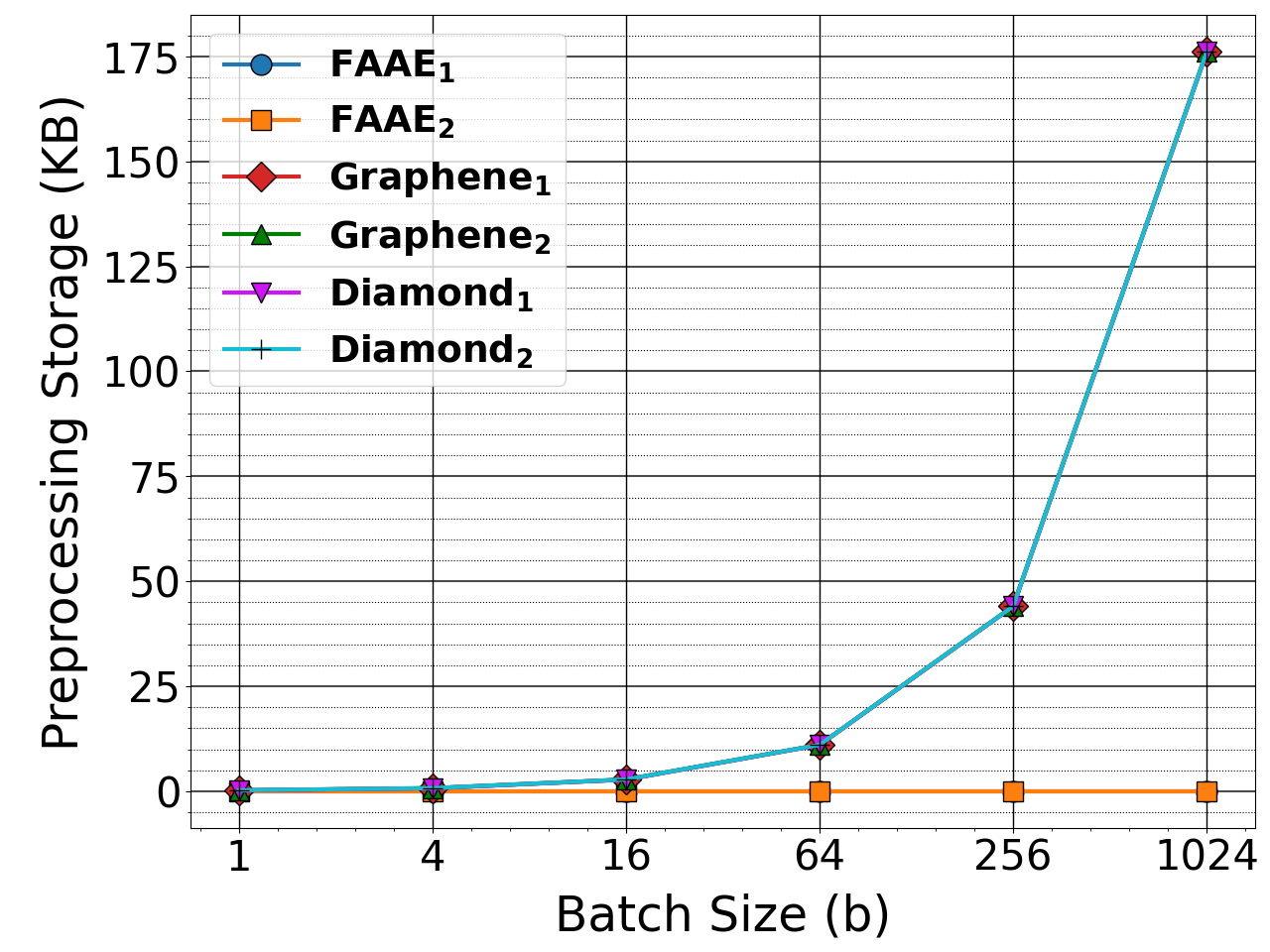}
		\caption{$\boldsymbol{m=128}$ Bytes}
	\end{subfigure} 
	\caption{Offline preprocessing storage overhead of \dmd~variants under different payload and batch sizes
	}
	\label{fig:storage_overhead}
\end{figure}

\begin{figure}[t]
	\centering
	
	\begin{subfigure}[b]{0.31\textwidth}
		\centering
		\includegraphics[width=\textwidth]{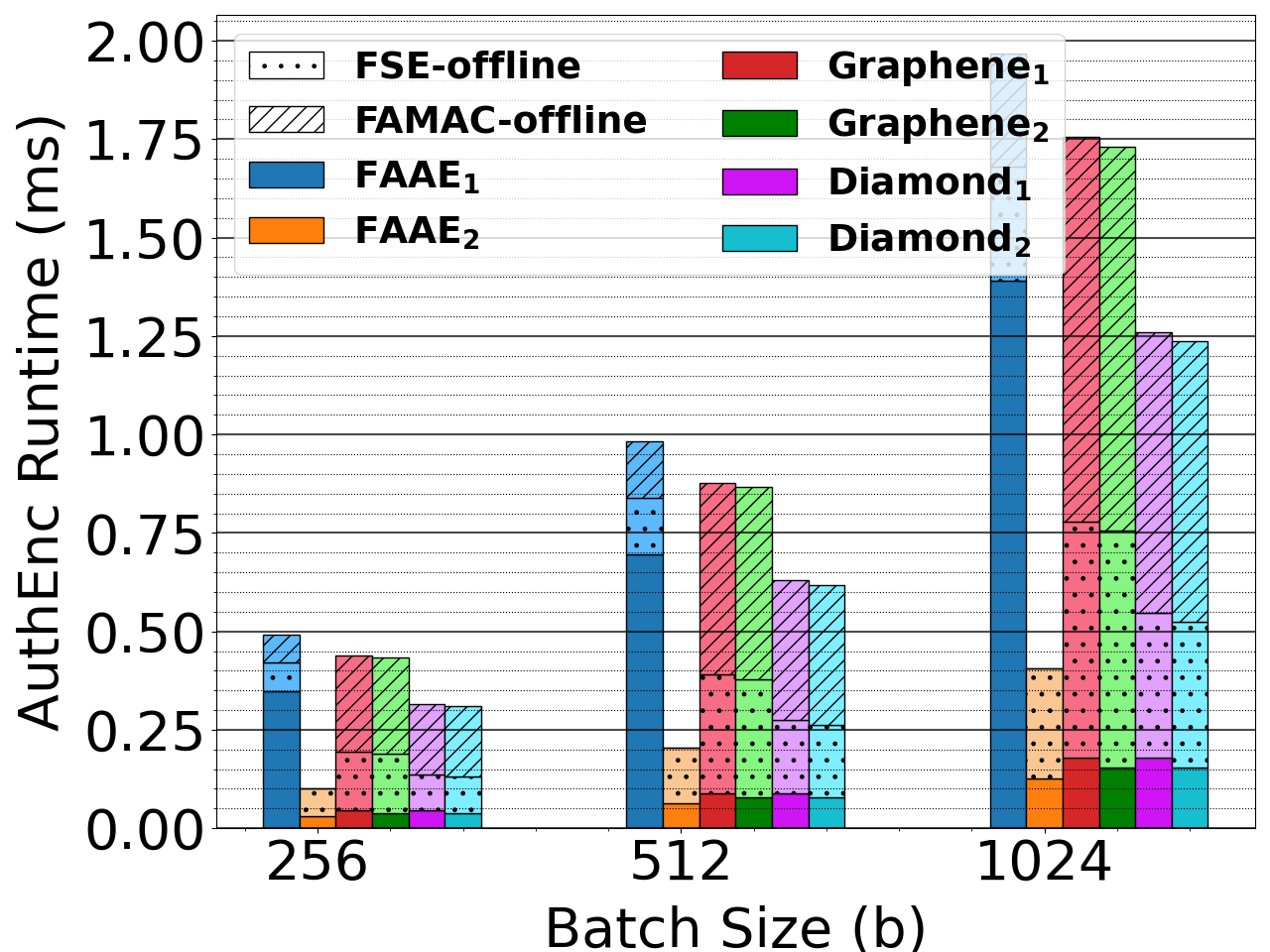}
		\caption{$\boldsymbol{m=16}$ Bytes}
	\end{subfigure} 
	\hfill
	\begin{subfigure}[b]{0.31\textwidth}
		\centering
		\includegraphics[width=\textwidth]{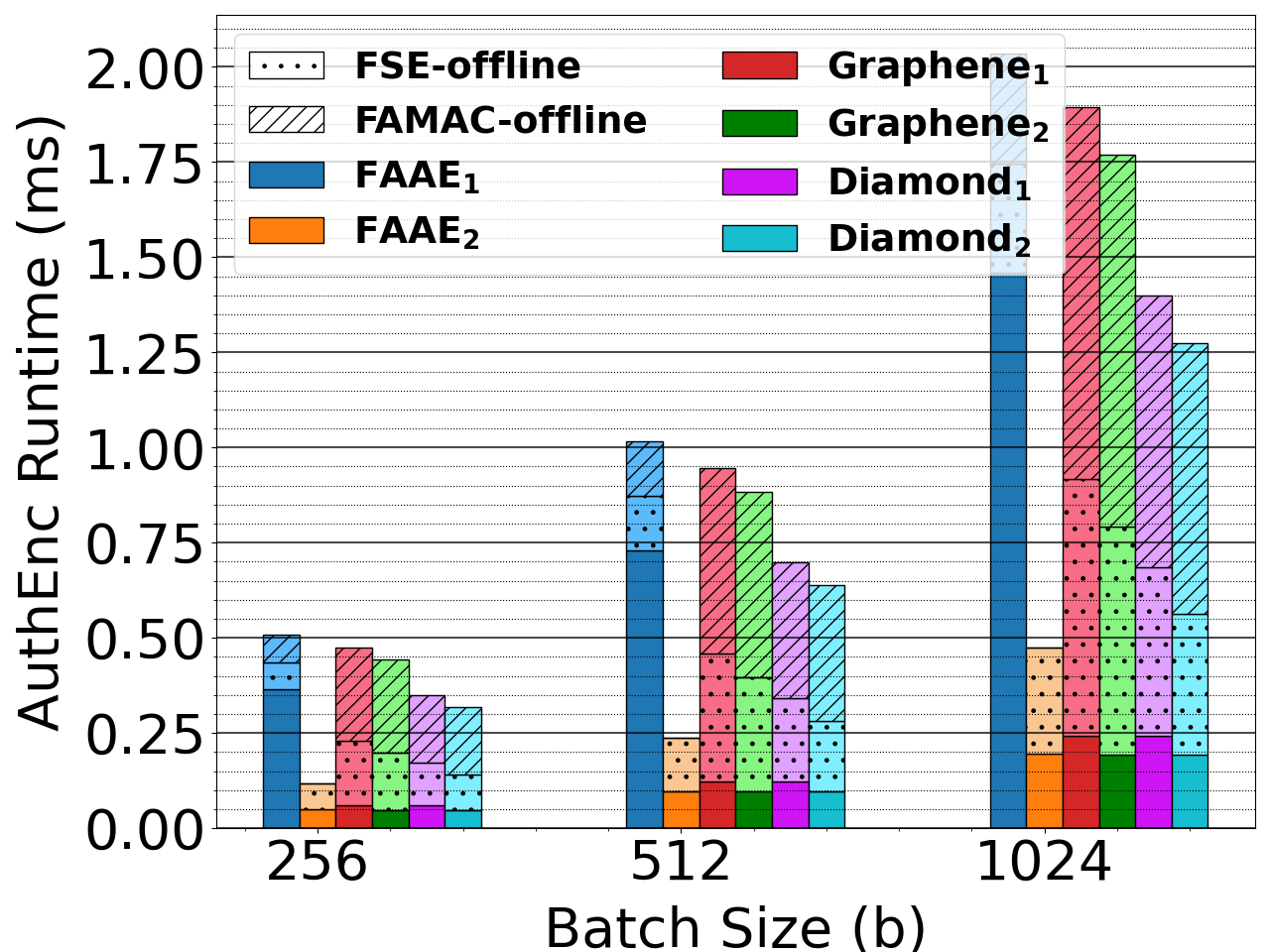}
		\caption{$\boldsymbol{m=64}$ Bytes}
	\end{subfigure} 
	\hfill
	\begin{subfigure}[b]{0.31\textwidth}
		\centering
		\includegraphics[width=\textwidth]{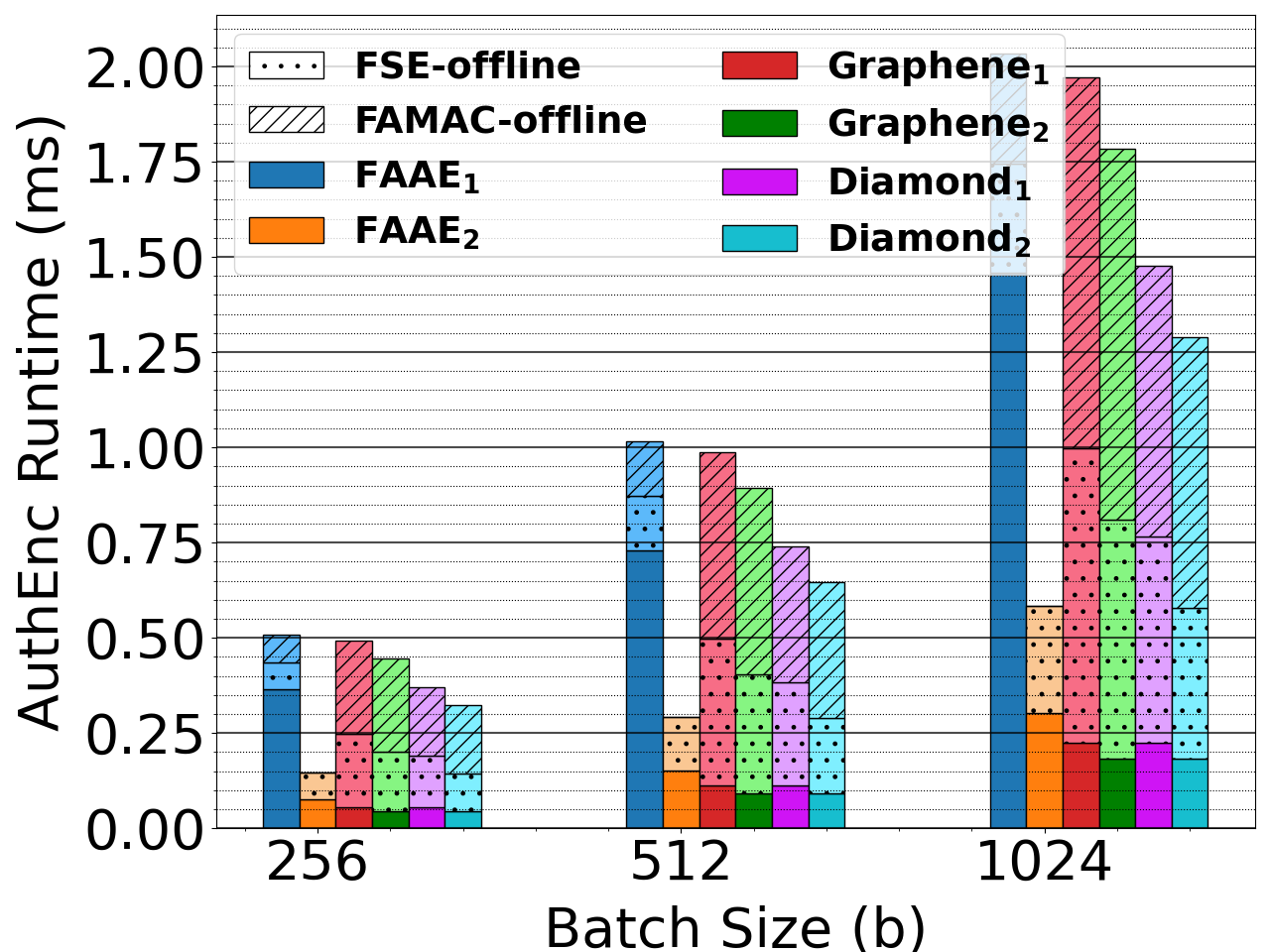}
		\caption{$\boldsymbol{m=128}$ Bytes}
	\end{subfigure} 
	
	\caption{Runtime of \dmd~variants under different payload and batch sizes on x86\_64 commodity platform
	}
	\label{fig:runtime_x86_64}
\end{figure}

\noindent \textbf{Storage Overhead.} At first, we analyze the storage overhead incurred during the {\em offline} stage of \dmdencmac~execution, under different message sizes (i.e., $m$) and batch sizes (i.e., $b$).
Fig. \ref{fig:storage_overhead} illustrates the offline preprocessing storage footprint of the proposed \dmd~variants and their baseline counterparts for varying batch sizes ($2^0$ to $2^{10}$). 
Specifically, \dmd~precomputes the pre-ciphertexts $\tilde{C}_i$ in the \fse~instantiation and the \prf~derived tags $\tilde{\sigma}_i$ under the \famac~construction. Recall that \graphene~and \dmd~incur identical storage overheads, as their distinguishing feature lies in the key evolution strategy (i.e., cryptographic hash update vs.~PRF-based \fprg). For a batch size of $2^{10}$, the total offline storage overhead reaches $60$~KB and $175$~KB when processing 16-byte and 128-byte message payloads, respectively.
This footprint is well-supported by our selected platforms. It constitutes less than 12\% of on-chip SRAM on the ARM Cortex-M4 and is negligible on the edge-class ARM Cortex-A72 systems. 
%
%
\faae~counterparts exhibit a minimal memory footprint, where they only store the \faae~secret key, which equals 32 and 16 bytes, for \faaea~and \faaeb. 
\vspace{2pt}

\noindent \textbf{x86\_64.} 
Fig.~\ref{fig:runtime_x86_64} illustrates the runtime of the cumulative authenticated encryption (\authenc) overhead for our benchmarked schemes evaluated under varying payload lengths and batch sizes. The runtime encapsulate both the {online} processing costs and the {\em offline} preprocessing for the \fse~and \famac~algorithms. 
Given that the experimental platform is a resourceful multi-core x86\_64 commodity hardware, the {\em online} execution footprints of the selected \faae{}, \graphene{}, and \dmd{} constructions are comparable for moderate message sizes except for the standard \faaea. Moreover, the online evaluation latency across all variants of \graphene~and \dmd~is indistinguishable under the selected parameters, as the core comparison is reduced to the evaluation of the key update mechanism.
Specifically, the \dmda~and \dmdb{} variants, optimized via \prf-based key update, demonstrate $1.43\times$ and $1.46\times$ speedups, respectively, over their initial \graphenea~and \grapheneb{} counterparts when instantiated over 128-byte payloads. This corresponds to achieving roughly a $30\%$ reduction in cycle count.
These savings translate directly into reduced energy usage and improved verification efficiency when payload lengths or batch sizes are further increased, making \dmd{} the best candidate for performance-critical deployments.
Compared to their \faaea~counterpart, \dmd~instantiations deliver substantial performance gains, achieving up to $9\times$ reduction in online \authenc~latency. Compared to NIST lightweight \faaeb, \dmdb~exhibits a nuanced cost profile by incurring moderately higher overhead on short 16-byte payloads (i.e., $1.22\times$ slower) while surpassing on larger 128-byte telemetry by delivering $1.66\times$ faster online \authenc~throughput. 
%
\vspace{2pt}

\begin{figure}[t!]
	\centering
	
	\begin{subfigure}[b]{0.31\textwidth}
		\centering
		\includegraphics[width=\textwidth]{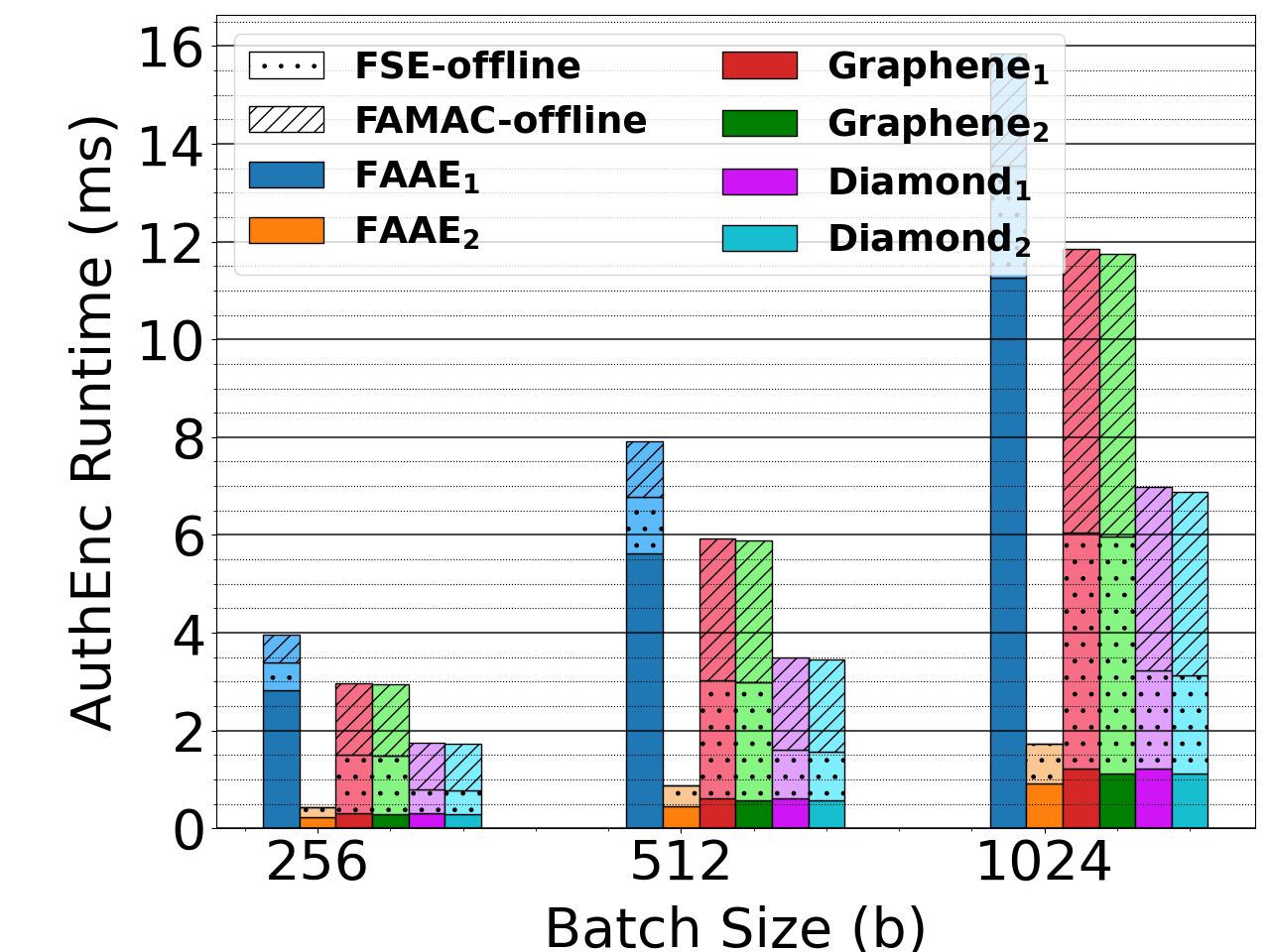}
		\caption{$\boldsymbol{m=16}$ Bytes}
	\end{subfigure} 
	\hfill
	\begin{subfigure}[b]{0.31\textwidth}
		\centering
		\includegraphics[width=\textwidth]{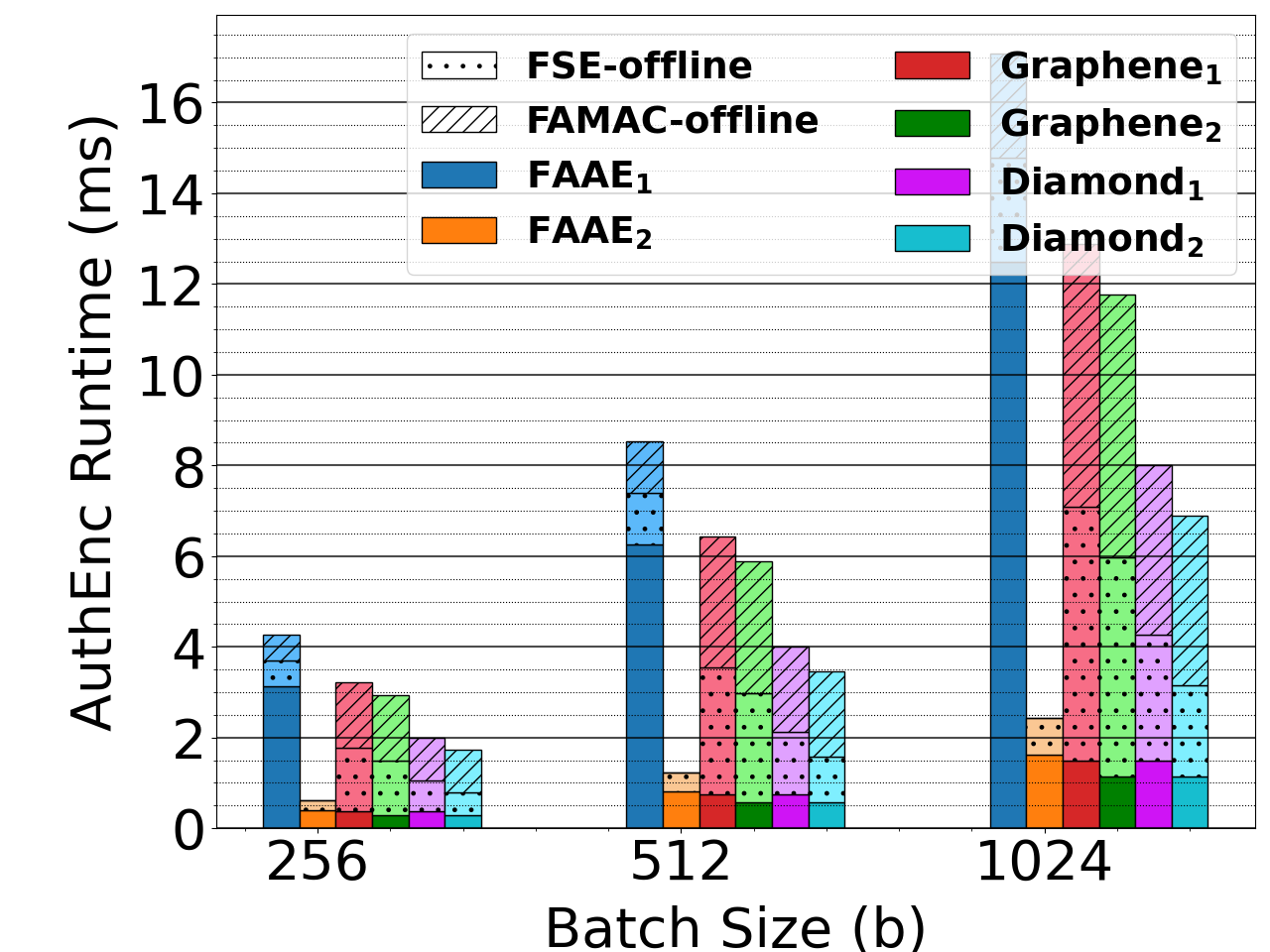}
		\caption{$\boldsymbol{m=64}$ Bytes}
	\end{subfigure} 
	\hfill
	\begin{subfigure}[b]{0.31\textwidth}
		\centering
		\includegraphics[width=\textwidth]{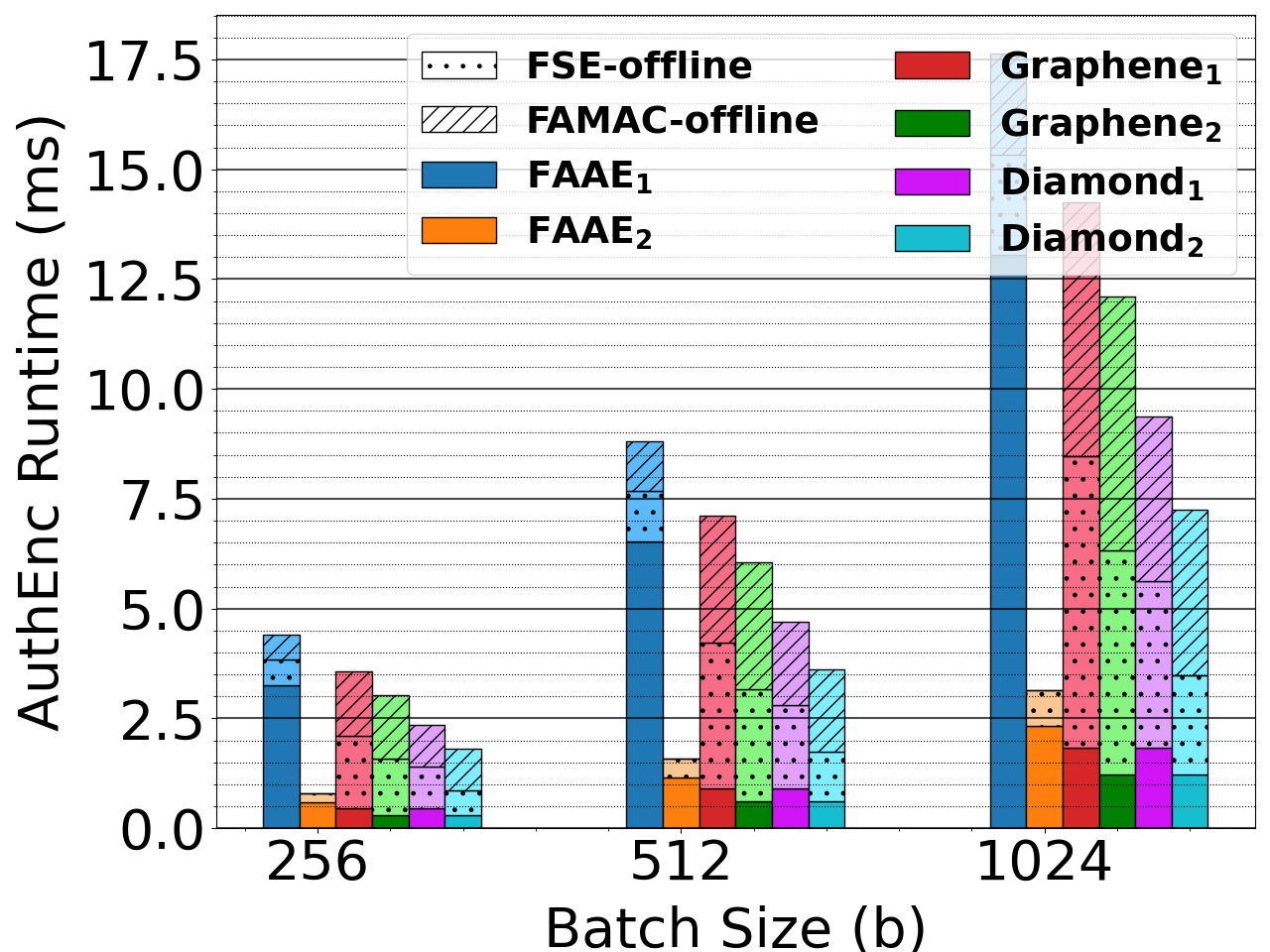}
		\caption{$\boldsymbol{m=128}$ Bytes}
	\end{subfigure} 
	
	\begin{subfigure}[b]{0.31\textwidth}
		\centering
		\includegraphics[width=\textwidth]{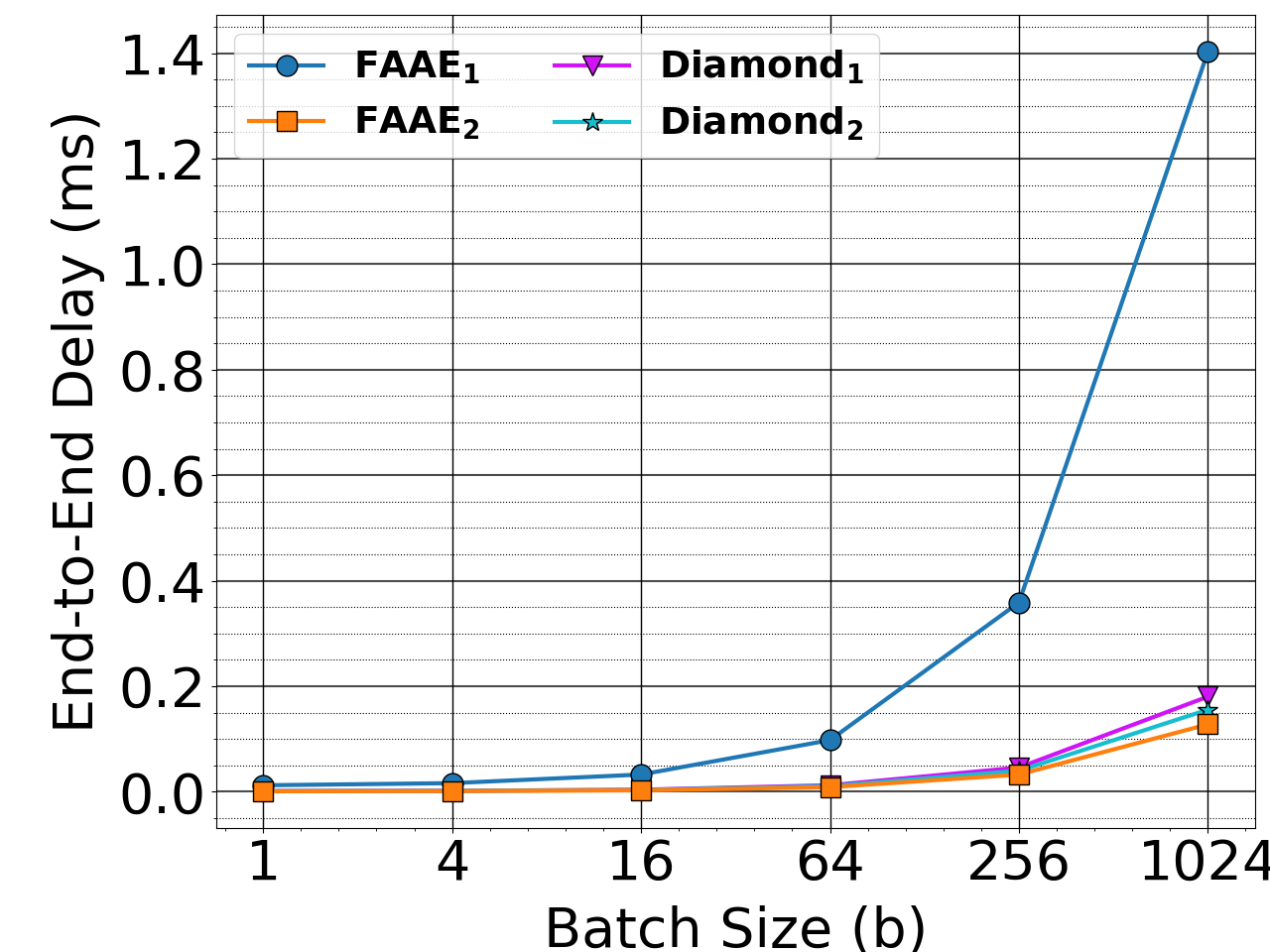}
		\caption{$\boldsymbol{m=16}$ Bytes}
		\label{subfig:arm_a72_msg16}
	\end{subfigure} 
	\hfill
	\begin{subfigure}[b]{0.31\textwidth}
		\centering
		\includegraphics[width=\textwidth]{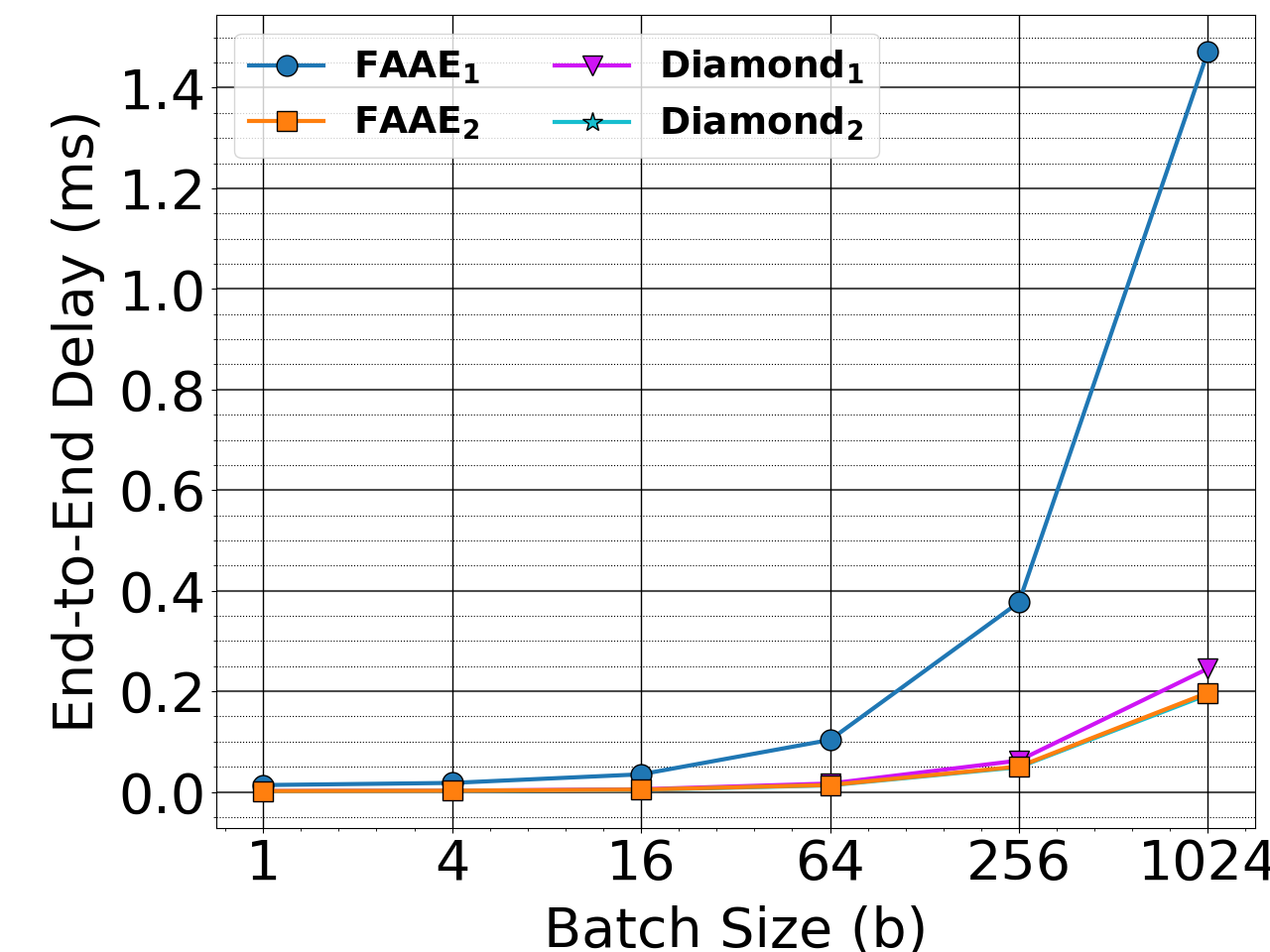}
		\caption{$\boldsymbol{m=64}$ Bytes}
		\label{subfig:arm_a72_msg64}
	\end{subfigure} 
	\hfill
	\begin{subfigure}[b]{0.31\textwidth}
		\centering
		\includegraphics[width=\textwidth]{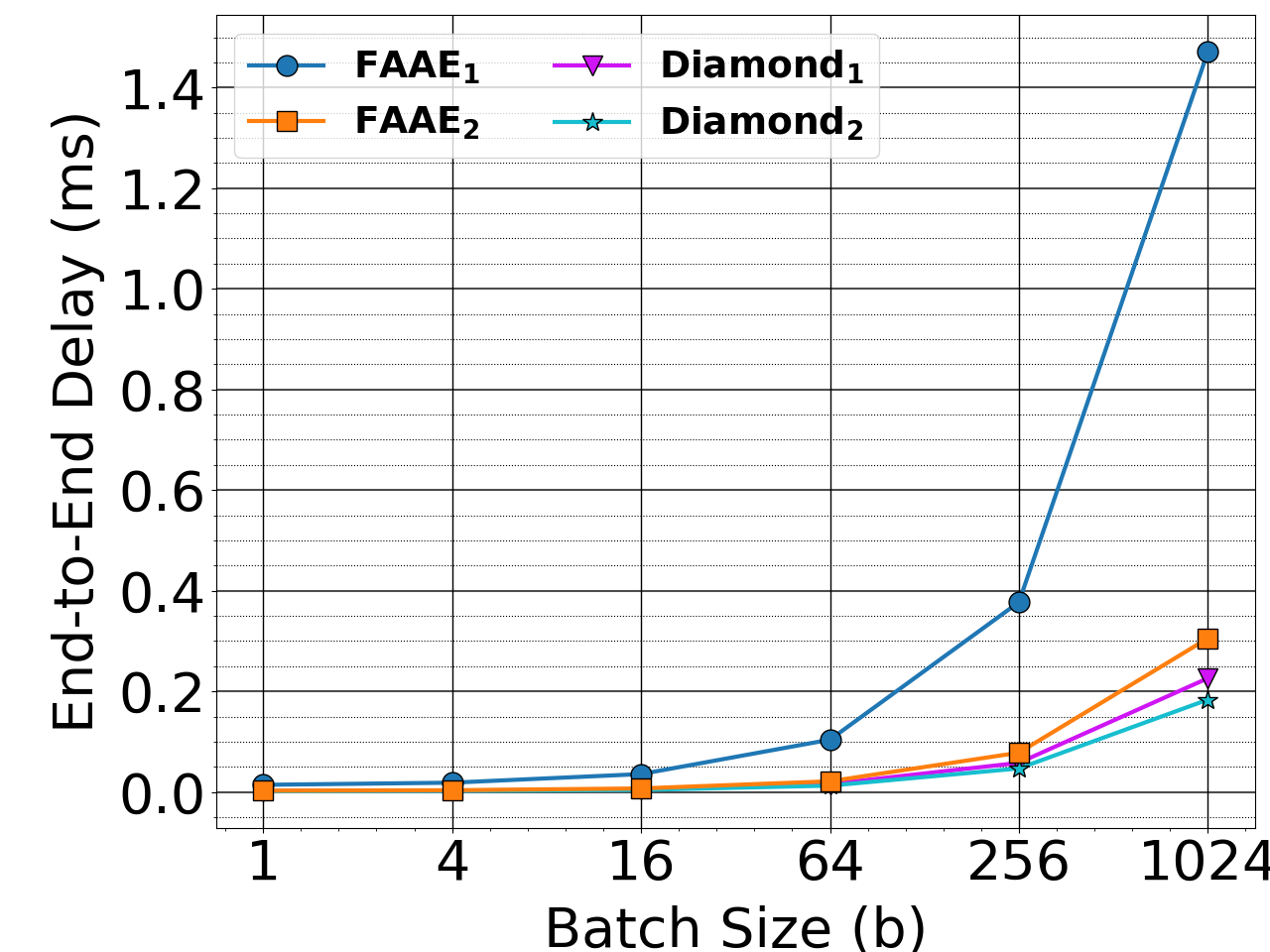}
		\caption{$\boldsymbol{m=128}$ Bytes}
		\label{subfig:arm_a72_msg128}
	\end{subfigure} 
	
	\caption{\authenc~Runtime overhead and End-to-End (E2D) delay of \dmd~variants under different input and batch sizes on Cortex ARM-A72 platform. E2E delay assumes a resourceful verifier (i.e., x86\_64 commodity).
	}
	\label{fig:runtime_arm_a72}
\end{figure}

\noindent \textbf{Cortex ARM A72.} 
Fig. \ref{fig:runtime_arm_a72} illustrates the runtime of authenticated encryption operations on the edge device, which does not possess specialized hardware acceleration, unlike x86\_64. Our experimental results show the high inefficiency of standard \faaea~where special AVX instructions are not available, while \faaeb~(employing the NIST lightweight \aee, Ascon128) exhibits slightly better efficiency compared to \graphene~and \dmd~instantiations for small 16-byte payloads. For example, the overall \authenc~runtime of \faaeb~is $4.67\times$ faster compared to the highly efficient \dmd~variants. However, for large (64-128 bytes) messages, the online \authenc~runtime of \dmd~outperforms that of \faaeb~by being $1.45\times$ faster. Comparing the instantiations of \dmd~against \graphene, the \dmda~and \dmdb~variants achieve $38.46\%$ and $40.17\%$ reduction in offline preprocessing runtime compared to that of the original \graphenea~and \grapheneb~instantiations, respectively, for a payload size of $m=64$ bytes. This is $\approx 8\%$ more computational savings compared to x86\_64.

To enable a holistic comparison of our selected performance schemes, we further plot the end-to-end (E2E) delay (see Fig. \ref{subfig:arm_a72_msg16}-\ref{subfig:arm_a72_msg128}) between the IoT device and resourceful verifier (e.g., edge device). As per Fig. \ref{fig:communication_flow}, the E2E verification delay is defined by a single \authenc~operation of the last message in the current batch at the IoT device and an online \averdec~of the received authenticated batch of ciphertexts at the IoT server.
We consider an x86\_64 commodity hardware to perform the verified decryption (\averdec) of received batches of encrypted payloads and their aggregate authentication tag. We omit the network delay incurred during the communication of cryptographic payloads, given that the transmission overhead of all benchmarked schemes is equal. 
Fig. \ref{subfig:arm_a72_msg16}-\ref{subfig:arm_a72_msg128} illustrates the advantage of \dmdb~(i.e., instantiated with Chacha20 and Poly1305) for large input sizes (64-128 bytes) by being $1.5\times$ faster E2E delay than \faaeb, while \faaeb~remains advantageous (i.e., $1.23\times$ faster than \dmdb) on small 16-byte payloads by having the lowest E2E delay.
\vspace{2pt}

\begin{figure*}[t]
	\centering
	
	\begin{subfigure}[b]{0.31\textwidth}
		\centering
		\includegraphics[width=\textwidth]{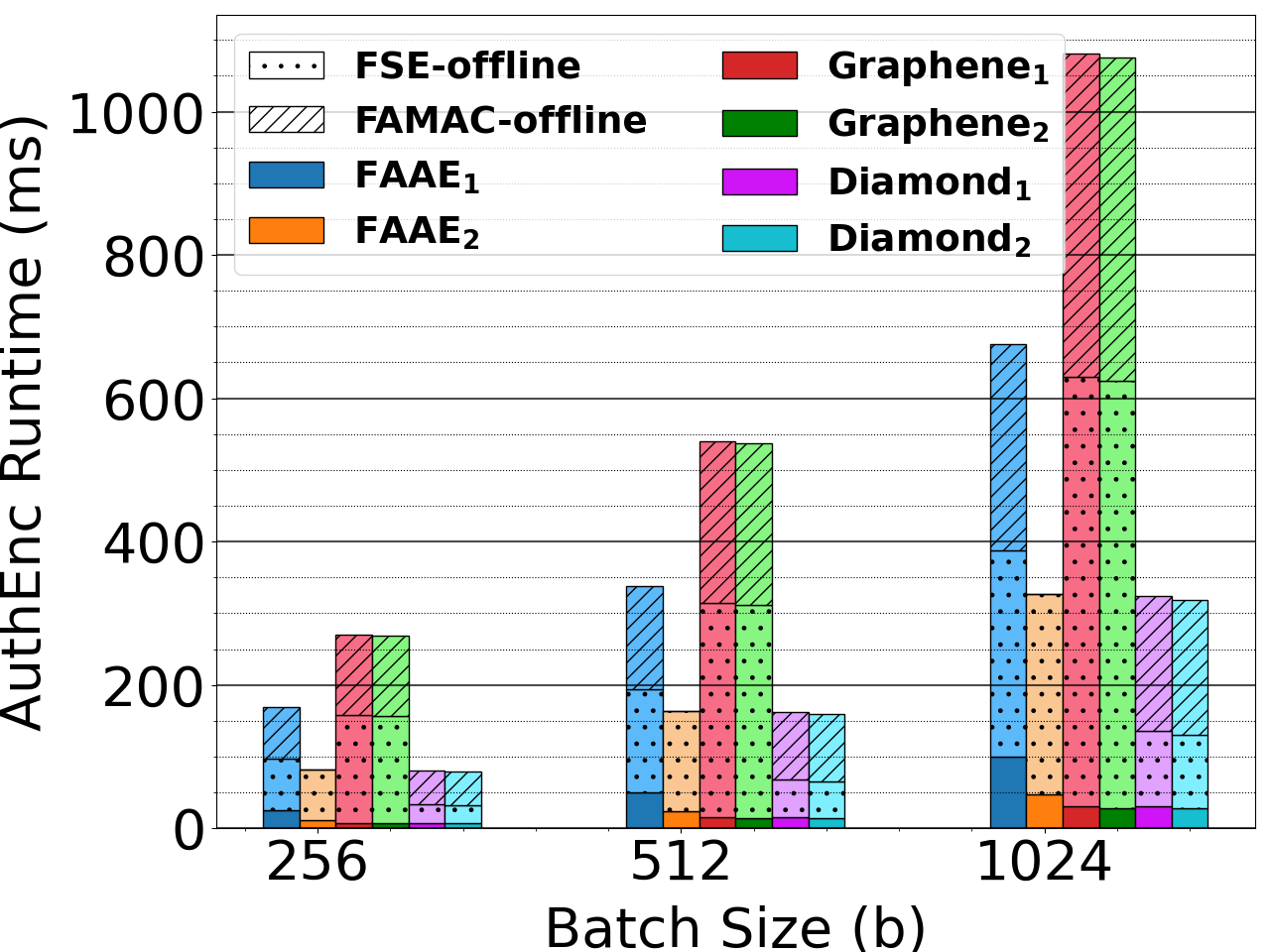}
		\caption{$\boldsymbol{m=16}$ Bytes}
	\end{subfigure} 
	\hfill
	\begin{subfigure}[b]{0.31\textwidth}
		\centering
		\includegraphics[width=\textwidth]{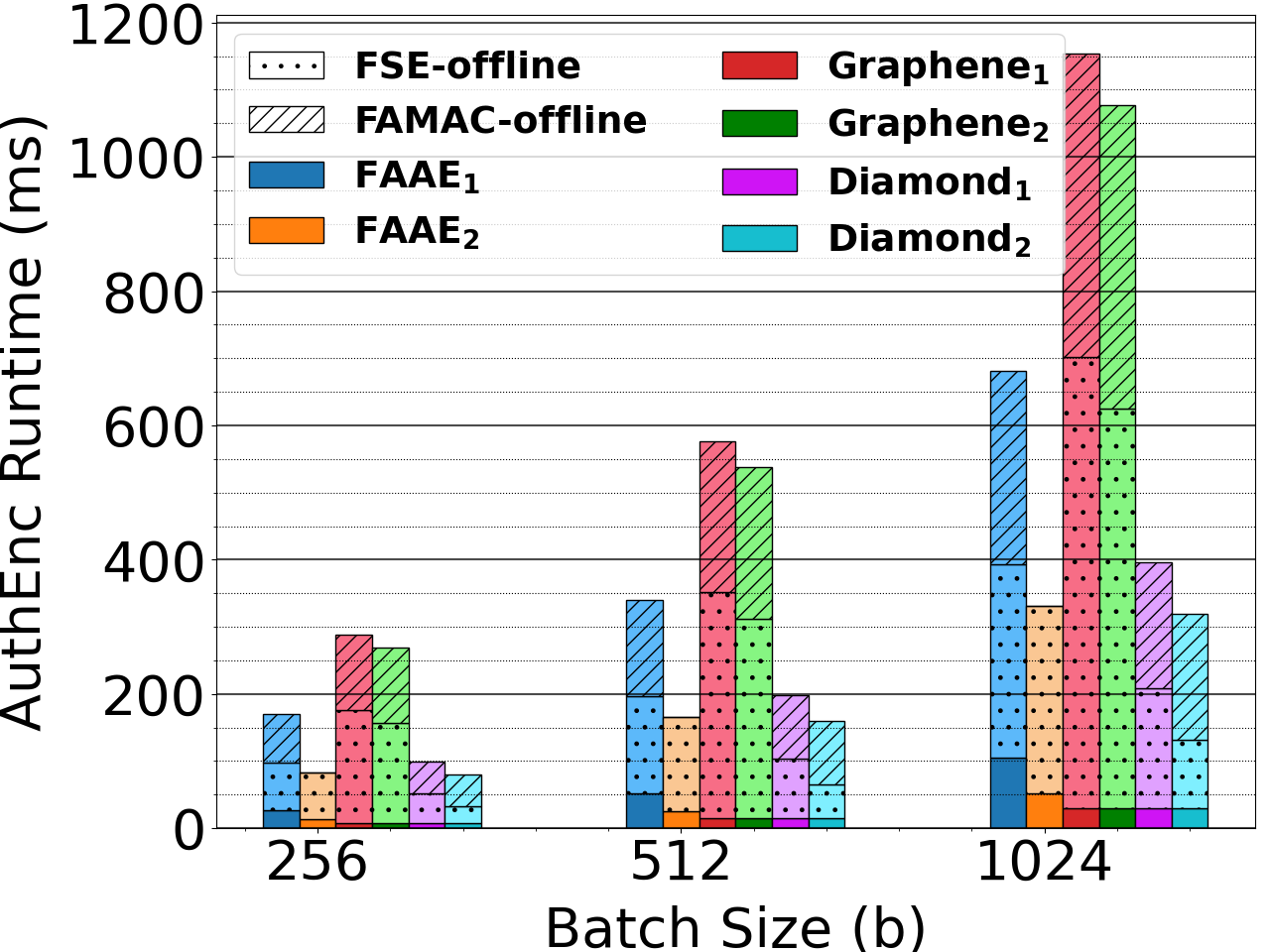}
		\caption{$\boldsymbol{m=64}$ Bytes}
	\end{subfigure} 
	\hfill
	\begin{subfigure}[b]{0.31\textwidth}
		\centering
		\includegraphics[width=\textwidth]{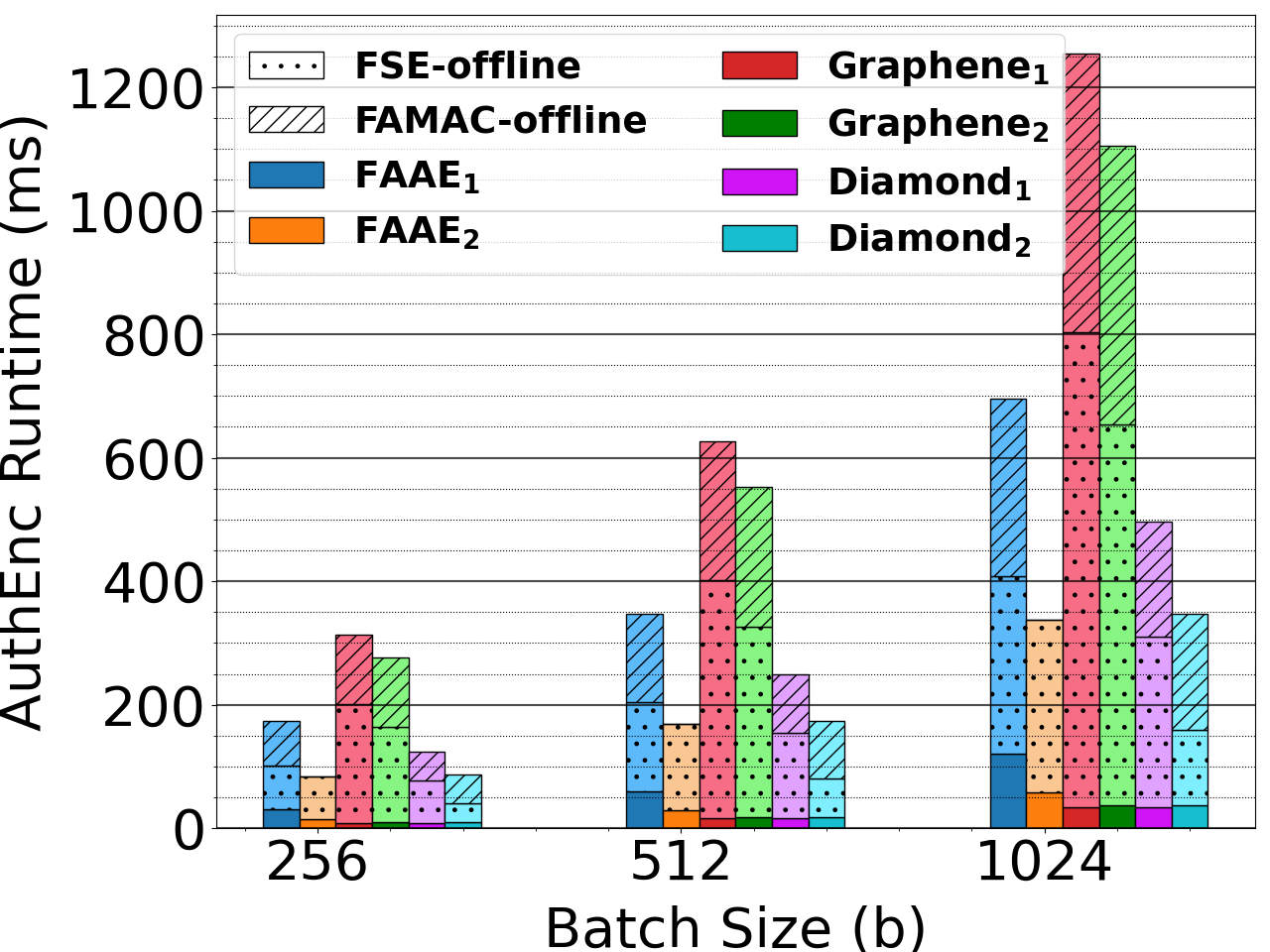}
		\caption{$\boldsymbol{m=128}$ Bytes}
	\end{subfigure} 

	\begin{subfigure}[b]{0.31\textwidth}
		\centering
		\includegraphics[width=\textwidth]{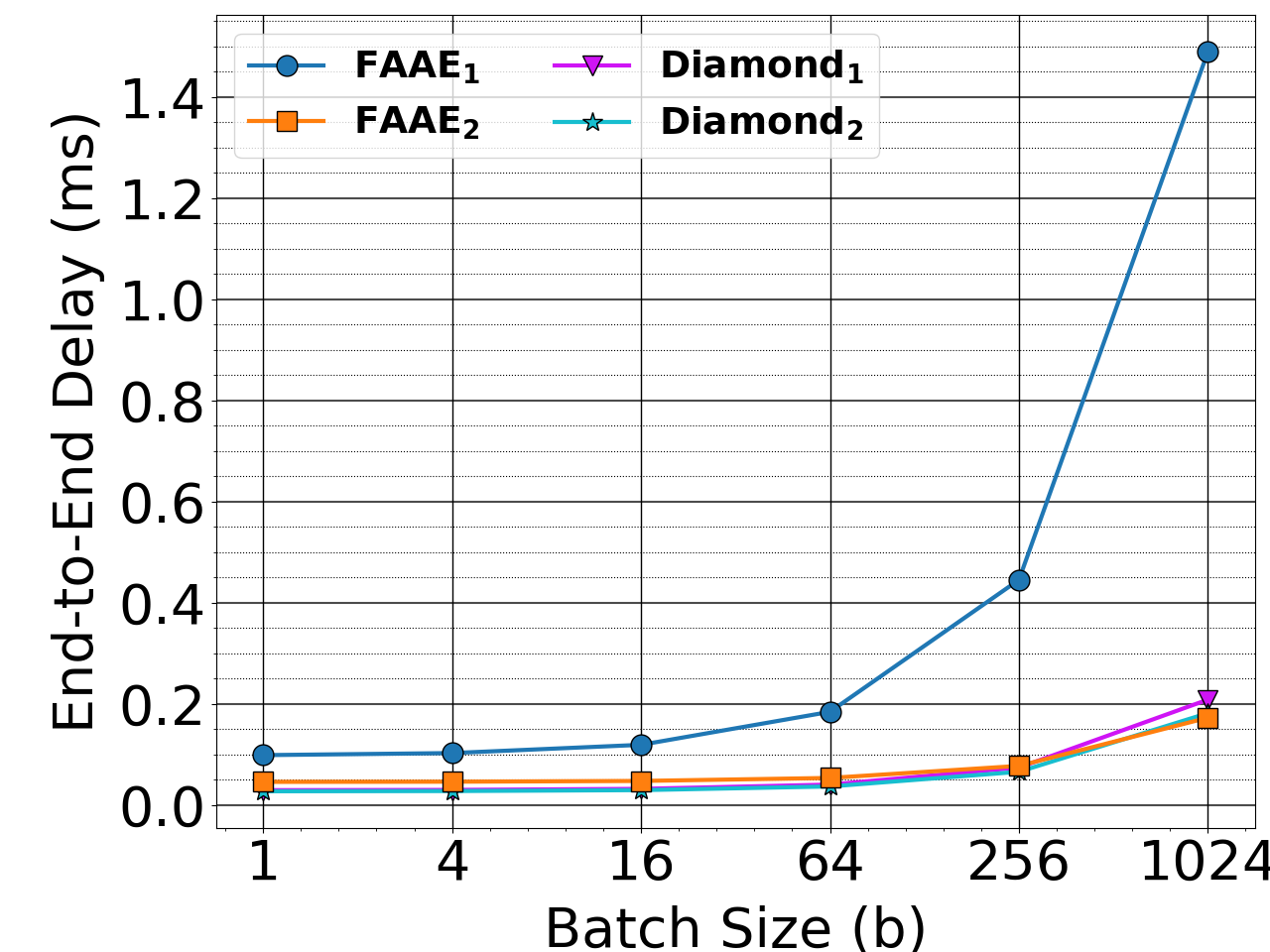}
		\caption{$\boldsymbol{m=16}$ Bytes}
		\label{subfig:arm_m4_msg16}
	\end{subfigure} 
	\hfill
	\begin{subfigure}[b]{0.31\textwidth}
		\centering
		\includegraphics[width=\textwidth]{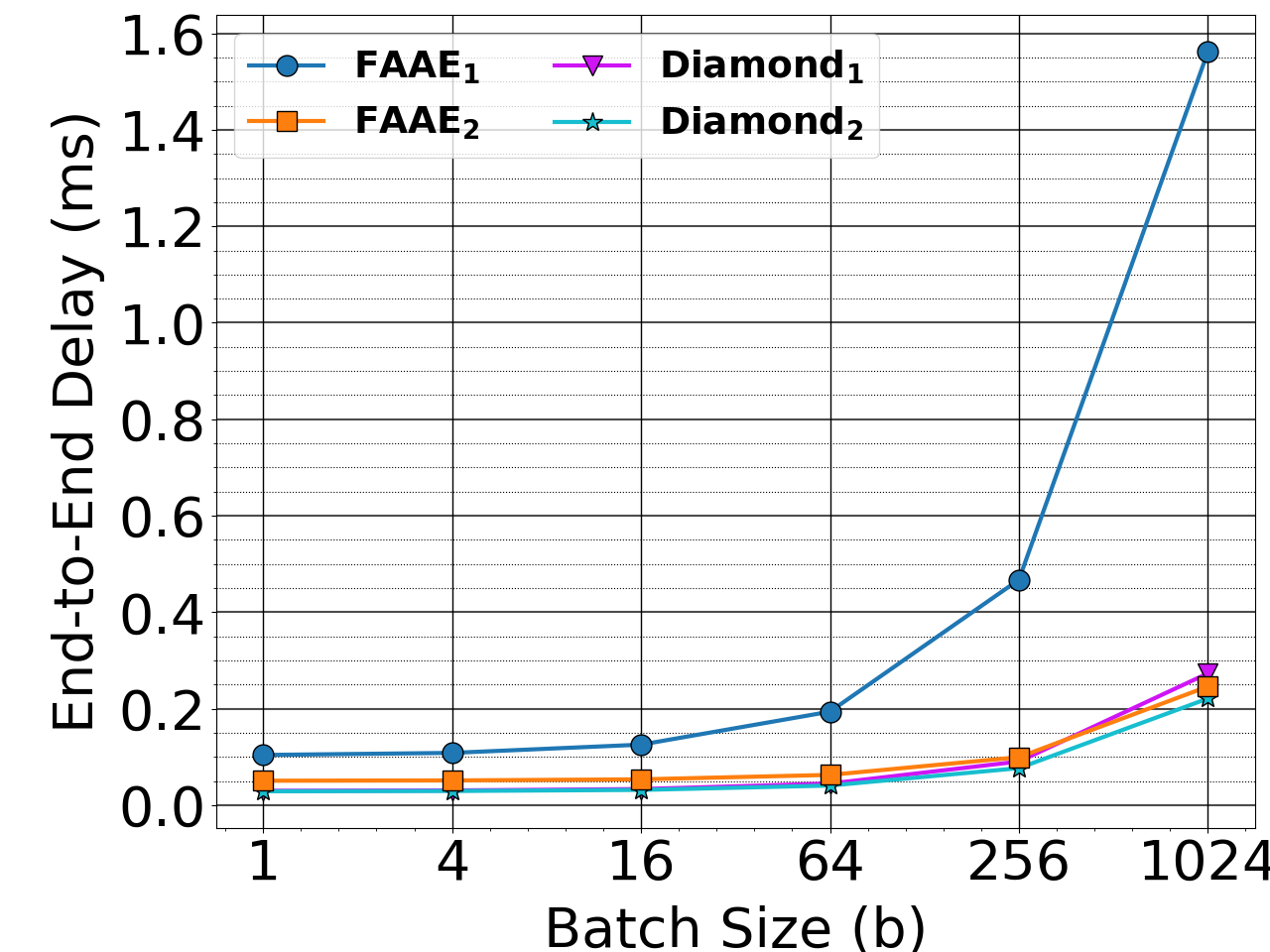}
		\caption{$\boldsymbol{m=64}$ Bytes}
		\label{subfig:arm_m4_msg64}
	\end{subfigure} 
	\hfill
	\begin{subfigure}[b]{0.31\textwidth}
		\centering
		\includegraphics[width=\textwidth]{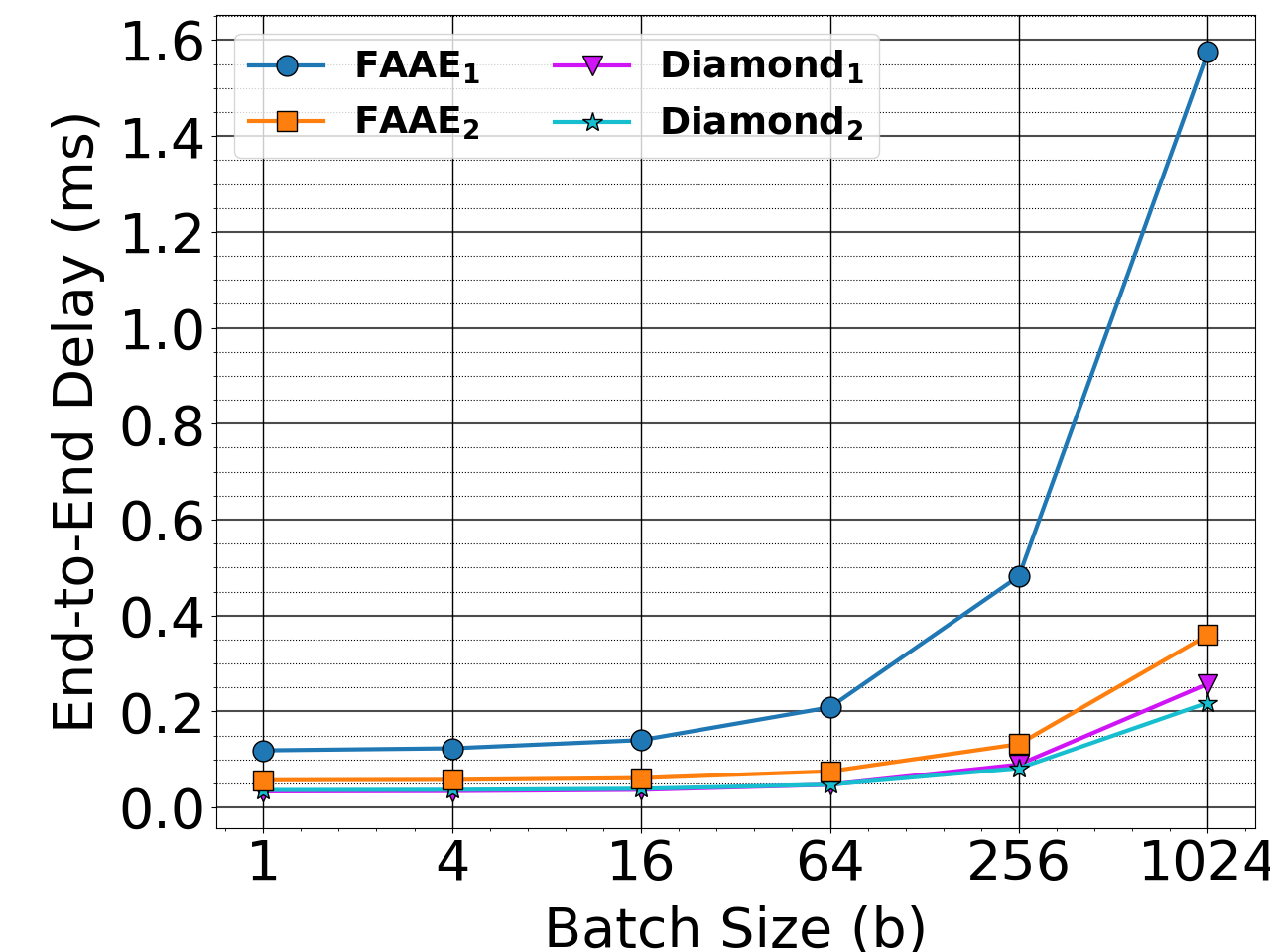}
		\caption{$\boldsymbol{m=128}$ Bytes}
		\label{subfig:arm_m4_msg128}
	\end{subfigure} 
	
	\caption{\authenc~Runtime overhead and End-to-End (E2D) delay of \dmd~variants under different input and batch sizes on Cortex ARM-M4 platform. E2E delay assumes a resourceful verifier (i.e., x86\_64 commodity).
	}
	\label{fig:runtime_arm_m4}
\end{figure*}

\noindent \textbf{Cortex ARM M4.} Fig. \ref{fig:runtime_arm_m4} illustrates the authenticated encryption runtime and end-to-end delay on a 32-bit Cortex-M4 MCU. On this platform, we enable optimized execution of cryptographic primitives (i.e., AES-128 and SHA-256) via the hardware acceleration peripheral.
When comparing \dmd~against \graphene~instantiations, the \authenc~overhead of \dmd~significantly reduces the offline \fse~and \famac~offline preprocessing cost by being $3.16\times$ faster thanks to the high efficiency of AES-based \prf~key updates compared to the original cryptographic hash operations. 
Compared to \faae~variants, the overall \authenc~of \dmd~variants outperforms \faaea~by a factor of $\approx 2\times$ while being comparable to \faaeb~based on the NIST lightweight Ascon. The online \authenc~runtime of \dmd~variants~is up to $3.5\times$ and $1.68\times$ faster than that of \faaea~and \faaeb, respectively. This underscores the high efficiency of online \dmd~variants on ARM Cortex-M4 controllers when harnessing their OO properties, compared to the most efficient \faae~counterparts.

\begin{figure*}[t]
	\centering
	
	\begin{subfigure}[b]{0.31\textwidth}
		\centering
		\includegraphics[width=\textwidth]{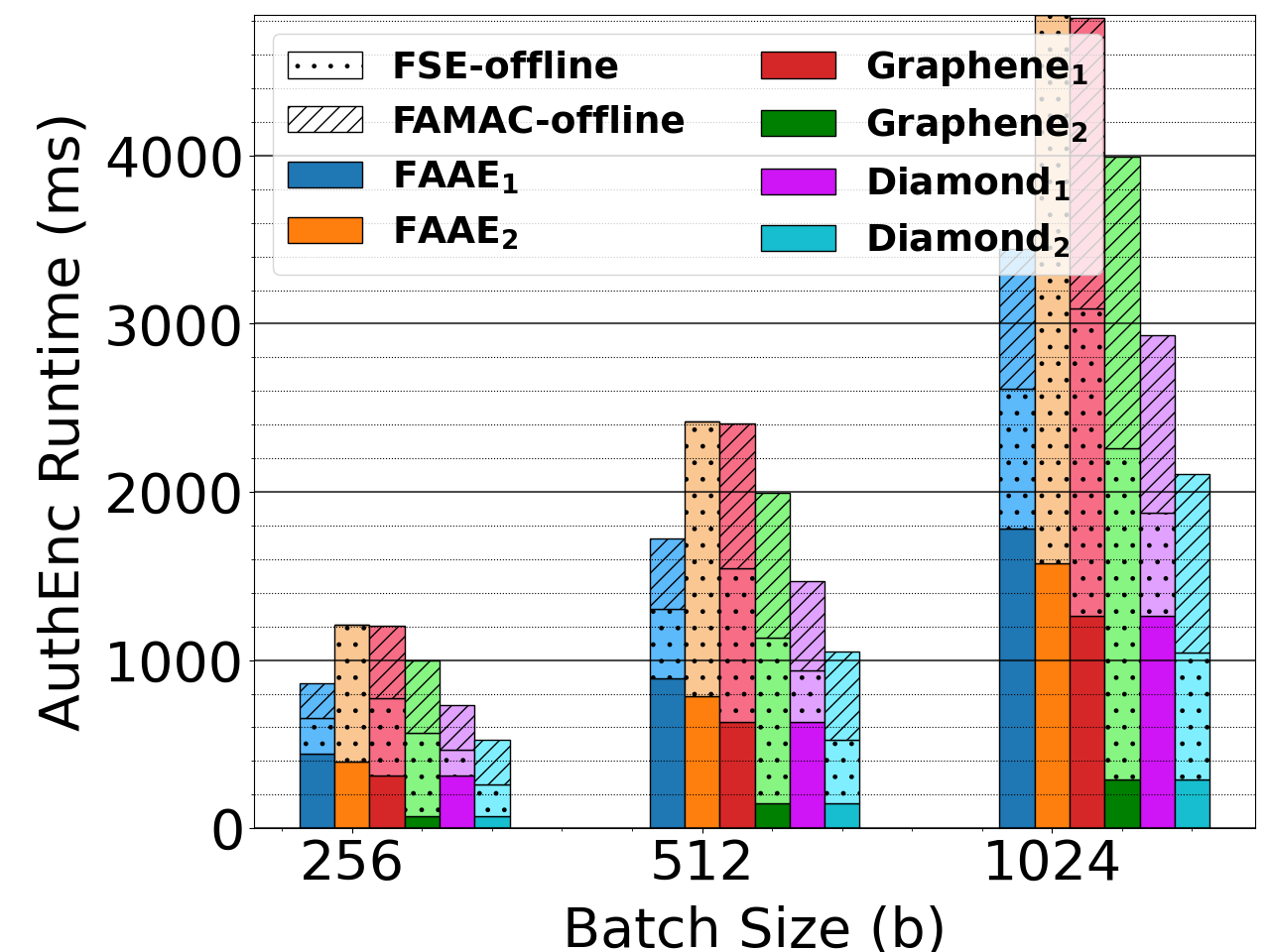}
		\caption{$\boldsymbol{m=16}$ Bytes}
	\end{subfigure} 
	\hfill
	\begin{subfigure}[b]{0.31\textwidth}
		\centering
		\includegraphics[width=\textwidth]{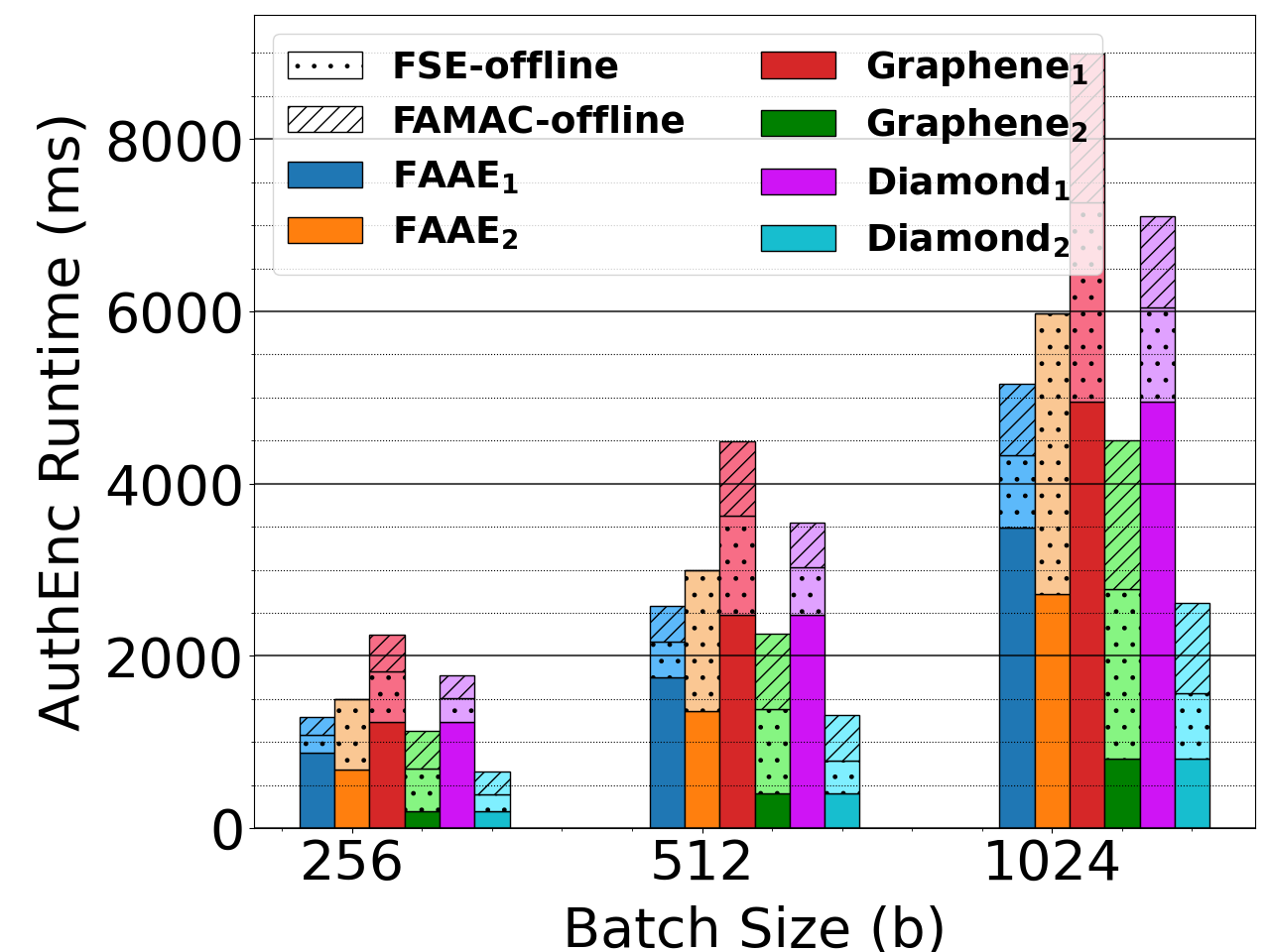}
		\caption{$\boldsymbol{m=64}$ Bytes}
	\end{subfigure} 
	\hfill
	\begin{subfigure}[b]{0.31\textwidth}
		\centering
		\includegraphics[width=\textwidth]{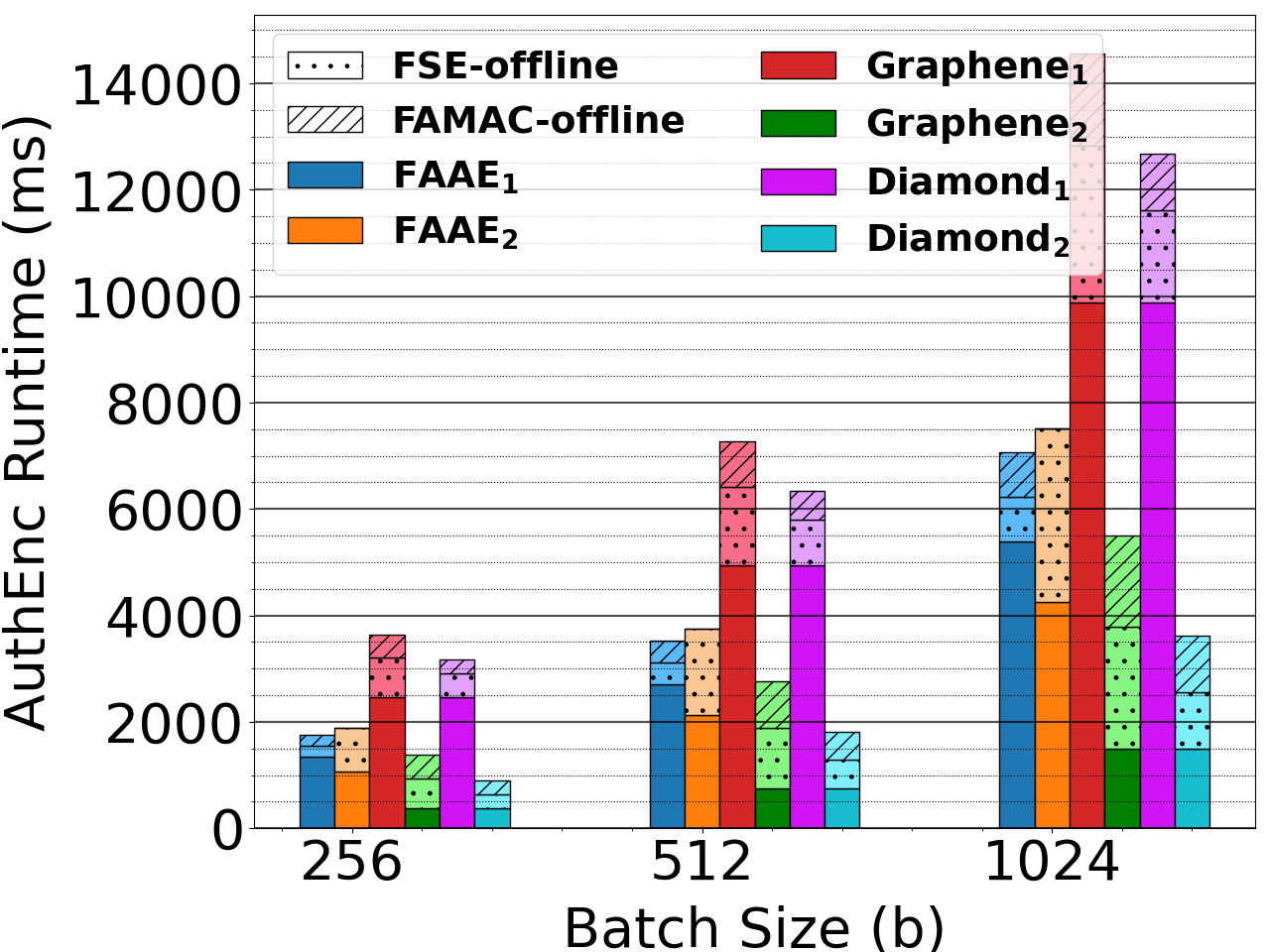}
		\caption{$\boldsymbol{m=128}$ Bytes}
	\end{subfigure} 

	\begin{subfigure}[b]{0.31\textwidth}
		\centering
		\includegraphics[width=\textwidth]{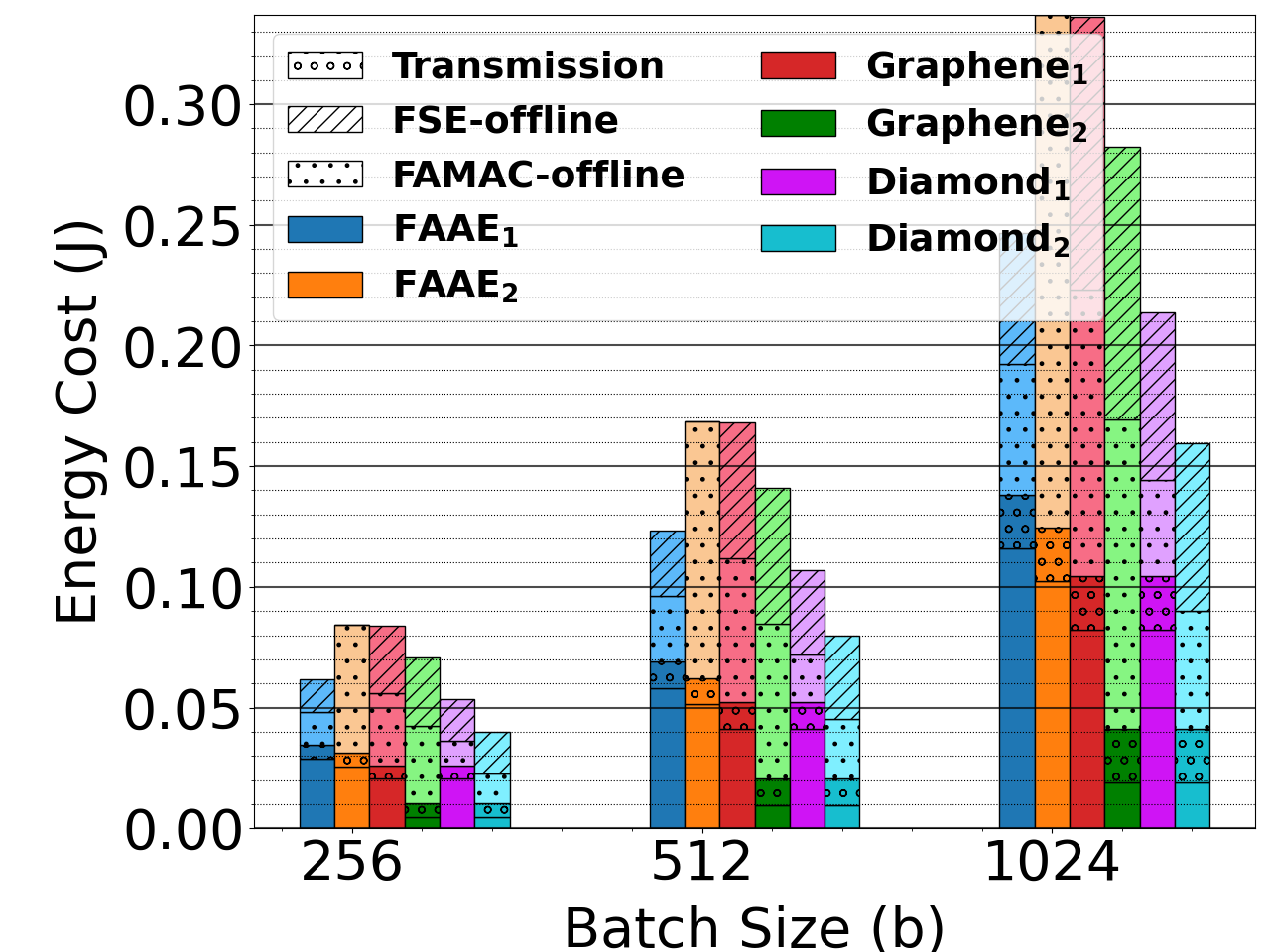}
		\caption{$\boldsymbol{m=16}$ Bytes}
		\label{subfig:avr_energy_m16}
	\end{subfigure} 
	\hfill
	\begin{subfigure}[b]{0.31\textwidth}
		\centering
		\includegraphics[width=\textwidth]{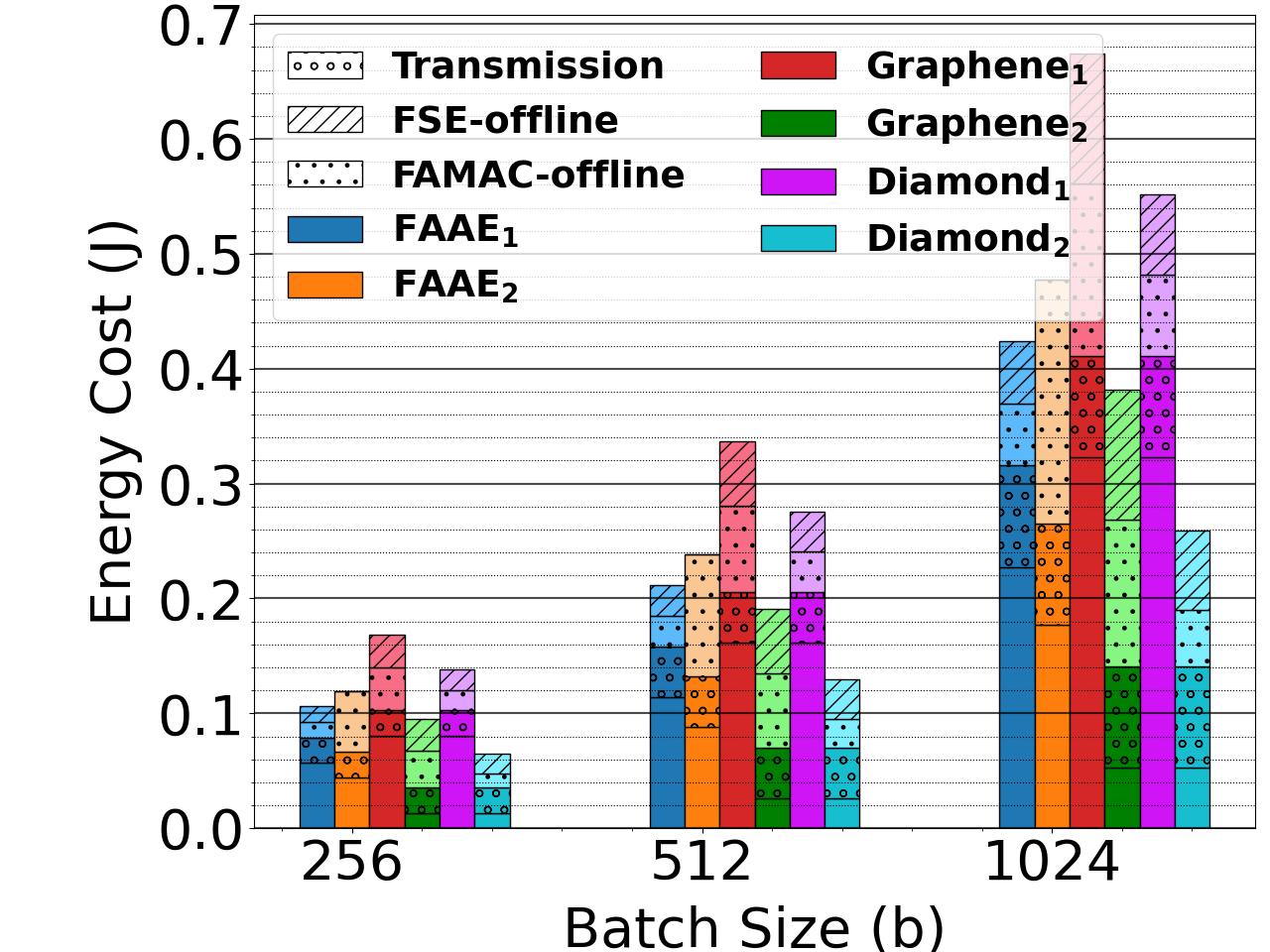}
		\caption{$\boldsymbol{m=64}$ Bytes}
		\label{subfig:avr_energy_m64}
	\end{subfigure} 
	\hfill
	\begin{subfigure}[b]{0.31\textwidth}
		\centering
		\includegraphics[width=\textwidth]{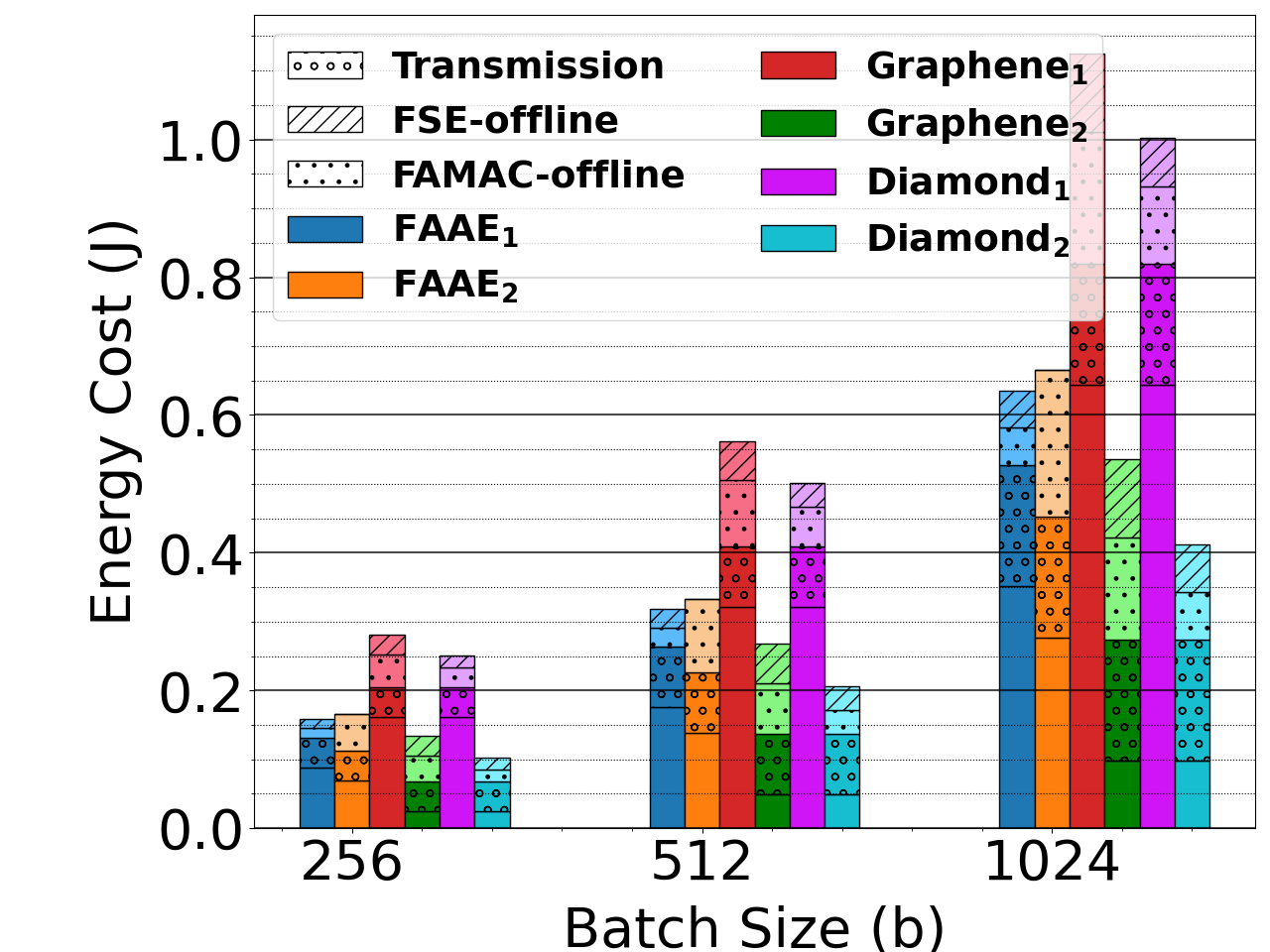}
		\caption{$\boldsymbol{m=128}$ Bytes}
		\label{subfig:avr_energy_m128}
	\end{subfigure} 

	\begin{subfigure}[b]{0.31\textwidth}
		\centering
		\includegraphics[width=\textwidth]{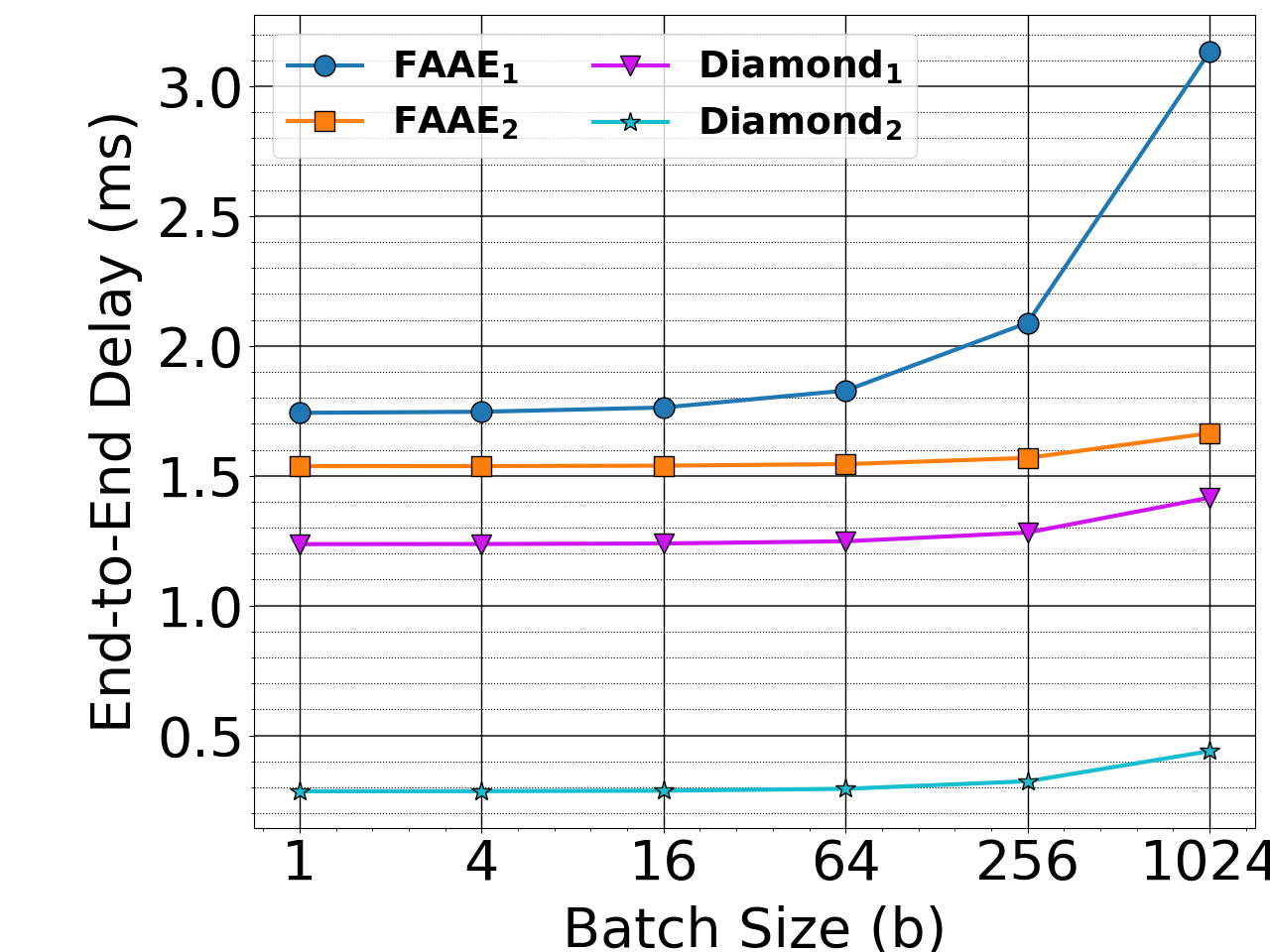}
		\caption{$\boldsymbol{m=16}$ Bytes}
            \label{subfig:avr_e2e_m16}
	\end{subfigure} 
	\hfill
	\begin{subfigure}[b]{0.31\textwidth}
		\centering
		\includegraphics[width=\textwidth]{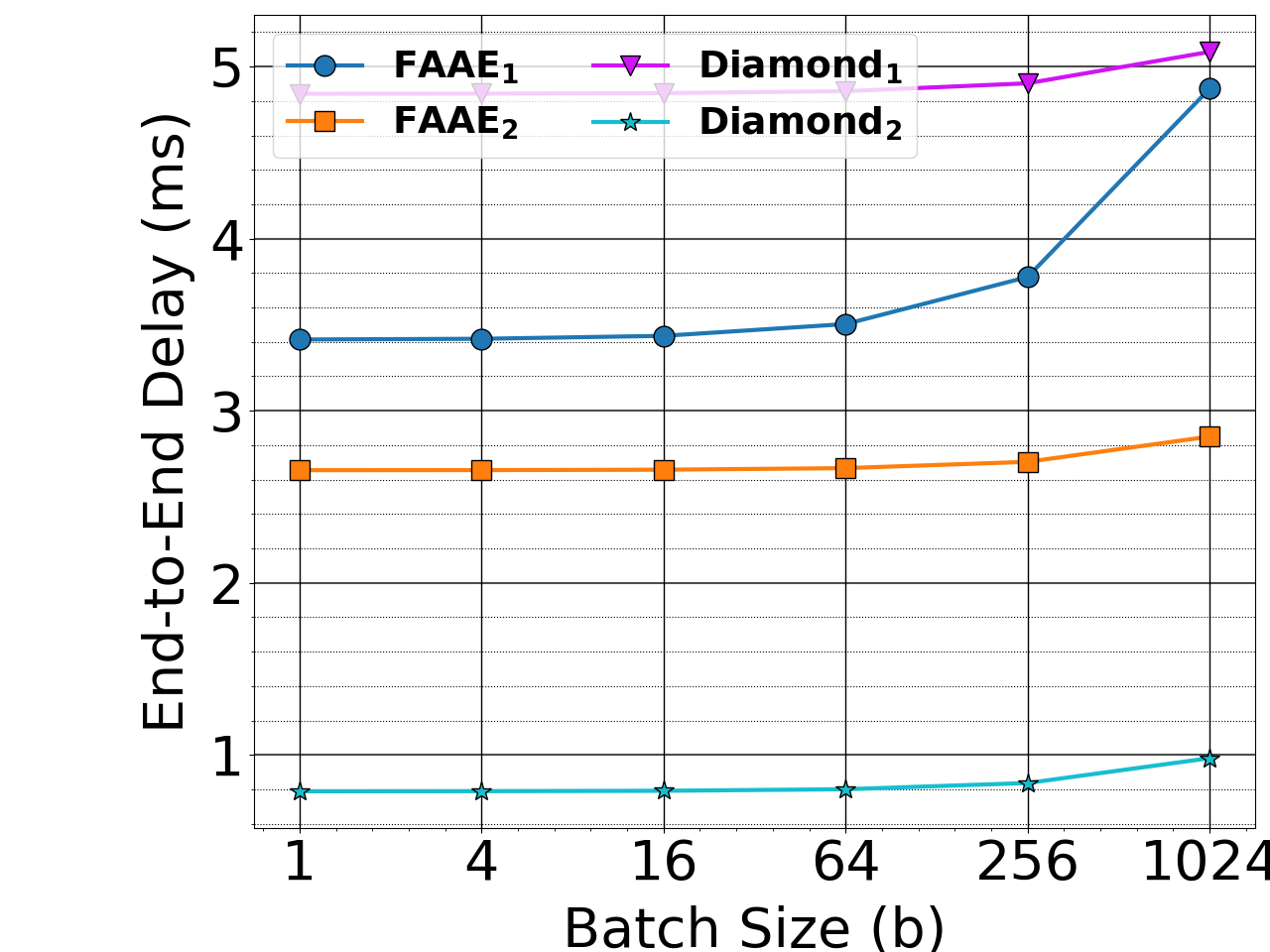}
		\caption{$\boldsymbol{m=64}$ Bytes}
            \label{subfig:avr_e2e_m64}
	\end{subfigure} 
	\hfill
	\begin{subfigure}[b]{0.31\textwidth}
		\centering
		\includegraphics[width=\textwidth]{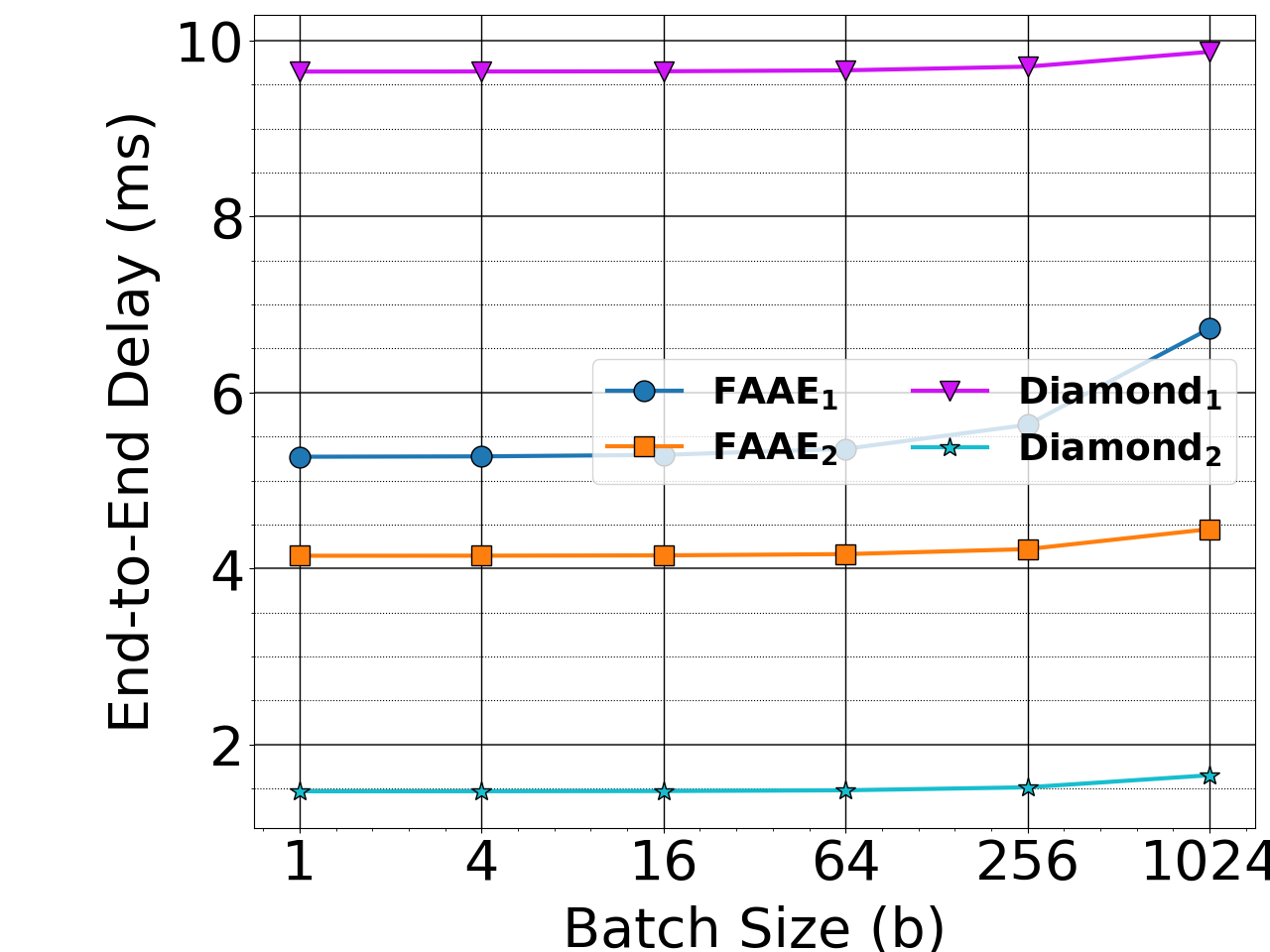}
		\caption{$\boldsymbol{m=128}$ Bytes}
            \label{subfig:avr_e2e_m128}
	\end{subfigure} 
	
	\caption{\authenc~Runtime overhead, energy cost of \authenc~and data transmission, and End-to-End (E2E) delay of \dmd~variants under different input and batch sizes on AVR ATmega2560 platform. E2E delay assumes a resourceful verifier (i.e., x86\_64).
	}
	\label{fig:runtime_avr_atmega}
\end{figure*}

Fig. \ref{subfig:arm_m4_msg16}-\ref{subfig:arm_m4_msg128} depicts the end-to-end delay when considering an IoT device equipped with a 32-bit ARM Cortex-M4 MCU and a resourceful server with x86\_64 commodity hardware. On 16-byte inputs, Fig. \ref{subfig:arm_m4_msg16} demonstrates that \dmdb~and \faaeb~achieve relatively equal E2E delay while being $1.2\times$ and $8.19\times$ faster than \dmda~and \faaea, respectively. On larger 128-byte payloads, \dmdb~exhibits the highest efficiency by being $7.2\times$ and $1.64\times$ faster than \faaea~and \faaeb, respectively, while also being $1.18\times$ faster than the AES-GCM-based instantiation \dmda.
%
\vspace{2pt}

\noindent \textbf{AVR ATmega2560.} Fig. \ref{fig:runtime_avr_atmega} illustrates the \authenc~operations and E2E delays given the ultra-low-power 8-bit AVR MCU in the order of seconds. Our experimental results showcase the high inefficiency of \graphenea~and \dmda~which stems from the high cycle count of the employed universal hash function GHASH. In contrast, the overall \authenc~runtime of \grapheneb~and its improved \dmdb~is lesser compared to our selected \faae~counterparts. For example, on 16-byte input sizes, the \authenc~of \dmdb~is $1.63\times$ and $2.29\times$ compared to \faaea~and \faaeb, respectively. Moreover, \dmdb~achieves $47.25\%$ reduction in total \authenc~computational overhead compared to \grapheneb~while also being $1.39\times$ faster than \dmda. 
On larger 128-byte payloads,  \dmdb~is still exhibiting high efficiency in terms of overall \authenc~runtime with $3.49\times$ faster than \dmda, and $\approx2\times$ faster than \faae~counterparts. The efficiency of the online \authenc~overhead of \dmdb~is more pronounced with $3.59\times$ and $2.82\times$ faster than \faaea~and \faaeb~counterparts. This showcases that on highly constrained devices, \dmdb~outperforms even \faaeb~from the NIST lightweight standard, Ascon, on both online and total \authenc~operations.

Across Fig. \ref{subfig:avr_e2e_m16}–\ref{subfig:avr_e2e_m128}, the end-to-end (E2E) latency exhibits near-constant behavior. This stems from the fact that the per-item online \authenc~cost on the IoT device dominates the amortized online verified-decryption cost (\averdec) over a large aggregated batch on the IoT server side. Across all evaluated payload sizes (16–128 bytes), \dmdb~consistently outperforms both \faae~variants and \dmda. Notably, \dmdb~achieves $3.79\times$ and $2.7\times$ lower E2E delay than the Ascon-based \faaeb—the most optimized competing design—on 16-byte and 128-byte messages, respectively. These results confirm that \dmdb, instantiated from ChaCha20 as the \prf~for both symmetric encryption and forward-secure key evolution and Poly1305 as the universal hash in \famac, constitutes the most suitable instantiation on resource-constrained 8-bit AVR microcontrollers.
\vspace{2pt}

%

\noindent \textbf{Energy Analysis on 8-bit AVR ATmega2560.}
To assess the impact of \dmd~on the energy usage on constrained devices, we estimated the total energy consumption per \authenc~operation using the MICAz energy model \cite{piotrowski2006public}. The MICAz node features an ATmega128L MCU operating at 16MHz, with 128KB flash and 4KB SRAM (i.e., similar computing capabilities to ATmega2560). Following the energy parameters in \cite{piotrowski2006public}, the ATmega128L consumes approximately $\text{E}_\text{CPU} = 4.07$nJ per clock cycle, while the CC2420 ZigBee transceiver incurs $\text{E}_\text{TX} = 0.168$ $\mu$J per transmitted bit.

Fig. \ref{subfig:avr_energy_m16}-\ref{subfig:avr_energy_m128} depicts the energy cost of \authenc~operation and transmission for batches of payloads and their authentication tags with different input sizes (16-128 bytes), with a breakdown illustrating the contribution of online \authenc, offline \fse~and \famac~computations, and the transmission overheads. 
For small telemetry payloads, the transmission cost dominates, often matching or exceeding the online \authenc~energy of the most efficient schemes. \dmdb~exhibits the lowest online \authenc~energy among all schemes, being approximately $36\%$ lower than \dmda, $54$–$67\%$ lower than \graphene~variants, and up to $4\times$ lower than \faae~variants.
For larger 128-byte inputs, \dmdb~requires $50$–$65\%$ less total energy compared to \graphenea~and \faaea~counterparts, and $35$–$45\%$ less energy compared to \grapheneb.
Across all input lengths, \dmdb~consistently yields the lowest total energy, with reductions ranging from 50-75\% compared to \faae~and 35–60\% compared to \graphene~variants.
The results confirm that \dmdb~is the most energy-optimal scheme, especially for compute- and energy-restricted IoT devices such as wearable devices, medical pacemakers, and resource-constrained sensor nodes.

%% file: security_analysis.tex
\section{Security Analysis}
\label{sec:security_analysis}
In this section, we present the security models of cryptographic primitives, used to construct our proposed scheme, \dmd. Then, we present security analysis of \fse, \famac, and \dmd.

\subsection{Security Models}
\label{subsec:secmodels}
\noindent \textbf{PRF.} The security of \prf~is defined by the following adversary's advantage:
	\[
	\advprf(t, q) = \max_{\A} |\Pr{[ f \Ra \mathcal{F} : \A^{f(.)} = 1 ]} - \Pr{[ K \Ra \kspace, g \as \prf_K : \A^{g(.)} = 1 ]}|,
	\]
	where the maximum is over all adversaries \A~making at most $q$ queries and running in time $t$.
 \prf~is secure iff $\advprf(t,q)$ is negligible.  
\vspace{2pt}

\noindent \textbf{PRG.} The security of \prg~is defined by the following adversary's advantage:
\[
\advprg(t, q=1) = \max_{\A} { |\Pr{[ K \Ra \kspace : \A(\prg(K)) = 1 ]} - \Pr{[ (K',Y) \Ra \kspace \times \ospace : \A(Y)=1 ]}| },
\]
where the maximum is over all adversaries \A~making at most one query and running in time $t$.
\vspace{2pt}

\noindent \textbf{FPRG.} Based on the seminal work of Bellare et al. \cite{bellare2003forward}, the security of \fprg~is defined as follows:
\[
\advfprg(t_{\fprg}) \le 2 \cdot n \cdot \advprf(t_{\prf}, q_{\prf}),
\]
where $q_\prf = \lceil \frac{(\tau' + \kappa)}{\tau} \rceil$ and $t_\prf = t_\fprg + \mathcal{O}(n \cdot (\kappa + \tau'))$.
\vspace{2pt}

\noindent \textbf{MAC.}
The standard security notion of \mac~is Existential Unforgeability under Chosen Message Attacks (\eucma) \cite{bellare2003forward}, which is defined based on the following experiment:

\textit{Experiment $\text{Expt}^{\eucma}_{\mac} (\A)$} \\
\hspace*{2em} $K \as \mackg(1^\kappa)$ \\ 
\hspace*{2em} $(M^*,\sigma^*) \Ra \A^{\macsign(K, .)}(find)$ \\
\hspace*{2em} \textbf{if} $M^*$ was not queried to $\macsign(K, .)$ \textbf{and} $\macver(K, M^*, \sigma^*) = 1$ \textbf{then return} 1 \textbf{else } 0.

In $\text{Expt}^{\eucma}_{\mac} $, \A~is given access to the signing oracle $\macsign(K,.)$ and aims to produce a forgery $(M^*, \sigma^*)$.
The advantage of \A~over \mac~is $\advmac(t,q) = \max_{\A} \Pr[ \text{Expt}^{\eucma}_{\mac}(\A)=1 ]$, where the maximum is over all adversaries \A~making at most $q$ queries and running in time $t$.
\vspace{2pt}

\noindent \textbf{FAMAC.}
The standard security notion of \famac~scheme is Forward-secure Aggregate \eucma~(\faeucma), which is defined based on the following experiment:
\vspace{2pt}

\textit{Experiment $\text{Expt}^{\faeucma}_{\famac} (\A)$} \\
\hspace*{2em} $K_1 \as \famackg(n)$ , $i \as 1$ , $phase \as find$\\ 
\hspace*{2em} \textbf{while} $i \le n$ \textbf{and} $phase \neq forge$ \\
\hspace*{3em} $phase \as \A^{\famacsign(K_i,.)}( find)$ \\
\hspace*{3em} $K_{i+1} \as \famacupd(K_i)$ \textbf{and} $i \as i+1$ \\
\hspace*{2em} $(\vect{M_{1,i^*}}, \sigma_{1,i^{*}}) \as \A( forge, K_{i})$ \\
\hspace*{2em} \textbf{if} $1 \le i^* < i$ \textbf{and} $\vect{M_{1,i^*}}$ is not queried to $\famacsign(K_i, .)$ oracle \textbf{and} $\famacaver(K_1,\vect{M_{1,i^*}}, \sigma_{1,i^*}) = 1$ \textbf{then return} 0 \textbf{else return} 1.

\vspace{2pt}
In $\text{Expt}^{\faeucma}_{\famac}$, \A~operates in two stages: 
{\em (1) find.} \A~makes at most $n$ queries to the signing oracle $\famacsign(K_i,.)$.
{\em (2) forge.} \A~is given the current private key $K_i$ during the break-in and asked to produce a non-trivial forgery $(\vect{M_{1,i^*}}, \sigma_{1,i^{*}})$ with index $i^* < i$.
The advantage of \A~over \famac~is $\advfamac(t,q) = \max_{\A} \Pr[ \text{Expt}^{\faeucma}_{\famac}(\A)=1 ]$, where the maximum is over all adversaries \A~making at most $q$ queries and running in time $t$. 
\vspace{2pt}

\noindent \textbf{SE.}
The standard security notion of an \encscheme~scheme is INDistinguishability under Chosen Ciphertext Attacks (\indcpa) \cite{bellare2003forward} which is defined based on the following experiment:
\vspace*{2pt}

\textit{Experiment $\text{Expt}^{\indcpa}_{\enc} (\A)$} \\
\hspace*{2em} $K \as \enckg(1^\kappa)$ and $b \Ra \{0,1\}$ \\ 
\hspace*{2em} $C_b \as \A^{\encenclr(K,.)}(find, M_0, M_1)$ \\
\hspace*{2em} $b' \as \A^{\encenclr(K,.)}(guess)$ \\
\hspace*{2em} \textbf{If} $|M_0|=|M_1|$ and $b=b'$ \textbf{then return} $1$ \textbf{else return} $0$.
\vspace{2pt}

In $\text{Expt}^{\indcpa}_{\encscheme}$, the adversary \A~operates in two phases:
{\em (1) Find.} \A~is given access to the encryption left or right oracle $\encenclr(K,.)$ which accepts two distinct plaintexts ($M_0,M_1$), and computes/returns to \A~the ciphertext $C_b$ based on a pre-defined bit $b$, generated during the setup phase. The challenger in the game computes the ciphertext $C_b$ with index equal to a bit $b$, uniformly generated at random. 
{\em (2) Guess.} \A~returns a bit $b'$. \A~succeeds if $b=b'$.
The advantage of \A~over \encscheme~is $\advenc(t,q) = \max_{\A} \{ | \Pr[ \text{Expt}^{\indcpa}_{\encscheme}(\A)=1 ] -\frac{1}{2} | \}$, where the maximum is over all adversaries \A~making at most $q$ queries and running in time $t$. 
\vspace{2pt}

\noindent \textbf{FSE.}
The standard security notion of an \fenc~scheme is Forward-secure \indcpa~(\findcpa), which is defined based on the following experiment:
\vspace*{2pt}

\textit{Experiment $\text{Expt}^{\findcpa}_{\fenc} (\A)$} \\
\hspace*{2em} $K_1 \as \fenckg(1^\kappa, n)$, $b \Ra \{0,1\}$, and $i \as 1$, $phase \as find$ \\ 
\hspace*{2em} \textbf{while} $i \le n$ \textbf{and} $phase \neq guess$ \\
\hspace*{3em} $(phase, C_b ) \as \A^{\fencenclr(K_i,.)}(find, M_0,M_1)$ \\
\hspace*{3em} $K_{i+1} \as \fencupd(K_i)$ \textbf{and} $i \as i+1$ \\
\hspace*{2em} $b' \as \A^{\fencenclr(K_i,.)}(guess, K_i)$ \\
\hspace*{2em} \textbf{if} $|M_0|=|M_1|$ and $b=b'$ \textbf{then return} $1$ \textbf{else return} $0$.
\vspace{2pt}

The adversary \A~in this experiment operates in two phases similar to \indcpa, whereas during the find phase, the challenger updates the private key $K_i$ after each query made by \A~to $\fencenclr(K_i,.)$. 
During the guess phase, $\A$ is given access to the current secret key and asked to return a bit $b'$. 
The advantage of \A~over \fenc~is $\advfenc(t,q) = \max_{\A} \{ |\Pr[\text{Expt}^{\findcpa}_{\fenc}(\A)=1 ] -\frac{1}{2} | \}$, where the maximum is over all \A s making at most $q$ queries and running time in $t$. 
\vspace{2pt}

\noindent \textbf{FAAE.}
The security model of an \faae~is based on the security of its underlying (authenticated) encryption and MAC components. We consider forward security, which implies periodic key updates, an essential feature in adversarial IoT environments against key compromise attacks.  For confidentiality, \dmd~adopts \findcpa. For integrity protection, we follow the security of \faae~\cite{SUHaSAFSS11, FssAggMAC_DiMa}, with forward-secure and aggregate existential unforgeability against chosen message attacks (\faeucma). These notions collectively ensure that even an adaptive PPT adversary $\mathcal{A}$ can neither decrypt previously and currently encrypted telemetry traffic nor forge valid aggregate and forward-secure tags without detection. 
The \findcpa~and \faeucma~security notions for \faae~are formalized by the following experiments:
%
\vspace{2pt}

\begin{minipage}[t]{0.52\textwidth}
	\textit{Experiment $\text{Expt}^{\findcpa}_{\faae} (\A)$} \\
	\hspace*{1em} $K_1 \as \faaekg(1^\kappa, n, b)$ and $c \Ra \{0,1\}$  \\ 
	\hspace*{1em} $i \as 1$ \textbf{and} $phase \as find$ \\ 
	\hspace*{1em} \textbf{while} $i \le n$ \textbf{and} $phase \neq guess$ \\
	\hspace*{1.5em} $(phase, C_c ) \as \A^{\authenclr(K_i,.)}(phase, M_{0}, M_{1})$ \\
	\hspace*{2em} $\vect{K_{i+1}} \as \faaeupd(\vect{K_i})$ \textbf{and} $i \as i+1$ \\
	\hspace*{1em} $c' \as \A^{\authenclr(K_i,.)}(phase, \vect{K_i})$ \\
	\hspace*{1em} \textbf{if} $|M_0|=|M_1|$ and $c=c'$ \textbf{then return} $1$ \textbf{else return} $0$.\\
\end{minipage}
\hfill
\vrule width0.5pt
\hfill
\begin{minipage}[t]{0.44\textwidth}
	\textit{Experiment $\text{Expt}^{\faeucma}_{\faae} (\A)$} \\
	\hspace*{1em} $\vect{K_1} \as \faaekg(1^{\kappa}, n, b)$\\
	\hspace*{1em} $i \as 1$ \textbf{and} $phase \as find$\\ 
	\hspace*{1em} \textbf{while} $i \le n-b+1$ \textbf{and} $phase \neq forge$ \\
	\hspace*{2em} $phase \as \A^{\faaeauthenc(\vect{K_i},.)}( phase)$ \\
	\hspace*{2em} $\vect{K_{i+1}} \as \faaeupd(\vect{K_i})$ \textbf{and} $i \as i+1$ \\
	\hspace*{1em} $(\vect{C_{i^*}}, \sigma_{i^{*}, i^{*}+b-1}\} \as \A( phase, \vect{K_{i}})$ \\
	\hspace*{1em} \textbf{if} $1 \le i^* < i$ and $(\vect{C_{i^{*}}},\sigma_{i^*,i^*+b-1}) \notin \mathcal{L} \as \{ (\vect{C_{i_1}},\sigma_{i_1,i_1+b-1}) | \forall 1 \le i_1 \le i^*-b+1 \}$  \textbf{and} $\vect{M_{i_1}} = \famacaver(\vect{K_{i_1}}, \vect{C_{i^*}}, \sigma_{i^*,i^*+b-1})$ \textbf{then return} 1 \textbf{else return} 0.\\
\end{minipage}

The experiment $\text{Expt}^{\findcpa}_{\faae}$ is similar to $\text{Expt}^{\findcpa}_{\enc}$. The adversary \A~in $\text{Expt}^{\faeucma}_{\faae}$ outputs, as forgery, a batch of ciphertexts $\vect{C_{i^*}}$ and an aggregate tag $\sigma_{i^{*}, i^*+b-1}$. \A~succeeds if the produced forgery is not a combination of previous outputs of the $\faaeauthenc(K_i, .)$ oracle during the find phase. 
The advantage of \A~over \faae~under \findcpa~and \faeucma~security notions is $\advfaaeindcpa(t,q)= \max_\A \{ |\Pr[\text{Expt}^{\findcpa}_{\faae}=1]-1/2 | \}$ and $\advfaaefaeucma(t,q)=\max_{\A} {\Pr[\text{Expt}^{\findcpa}_{\faae}=1]}$, respectively, making at most $q$ queries and in running time $t$. 

Note that \faeucma~captures integrity of ciphertexts (\intctxt), since
$\mathcal{L}$ records all previously authenticated ciphertext-tag pairs
and the adversary must produce a fresh, verifying batch outside $\mathcal{L}$.
By the Bellare--Namprempre composition theorem~\cite{bellare2000authenticated},
an \faae~scheme that is simultaneously \findcpa- and \intctxt-secure
achieves \indcca~security.

\subsection{Security Proof}

\setcounter{theorem}{0}

Let $\fenc$ be the CTR-based forward-secure symmetric encryption scheme defined in Fig.~\ref{alg:fenc}, where $\prf_1$ is used in CTR mode to generate the keystream blocks and $\prf_2$ is used to instantiate the forward-secure pseudo-random generator $\fprgupd$ to update the states and keys.
We prove that if both $\prf_1$ and $\prf_2$ are secure pseudo-random functions, then $\fenc$ is $\findcpa$-secure.

\begin{theorem}
	If $\fenc$ is constructed as above, then for every adversary
	$\mathcal{A}$ running in time $t_{\fenc}$ and making at most
	$q_{\fenc}$ encryption queries, there exist distinguishers
	$\mathcal{B}_1$ and $\mathcal{B}_2$ such that:
	\[
	\adv_{\fenc}^{\findcpa}(t_{\fenc},q_{\fenc})
	\le
	n\!\cdot\!\adv_{\prf_1}(t_{\prf_1},q_{\prf_1})
	+
	2\!\cdot\!n\!\cdot\!\adv_{\prf_2}(t_{\prf_2},q_{\prf_2}),
	\]
	where
	\[
	\begin{aligned}
		q_{\prf_1} &= q_{\fenc}\!\cdot\!\Big\lceil\frac{m}{\ell}\Big\rceil, &
		t_{\prf_1} &= t_{\fenc}+O\!\Big(q_{\fenc}\!\cdot\!\Big\lceil\frac{m}{\ell}\Big\rceil\Big),\\[6pt]
		q_{\prf_2} &\le 2\!\cdot\!n, &
		t_{\prf_2} &= t_{\fenc}+O(n+\kappa).
	\end{aligned}
	\]
\end{theorem}

\begin{proof}
	We define a series of hybrid games $H_0 \Rightarrow H_1^{(1)} \Rightarrow \dots \Rightarrow H_n^{(1)} \Rightarrow H_1^{(2)} \Rightarrow \dots \Rightarrow H_n^{(2)}$
	
	
	
	\paragraph{Game $\mathbf{H_0}$ (Real experiment).}
	This is the real $\findcpa$ experiment for $\fenc$:
	both $\prf_1$ and $\prf_2$ are instantiated as true \prf~functions.
	Let $\adv_0=\big|\Pr[H_0(\mathcal A)=1]-\tfrac12\big|$ denote $\mathcal A$’s advantage.
	
	\paragraph{Game $\mathbf{H_i^{(1)}}$ (Replacing $\prf_1$ in period $i$).}
	For $i \in \{1,\ldots,n\}$, we define $H_i^{(1)}$ to be identical to $H_{i-1}^{(1)}$
	except that in period $i$ all values $\tilde{C}_j \as \prf_1(K_i,ctr+j)$ used to generate CTR keystream blocks are replaced by independent uniform random $\ell$-bit strings (i.e., $\tilde{C}_j \Ra \{0,1\}^\ell$). All other computations remain unchanged. 
	Let $\Delta_i^{(1)}
	=\big|\Pr[H_{i-1}^{(1)}(\mathcal A)=1]-\Pr[H_i^{(1)}(\mathcal A)=1]\big|$.
	
	
		The distinguisher $\mathcal{B}_1$ is given oracle access to a function $\mathcal{O}(\cdot)$ which is either $\prf_1(K^\star,\cdot)$ for $K^\star \Ra \{0,1\}^\kappa$ or a true random function. It simulates the $\findcpa$ experiment for $\mathcal{A}$ as follows:
		
		\begin{enumerate}[1.]
			\item For the target period $i$, $\B_1$~computes the keystream blocks $\{\tilde{C}_j' \as \mathcal{O}(ctr+j \bmod 2^\kappa)\}_{j=1,\ldots,\lceil m/\ell \rceil}$ and returns $C_i \as \tilde{C}_i \oplus M_i$ to $\mathcal{A}$ where $\tilde{C}_i \as \tilde{C}_1' \| \ldots \| \tilde{C}'_{\lceil m/\ell \rceil}$. 
			\item For all other periods $j \ne i$, $\mathcal{B}_1$ samples uniformly random keys 
			$K_j \Ra \{0,1\}^\kappa$ and computes $\tilde{C}_j' \as \prf_1(K_j,ctr+j \bmod 2^\kappa)$ and returns the final ciphertext $C_i$ to \A~following \fencenc.
			\item All $\prf_2$ evaluations in \fprg~are computed honestly using uniformly random keys.
		\end{enumerate}
		
		If $\mathcal{O}$ is instantiated with $\prf_1(K^\star,\cdot)$, the simulation is identical to hybrid $H_{i-1}^{(1)}$; if $\mathcal{O}$ is random, the view matches $H_i^{(1)}$. Therefore, $\Delta_i^{(1)}$ is bounded by the distinguishing advantage of $\mathcal{B}_1$ (i.e., $\Delta_i^{(1)} \le \adv_{\prf_1}(t_{\prf_1}, q_{\prf_1})$).
	%
	Then, we sum $\Delta_i^{(1)}$ for all $i\in \{1,\dots,n\}$ to obtain:  
	\[
	\big|\Pr[H_0^{(1)}(\A)=1]-\Pr[H_n^{(1)}(\A) =1]\big| 	 = | \sum_{i=1}^n \Delta_i^{(1)}  | 
	\le \sum_{i=1}^n | \Delta_i^{(1)}  | 
	\le  n\cdot\adv_{\prf_1}(t_{\prf_1},q_{\prf_1}).
	\]
	
	\paragraph{Game $\mathbf{H_i^{(2)}}$ (Replacing $\prf_2$ in \fprg).}
	Let $H_0^{(2)}$  be $H_n^{(1)}$, for each $i=1,\ldots,n$, let $H_i^{(2)}$ be identical to $H_{i-1}^{(2)}$
	except that $i^\text{th}$ key update
	$(S_{i+2},K_{i+1})\leftarrow\fprgupd(S_{i+1})$
	is replaced by random generation ($S_{i+2} \Ra \{0,1\}^\kappa,K_{i+1} \as \{0,1\}^\kappa$).
	We denote $\Delta_i^{(2)} =\big|\Pr[H_{i-1}^{(2)}(\mathcal A)=1]-\Pr[H_i^{(2)}(\mathcal A)=1]\big|$.
	
	
		The distinguisher $\B_2$ is given oracle access to a function $\mathcal O(\cdot)$ which is either $\prf_2(K^\star,\cdot)$ where $K^\star \as \{0,1\}^\kappa$ or a random function.
		It simulates the \findcpa~experiment for $\mathcal A$ as follows:
		
		\begin{enumerate}[1.]
			\item For the target update $i$, $\B_1$ computes
			$(S_{i+2},K_{i+1})$ by querying
			$\mathcal O(S_{i+1}\|0)$ and $\mathcal O(S_{i+1}\|1)$.
			\item For $j\neq i$, it computes $(S_{i+2},K_{i+1}) \as (\prf_2(S_{i+1}, 0), \prf_2(S_{i+1}, 1))$ where $S_{i+1} \Ra \{0,1\}^\kappa$.
		\end{enumerate}
		
		If $\mathcal O$ is instantiated with $\prf_2(S_{i+1},.)$, the simulation is identical to $H_{i-1}^{(2)}$. Otherwise, if random, the view matches $H_i^{(2)}$.
		Therefore, the distinguishing advantage of $\B_2$ equals $\Delta_i^{(2)}$ (i.e., $\Delta_i^{(2)} \le 2\cdot\adv_{\prf_2}(t_{\prf_2},q_{\prf_2})$). 
	%
	Then, we sum $\Delta_i^{(2)}$ for all $i \in \{1,\dots,n \}$ to obtain: 
	\[
	\big|\Pr[H_n^{(1)} (\A)]-\Pr[H_n^{(2)} (\A)]\big| = | \sum_{i=1}^n \Delta_i^{(2)}  | \le \sum_{i=1}^n | \Delta_i^{(2)}  | 
	\le 2 \! \cdot n\cdot\adv_{\prf_2}(t_{\prf_2},q_{\prf_2}).
	\]
	
	\paragraph{Final game $\mathbf{H_n^{(2)}}$.}
	In $H_n^{(2)}$, every $\prf_{1,2}$~outputs is replaced by uniform random values. Hence, ciphertexts are generated by a fresh one-time pad, and all keys are uniformly random values. Therefore, the adversary’s success probability $\Pr [H_n^{(2)}(\A)=1]$ equals the random guess probability $1/2$. 
	Notice that $\advfenc(t_\fse,q_\fse) = \max_\A |\Pr[ \text{Expt}^{\findcpa}_{\fenc}(\A) =1 ] - 1/2|$ which is equivalent to $\advfenc(t_\fse,q_\fse)  =\max_\A |\Pr[H_0^{(1)} (\A)=1] - \Pr[H_n^{(2)}(\A)=1] |$. That is, 
	{\small
		\[
		\advfenc(t_\fse,q_\fse) \le  |\Pr[H_0^{(1)}(\A)=1]-\Pr[H_n^{(1)}(\A) =1]| +  |\Pr[H_n^{(1)}(\A)=1]-\Pr[H_n^{(2)}(\A) =1]|
		\]
	}
	
	By combining the results of the hybrid games, we obtain the desired bound and conclude the proof: 
	\[
	\adv_{\fenc}^{\findcpa}(t_{\fenc},q_{\fenc})
	\le
	n\cdot\adv_{\prf_1}(t_{\prf_1},q_{\prf_1})
	+
	2 \! \cdot n\cdot\adv_{\prf_2}(t_{\prf_2},q_{\prf_2}).
	\]
\end{proof}

\label{subsec:secproof:famac}

\begin{theorem}
	Let $\famac$ be the forward-secure aggregate \mac~scheme in Fig.~4, instantiated from an $\varepsilon$-almost-universal hash $\uhash$ and a pseudo-random function 
	$\prf_2$.  
	Let $\mathcal{A}$ be any adversary running in time $t_{\famac}$ and making at most $q_{\famac}$ 
	signing queries, where the epoch size is $b$ and the tag space $\mathcal{T}$ is of size $\tau$.  
	Then there exists a distinguisher $\mathcal{B}_{\prf}$ against $\prf_2$ such that
	\[
	\adv_{\famac}^{\faeucma}(t_{\famac},q_{\famac})
	\;\le\;
	n\cdot\adv_{\prf_2}(t_{\prf},q_{\prf})
	\;+\; q_{\famac}\, \cdot b\, \cdot \varepsilon
	\;+\; \tfrac{q_{\famac}\,\cdot b}{2^\tau},
	\]
	where $q_{\prf}=O(q_{\famac}\,\cdot b)$ and $t_{\prf}=t_{\famac}+O(q_{\famac}\,\cdot b)$.
\end{theorem}

\begin{proof}
	We prove the theorem by a sequence of hybrid games, as follows.
	
	\paragraph{Game $\mathbf{H_0}$ (Real experiment).}
	This is the real $\faeucma$ experiment where each offline mask 
	$\tilde{\sigma}_i \as \prf_2(K_i^{\prf_2},i)$ and \uhash~term $\bar{\sigma}_i \as \uhash(K_i^{\uhash},M_i)$ are aggregated additively ($\sigma_i \as \tilde{\sigma}_i + \bar{\sigma}_i$).  
	
	\paragraph{Game $\mathbf{H_i^{(1)}}$ (Replace $\prf_2$ output for epoch $i$).}
	For $i=1,\ldots,n$, we define $H_i^{(1)}$ as $H_{i-1}^{(1)}$ except that in epoch $i$
	all offline tags $\tilde{\sigma}_i$ derived from $\prf_2$ are replaced by uniformly random elements of $\mathcal{T}$.  
	We define $\Delta_i^{(1)} 
	= \big| \Pr[H_{i-1}^{(1)}(\mathcal{A})=1]-\Pr[H_i^{(1)}(\mathcal{A})=1]
	\big|$ to be the difference in the advantage of \A~in two consecutive games $H_{i-1}^{(1)}$ and $H_i^{(1)}$. 
	The distinguisher $\B_{\prf}$ is given access to an oracle $\mathcal{O}(\cdot)$, which is either $\prf_2(K_i,\cdot)$ or a truly random function.  $\B_1$ chooses the target epoch $i$ and simulates \faeucma~for \A~as follows:
	\begin{enumerate}[1.]
		\item For epoch $i$, $\B_1$ queries $\bar{\sigma}_i \as O(M_i) $ and outputs $\sigma_i \as \bar{\sigma}_i + \tilde{\sigma}_i$ to \A~where $\tilde{\sigma}_i \as \uhash(K_i^\uhash, i)$ . 
		
		\item For $j\neq i$, $\B_1$ computes $\sigma_i$ following the \famac~signing algorithm using uniformly random keys. 
	\end{enumerate}
	
	If $\mathcal{O}$ is instantiated with $\prf_2$, the simulation equals $H_{i-1}^{(1)}$;  
	if random, the view matches $H_i^{(1)}$.  Hence $\Delta_i^{(1)}=\adv_{\prf_2}(t_{\prf},q_{\prf})$.  
	Then, we sum $\Delta_i^{(1)}$ over $i \in \{1,\ldots, n\}$ to obtain:
	\[
	\big|
	\Pr[H_{0}(\A)=1] - \Pr[H_{n}^{(1)}(\A)=1]
	\big|
	\;\le\;
	n\cdot\adv_{\prf_2}(t_{\prf},q_{\prf}).
	\tag{1}
	\]
	
	\paragraph{Game $\mathbf{H_n^{(1)}}$ (All offline masks random).}
	In this game, all $\tilde{\sigma}_i$ are chosen uniformly at random, therefore independent of $\mathcal{A}$’s view.
	For each authentication tag $\sigma_i = \tilde{\sigma}_i + \bar{\sigma}_i$,
	a successful forgery requires $\sigma' - \bar{\sigma}' = \tilde{\sigma}$.
	Since $\tilde{\sigma}$ is uniformly random from $\mathcal{T}$, 
	this equality holds with probability at most $\frac{1}{\tau}$ for any $(M',\sigma')$ pair.
	The collisions in $\uhash$ outputs provide another path to forgery.
	Since $\uhash$ is an $\varepsilon$-almost-universal hash function, any pair of distinct messages
	collide with probability at most $\varepsilon$.  
	By a union bound over $q_{\famac} \cdot b$ total attempts:
	\[
	\Pr[H_n^{(1)}(\A)=1]
	\;\le\;
	q_{\famac}\cdot b \cdot \varepsilon
	\;+\;
	\tfrac{q_{\famac}\, \cdot b}{\tau}.
	\tag{2}
	\]
	
	\paragraph{Combining (1) and (2).}
	We obtain the desired reduction and therefore concludes the proof.
	\[
	\adv_{\famac}^\faeucma(\A) 
	\;\le\;
	n\cdot\adv_{\prf_2}(t_{\prf},q_{\prf})
	\;+\;
	q_{\famac}\,\cdot b \cdot \,\varepsilon
	\;+\;
	\tfrac{q_{\famac}\,\cdot b}{\tau},
	\]
\end{proof}

\begin{theorem}
	Let $\dmd$ be a forward-secure and aggregate authenticated encryption (\faae) scheme constructed from a CTR-based forward-secure encryption (\fse) and a universal forward-secure aggregate MAC (\famac), following Fig.\ref{alg:fenc} and Fig.\ref{alg:umac}, respectively. 
	
	Let $\A$ be an adversary running in time $t$ and making at most $q$ queries to the $\dmdae$ oracle. Then, there exist adversaries $\A_{\fenc}$ and $\A_{\famac}$ against the underlying \fenc~and \famac~such that
	\[
	\adv_{\dmd}^{\findcpa}(t,q)
	\le b \cdot \adv_{\fenc}^{\findcpa}(t_\fenc,q_\fenc),
	\]
	\[
	\adv_{\dmd}^{\faeucma}(t,q)
	\le \adv_{\famac}^{\faeucma}(t_\famac,q_\famac),
	\]
	where $t_\fenc,t_\famac \le t$, and $q_\fenc = q_\famac = q + \mathcal{O}(1)$.
\end{theorem}

\begin{proof}
	We prove that the security of $\dmd$ under the \findcpa~and \faeucma~notions reduces to that of its underlying components: the forward-secure encryption $\fenc$ and the forward-secure aggregate message authentication $\famac$.
	
	\paragraph{(1) Confidentiality Reduction to $\fenc$.}
	In the following, we construct an adversary $\A^{\fenc}$ that leverages $\A$ to break the \findcpa~security of $\fenc$ scheme.
	
	\noindent\textit{Setup.} 
	$\A^{\fenc}$ begins by sampling an initial \famac~key $K^{\famac}_1 \Ra \{0,1\}^{|K^{\famac}_1|}$ and internally runs $\A$.
	
	\noindent\textit{Find.} 
	Whenever $\A$ issues a left-or-right encryption query $(M_0, M_1)$ to $\authenclr(K_i, \cdot)$, $\A_{\fenc}$ forwards it to its own $\fencenclr(K_i^\fenc, \cdot)$ oracle and receives back a ciphertext $C_b$. 
	It then computes $\sigma_i \gets \famacsign(K_i^{\famac}, C_i^b)$ and updates the \famac~key as $K_{i+1}^{\famac} \gets \famacupd(K_i^{\famac})$. 
	Finally, $\A^{\fenc}$ returns $(C_i^b, \sigma_i)$ to $\A$.
	
	\noindent\textit{Guess.} 
	Whenever $\A$ outputs its final bit $b'$, $\A^{\fenc}$ also outputs $b'$.
	
	$\A^{\fenc}$ simulates the $\dmd$ encryption oracle for $\A$. 
	$\A^{\fenc}$ succeeds in the \findcpa~experiment of $\fenc$ whenever $\A$ succeeds in guesses the left-or-right bit (i.e., $b=b'$). 
	Therefore,
	\[
	\adv_{\dmd}^{\findcpa}(t,q) \le b \cdot \adv_{\fenc}^{\findcpa}(t_\fenc,q_\fenc).
	\]
	
	\paragraph{(2) Integrity Reduction to $\famac$.}
	We now construct an adversary $\A^{\famac}$ that leverages $\A$ to break the \faeucma~security of $\famac$:
	
	\noindent\textit{Setup.} 
	$\A^{\famac}$ generates an initial $\fenc$ key $K_1^{\fenc} \Ra \{0,1\}^{|K_1^{\fenc}|}$ to locally simulate the encryption component. It then launches $\A$ as a subroutine.
	
	\noindent\textit{Find.} 
	When $\A$ queries $\dmdae(K_i,\cdot)$ on message $M_i$, $\A^{\famac}$ computes 
	$C_i \gets \fseenc(K_i^{\fenc}, M_i)$ 
	and then queries its own $\famacsign$ oracle on $C_i$ to obtain $\sigma_i$. 
	It updates the encryption key $K_{i+1}^{\fenc} \gets \fencupd(K_i^{\fenc})$ and returns $(C_i, \sigma_i)$ to $\A$.
	
	\noindent\textit{Forge.} 
	Eventually, $\A$ proceeds to the forge phase and outputs a non-trivial forgery $(\vect{C_{i^*}}, \sigma_{i^*, i^*+b-1})$, following the $\faeucma$ experiment in Sec.~\ref{subsec:secmodels}. 
	$\A^{\famac}$ also receives the current secret key $K_i^{\famac}$ (i.e., after a break-in). 
	It verifies $\A$’s forgery using its verification oracle
	(i.e., $
	b \gets \famacaver(K_i^{\famac}, \vect{C_{i^*}}, \sigma_{i^*, i^*+b-1}).
	$).
	If $b = 1$, $\A^{\famac}$ decrypts each ciphertext as
	(i.e., $
	\vect{M_{i^*}} \gets \{ \fencdec(K_j^{\fenc}, C_j) \}_{j=i^*}^{i^*+b-1}.
	$)
	Finally, $\A^{\famac}$ outputs success if $\A$’s forgery passes verification and is non-trivial (i.e., not a recombination of previously authenticated ciphertexts).
	
	$\A^{\famac}$ wins the $\faeucma$ experiment of $\famac$ whenever $\A$ forges a valid ciphertext–tag pair under $\dmd$. Hence, we obtain the security bound which concludes the proof.
	\[
	\adv_{\dmd}^{\faeucma}(t,q) \le \adv_{\famac}^{\faeucma}(t_\famac,q_\famac).
	\]
	
\end{proof}

%% file: conclusion.tex
\section{CONCLUSION} \label{sec:conclusion}

We presented \dmd, a provable secure symmetric-key Forward-secure and Aggregate Authenticated Encryption (\faae) framework tailored for the stringent constraints of contemporary IoT platforms. \dmd~addresses foundational limitations in existing \aee~and \mac~constructions by synergizing efficient forward-secure key evolution, OO cryptographic optimization, and compact tag aggregation. \dmd~guarantees breach-resilient confidentiality and authenticity, while ensuring near-optimal online latency across heterogeneous hardware classes. Through a comprehensive experimental evaluation on 64-bit ARM Cortex-A72, 32-bit ARM Cortex-M4, and 8-bit AVR microcontrollers, we demonstrated that Diamond’s instantiations deliver substantial reductions in offline preprocessing, order-of-magnitude improvements in online \authenc~and verification throughput, and significantly lower end-to-end latency for large telemetry batches compared to \faae~baselines and NIST lightweight \aee~candidates. \dmd’s modular design, extensibility to diverse universal \amac~constructions, and compatibility with commodity toolchains make it a practical, deployment-ready \faae~primitive for mission-critical IoT domains. To encourage reproducibility and adoption, we release our open-source full-fledged implementation.
